\documentclass[twocolumn,tighten]{aastex62}
\hypersetup{linkcolor=red,citecolor=blue,filecolor=cyan,urlcolor=magenta}
\usepackage{amsmath}
\usepackage{natbib}
\usepackage{color}
\usepackage{hyperref}
\usepackage{url}

\newcommand{\mh}{\mbox{$\mathrm{[M/H]}$}}
\newcommand{\am}{\mbox{$[\alpha/\mathrm{M}]$}}
\newcommand{\afe}{\mbox{$[\alpha/\mathrm{Fe}]$}}

\newcommand{\epsilonMH}{\mbox{$\epsilon_{\mathrm{[M/H]}}$}}
\newcommand{\epsilonAM}{\mbox{$\epsilon_{[\alpha/\mathrm{M}]}$}}

\newcommand{\deltaMH}{\mbox{$\delta_{\mathrm{[M/H]}}$}}
\newcommand{\deltaAM}{\mbox{$\delta_{[\alpha/\mathrm{M}]}$}}

\newcommand{\logg}{\mbox{log~{\it g}}}

\newcommand{\teff}{\mbox{$T_{\rm eff}$}}

\def\vector#1{\mbox{\boldmath $#1$}}

\newcommand{\kpc}{\ensuremath{\,\mathrm{kpc}}}

\newcommand{\kms}{\ensuremath{\,\mathrm{km\ s}^{-1}}}

\newcommand{\eq}[1]{\begin{align}#1\end{align}}

\newcommand{\revise}[1]{{#1}}

\begin{document}

\title{%
Metallicity and $\alpha$-abundance for 48 million stars 
in low-extinction regions in the Milky Way 
}

\shorttitle{Gaia XP spectra and chemistry}
\shortauthors{K. Hattori}

\author[0000-0001-6924-8862]{Kohei~Hattori}
\affiliation{National Astronomical Observatory of Japan, 2-21-1 Osawa, Mitaka, Tokyo 181-8588, Japan}
\affiliation{The Institute of Statistical Mathematics, 10-3 Midoricho, Tachikawa, Tokyo 190-8562, Japan}
\affiliation{Department of Astronomy, University of Michigan,
1085 S.\ University Avenue, Ann Arbor, MI 48109, USA}
\email{Email:\ khattori@ism.ac.jp}

\begin{abstract}

We estimate ([M/H], [$\alpha$/M]) for 48 million giants and dwarfs 
in low-dust extinction region 
from the Gaia DR3 XP spectra  
by using tree-based machine-learning models 
trained on APOGEE DR17 and metal-poor star sample \revise{from} Li et al. 
The root mean square error of our estimation 
is 0.0890 dex 
for [M/H] and 0.0436 dex 
for [$\alpha$/M], 
when we evaluate our models \revise{on} the test data 
that are not used in training the models. 
Because the training data is dominated by giants, 
our estimation is most reliable for giants. 
The high-[$\alpha$/M] stars  
and low-[$\alpha$/M] stars  
selected by our ([M/H], [$\alpha$/M]) 
show different kinematical properties 
for giants and low-temperature dwarfs. 
We further investigate 
how our machine-learning models 
extract information on ([M/H], [$\alpha$/M]). 
Intriguingly, we find that our models seem to 
extract information on [$\alpha$/M] 
from  Na D lines (589 nm) and Mg I line (516 nm). 
This result is understandable 
given the observed correlation between Na and Mg abundances in the literature. 
The catalog of 
([M/H], [$\alpha$/M]) 
as well as their associated uncertainties 
is publicly available online.

\end{abstract}
\keywords{ 
Spectroscopy (1558), 
Stellar abundances (1577), 
Milky Way disk (1050), 
Milky Way stellar halo (1060),
Astroinformatics (78)
}

\section{Introduction}

\subsection{Spectroscopic surveys and chemical abundances}

The chemical abundances of stars imprint the chemistry of the gas from which they were formed. 
Determining the stellar chemical abundances is, therefore, 
an important task in understanding the history of the Milky Way. 
Many spectroscopic surveys 
conducted by ground-based telescopes 
(e.g.,  
RAVE, \citealt{Steinmetz2006AJ....132.1645S}; 
SEGUE, \citealt{Yanny2009AJ....137.4377Y}; 
APOGEE, \citealt{Majewski2016AN....337..863M}; 
LAMOST, \citealt{Zhao2012RAA....12..723Z}; 
GALAH, \citealt{DeSilva2015MNRAS.449.2604D}; 
Gaia-ESO, \citealt{Gilmore2012Msngr.147...25G})  
have obtained 
chemical abundances of millions of stars in the Milky Way 
and other local group galaxies 
based on 
the high-resolution ($\lambda/\Delta\lambda \gtrsim 20,000$) 
or low/medium-resolution ($\lambda/\Delta\lambda \simeq 2,000$--$20,000$) 
spectra taken from many years of observations.

\subsection{Data mining of Gaia XP spectra}

Recently, 
Gaia Data Release 3 (DR3) provided extremely low-resolution BP/RP spectra 
(hereafter XP spectra following the convention)
with $\lambda/\Delta\lambda \sim 50$-$100$ 
for 219 million stars \citep{DeAngeli2023A&A...674A...2D, Montegriffo2023A&A...674A...3M, 
Gaia2016A&A...595A...1G, 
Gaia2021A&A...649A...1G, 
GaiaCollaboration2023A&A...674A...1G}. 
Although the XP spectra have much lower spectral resolution 
than the spectra of other spectroscopic surveys, 
the size and homogeneity of this data set 
has opened a new possibility to investigate 
the stellar atmospheric parameters 
(such as $\teff$, $\logg$) 
and stellar chemical abundances 
(such as \mh, \am, [C/Fe], [N/Fe], [O/Fe]) 
for many stars by using machine learning (ML) models.

\subsubsection{Before the publication of Gaia DR3}
\label{sec:beforeDR3}

Before the launch of Gaia, 
\cite{BailerJones2010MNRAS.403...96B} 
explored a theoretical framework to infer the stellar atmospheric parameters, 
dust extinction, and metallicity [Fe/H] of stars with Gaia XP spectra. 
The author used synthetic spectra to test the method, 
but did not try to estimate \afe\ from Gaia XP spectra. 

When Gaia XP spectra were analyzed internally by the Gaia team, 
\cite{Gavel2021A&A...656A..93G} 
used an ML algorithm called ExtraTrees 
to try to estimate \afe\ from synthetic Gaia XP spectra 
and the actual, unpublished Gaia XP spectra. 
When they used a model which was trained on synthetic spectra, 
they were able to estimate \afe\ of synthetic spectra, 
but were unable to estimate \afe\ of Gaia XP spectra. 
When they used a model which was trained on Gaia XP spectra, 
they were able to estimate \afe\ from Gaia XP spectra for cool stars ($\teff < 5000 \;\mathrm{K}$ or equivalently, $1.1<(G_\mathrm{BP}-G_\mathrm{RP})$), 
but were unable to estimate \afe\ from Gaia XP-like synthetic spectra. 
Their finding indicates that estimating \afe\ 
is difficult for stars with 
$5000 \;\mathrm{K} < \teff$ or $(G_\mathrm{BP}-G_\mathrm{RP})<1.1$. 
Based on these findings, they inferred that their models appeared to estimate \afe\ (of cool stars) by using indirect correlations between \afe\ and other stellar properties including but not limited to the metallicity [Fe/H].

\cite{Witten2022MNRAS.516.3254W} investigated 
the information content of Gaia XP spectra 
and showed that the Gaia XP-like synthetic spectra 
of Solar-metallicity stars with $G=16$ 
do not have enough information to reliably estimate 
the \afe\ abundance, unless $\teff < 5000 \; \mathrm{K}$ is satisfied, 
supporting the result in \cite{Gavel2021A&A...656A..93G}.

The results of \cite{Gavel2021A&A...656A..93G} and 
\cite{Witten2022MNRAS.516.3254W} 
indicate that extracting information on \am\ 
from Gaia XP spectra is challenging. 
However, we dare to tackle this problem in this paper 
due to the following reasons. 
\revise{First},
\cite{Gavel2021A&A...656A..93G} 
used only ExtraTrees algorithm. 
We note that other ML algorithms 
might be more suited to extract \am\ information. 
\revise{Second},
\cite{Witten2022MNRAS.516.3254W} 
pointed out the difficulty of estimating \afe\ 
based on their analysis of 
synthetic spectra of stars with $G=16$ (see their Fig.~9), 
and it is unclear whether 
the same argument is valid for brighter stars. 
For example, their Fig.~3 indicated that 
the uncertainty in [Fe/H] for a star with $G=13$ 
is $\sim 4$ times smaller than that for a star with $G=16$. 
Thus, 
the uncertainty in \afe\ \revise{for} a $G=13$ star 
might be a few times smaller than that \revise{for} a $G=16$ star. 
In such a case, 
trying to infer \afe\ (or \am) is still meaningful.\footnote{
\revise{
It may be worth noting that Gaia XP spectra are of prime quality for stars with $G \simeq 13$-$14$. 
For brighter stars, the instrumental settings are changed (to avoid saturation) and the calibration of Gaia XP spectra becomes tedious 
\citep{Montegriffo2023A&A...674A...3M}, 
which result in lower signal-to-noise ratio of Gaia XP spectra. 
}
}

\subsubsection{After the publication of Gaia DR3}

After the publication of Gaia DR3, 
many authors have used Gaia XP spectra 
to infer the stellar properties, 
including the chemical abundances 
\citep{Rix2022ApJ...941...45R, Andrae2023ApJS..267....8A, Zhang2023MNRAS.524.1855Z, Bellazzini2023A&A...674A.194B, Sanders2023MNRAS.521.2745S, Yao2024MNRAS.52710937Y, Martin2023arXiv230801344M, Xylakis-Dornbusch2024arXiv240308454X}.

\cite{Rix2022ApJ...941...45R} did a pioneering work 
to estimate $(\teff, \logg, \mh)$ of 2 million giants 
within 30$^\circ$ of the Galactic center. 
They used the reliable stellar parameters 
from APOGEE DR17 \citep{Abdurrouf2022ApJS..259...35A} 
and trained the ML model called XGboost to infer the stellar parameters. 
Importantly, they used external catalog (AllWISE photometry; \citealt{Cutri2014yCat.2328....0C}) 
to aid their model to infer the stellar parameters 
for stars with non-negligible dust extinction. 
Following the success of \cite{Rix2022ApJ...941...45R}, 
\cite{Andrae2023ApJS..267....8A} 
estimated $(\teff, \logg, \mh)$ of 175 million stars across the sky. 
Their mean stellar parameter precision is 50 K in \teff,  0.08 dex in \logg, and 0.1 dex in \mh. 
In their work, they used a metal-poor star sample in \cite{Li2022ApJ...931..147L} 
in addition to the APOGEE sample 
so that their training data cover a wide range of \mh, 
which enhanced the reliability of \mh\ at the low-\mh\ region. 
\cite{Zhang2023MNRAS.524.1855Z} 
used a ML model based on a neural network 
to infer the stellar parameters $(\teff, \logg, \mh)$, 
parallax, 
and the dust extinction 
for 220 million stars with Gaia XP spectra. 
They used the LAMOST DR8 sample \citep{Wang2022ApJS..259...51W} 
as the training data, 
because it covers a wider parameter space than APOGEE sample. 
An interesting part of their model is that 
their model can predict the XP spectra 
given the input stellar parameters. 
More recently, 
\cite{Leung2023MNRAS.tmp.2896L} 
used a modern ML models based on Transformer model 
(which is also used in Large Language Models) 
and showed a way to infer the stellar labels 
$(\teff, \logg, \mh)$ with high accuracy. 
Also, there is an attempt to produce a generative model 
of Gaia XP spectra \citep{Laroche2023arXiv230706378L}, 
which may be useful to interpret 
the observed Gaia XP spectra directly, 
without estimating the stellar chemical abundances.

We note that 
some authors inferred the 
chemical abundances of stars from the XP spectra 
but did not publish the process as the main part of their paper 
\citep{Belokurov2023MNRAS.518.6200B, Chandra2023ApJ...951...26C}.

\subsection{Scope of this paper: Estimation of \am\ from Gaia XP spectra}

The previous works mentioned above have estimated \mh\ (or [Fe/H]), 
but none of them estimated \am\ (or [Mg/Fe]). 
This is partly because of the difficulty of estimating \am, 
as presented by 
\cite{Gavel2021A&A...656A..93G} 
and 
\cite{Witten2022MNRAS.516.3254W}. 
However, 
as mentioned earlier (Section \ref{sec:beforeDR3})
we dare to tackle this problem 
using a classical ML model 
that is different from the ExtraTrees model 
used in \cite{Gavel2021A&A...656A..93G}. 
The simpleness of our models enables 
us to investigate how the ML models 
infer the chemical abundances from the XP spectra.

While we were preparing our manuscript, 
we noticed that independent groups had 
tackled this problem with the same aim of estimating \am\ 
\citep{Guiglion2024A&A...682A...9G, Li2023arXiv230914294L}. 
We note that these papers used a modern ML architecture, 
while we use a classical ML model, 
and thus the interpretation of the ML model 
is more straightforward in this paper. 
\cite{Guiglion2024A&A...682A...9G} 
used the medium-resolution spectra from Gaia RVS 
(with $\lambda/\Delta \lambda \simeq 10,000$), 
in addition to the XP spectra. 
\revise{
Because \cite{Guiglion2024A&A...682A...9G} combined XP and RVS spectra, their method to derive \am\ was applicable to only $\sim 1$ million stars. 
}
Also, \cite{Li2023arXiv230914294L} 
focused on giant stars,
while we do not specifically restrict our analysis to giants.

\subsection{Structure of this paper}

Our primary goal in this paper is 
to estimate \mh\ and \am\ from Gaia XP spectra 
with classical ML models. 
This paper is organized as follows. 
In Section \ref{sec:data}, we introduce the data set we used. 
In Section \ref{sec:model}, we describe how we construct our ML models. 
In Section \ref{sec:validation}, we validated our estimation of \mh\ and \am. 
In Section \ref{sec:results}, we describe the catalog of (\mh, \am) derived from our analysis. 
In Section \ref{sec:interpret}, we try to interpret how our ML models infer (\mh, \am), 
by quantifying which wavelength ranges of the XP spectra are important. 
In Section \ref{sec:conclusion}, we summarize this paper.

\begin{figure*}
\centering 
\includegraphics[width=0.98\textwidth ]{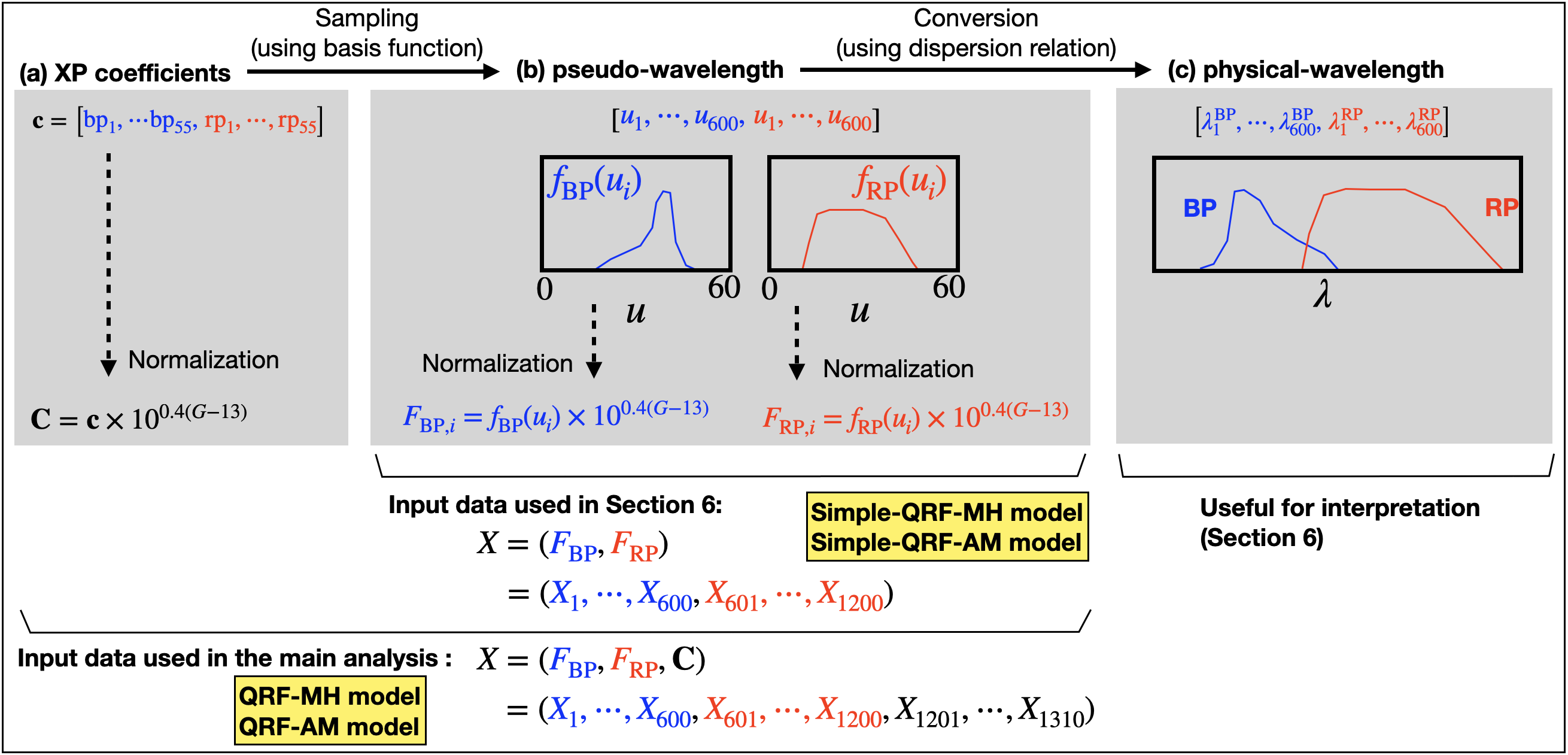}
\caption{Schematic diagram of the data used in this paper. 
}
\label{fig:XPschematic}
\end{figure*}

\section{Data} \label{sec:data}

Here we describe the data sets used in this paper.

\subsection{Sample stars with Gaia XP spectra}
\label{sec:XPdata}

From Gaia DR3, 
we select 219 million stars for which 
the mean BP/RP spectrum (so-called XP spectrum) is available 
(\texttt{has\_xp\_continuous}=\texttt{True} in Gaia DR3). 
For each of these stars, Gaia provides 
110 coefficients 
\eq{
\vector{c} = (bp_1, \cdots, bp_{55}, rp_1, \cdots, rp_{55})
}
that represent the mean BP/RP spectra. 
Gaia also provides the uncertainty in $\vector{c}$ and their correlations, 
but we neglect these quantities to simplify our analysis.

As shown in Figure \ref{fig:XPschematic}, 
the coefficients $\vector{c}$ can be converted into 
the mean BP and RP spectra $(f_\mathrm{BP}, f_\mathrm{RP})$ in terms of 
the pseudo-wavelength $u$, 
\revise{
by using a Python package \texttt{GaiaXPy}
}
\citep{Montegriffo2023A&A...674A...3M}.\footnote{
\revise{
\url{https://gaia-dpci.github.io/GaiaXPy-website/}
}
}
Here, $0<u<60$ is a dimensionless quantity 
and it is differently defined for BP domain and RP domain. 
For example, $u=30$ defined in the BP domain 
and that in RP domain correspond to different wavelengths. 
By using $\vector{c}$, we can express 
$f_\mathrm{BP} = f_\mathrm{BP}(u)$ and 
$f_\mathrm{RP} = f_\mathrm{RP}(u)$ for each star. 
In our analysis, 
we define equally-spaced 600 points 
$u_i = 60 \times (i-1)/599$, $(i=1,\cdots 600)$ 
within the range of $0\leq u \leq 60$. 
\revise{
(This is the default sampling scheme in \texttt{GaiaXPy}. 
It turns out that this sampling scheme is good enough to 
interpret which part of the spectrum is informative in extracting \mh\ and \am, 
as we will discuss in Section \ref{sec:interpret}.)
}
We evaluate $f_\mathrm{BP}(u_i)$ and $f_\mathrm{RP}(u_i)$, 
and save these 1200 quantities for each star.

In the main analysis of this paper, 
we use the 1310-dimensional information 
consisting of the 1200-dimensional flux data 
($f_\mathrm{BP}(u_i), f_\mathrm{RP}(u_i)$) 
and the 110-dimensional coefficients $\vector{c}$.

\begin{figure*}
\centering 
\includegraphics[width=0.95\textwidth ]{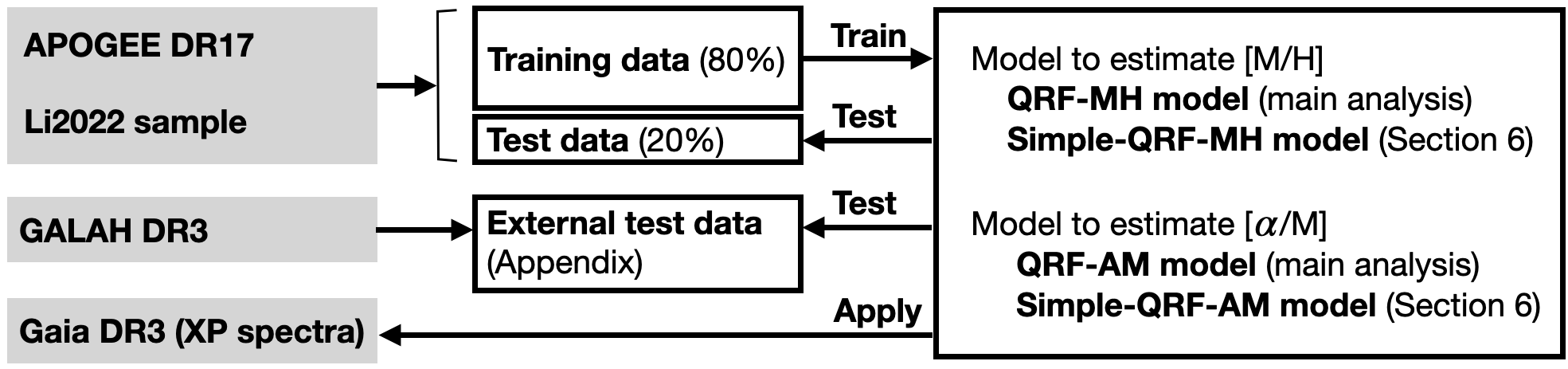}
\caption{
Summary of the data sets and models. 
We use the combined data from APOGEE DR17 and \cite{Li2022ApJ...931..147L} 
to train and test our models 
(QRF-MH model to estimate \mh, 
and QRF-AM model to estimate \am). 
Because we have small number of dwarf stars in the training data, 
we additionally use GALAH DR3 data as the external data 
to test our models. 
}
\label{fig:data_and_model}
\end{figure*}

\begin{figure*}
\centering 
\includegraphics[width=0.24\textwidth ]{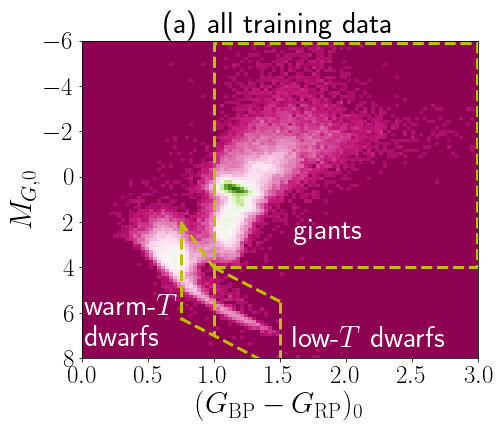} 
\includegraphics[width=0.24\textwidth ]{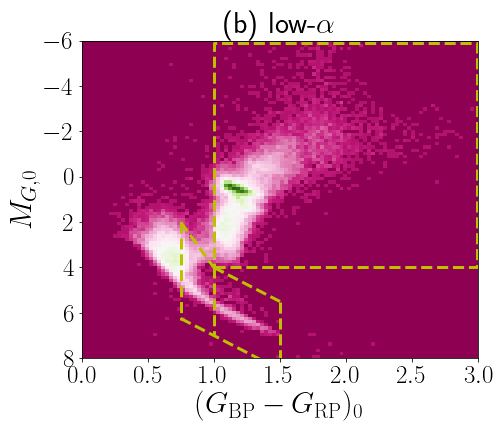} 
\includegraphics[width=0.24\textwidth ]{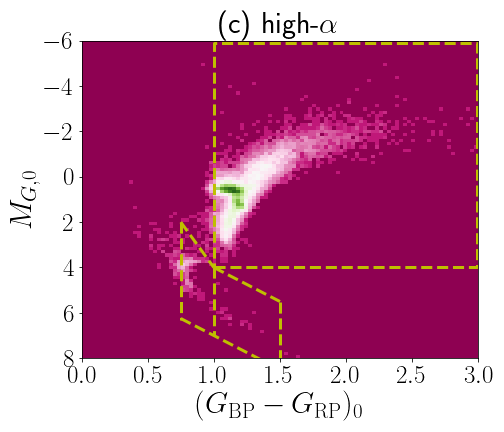} 
\includegraphics[width=0.24\textwidth ]{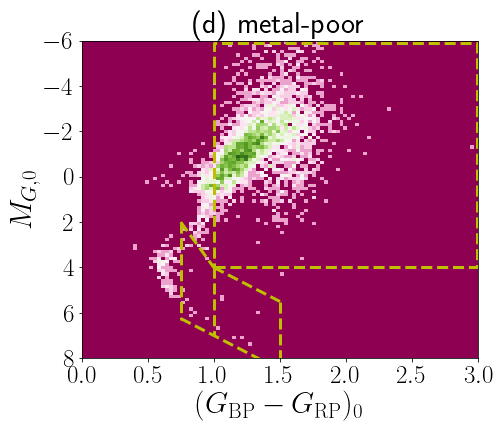} \\
\includegraphics[width=0.24\textwidth ]{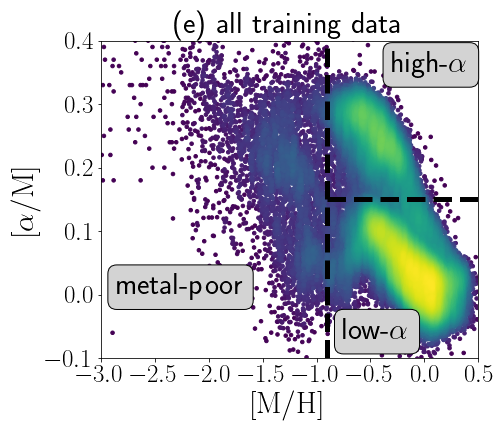} 
\includegraphics[width=0.24\textwidth ]{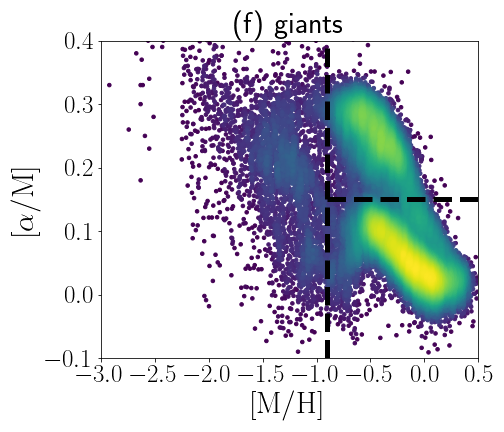} 
\includegraphics[width=0.24\textwidth ]{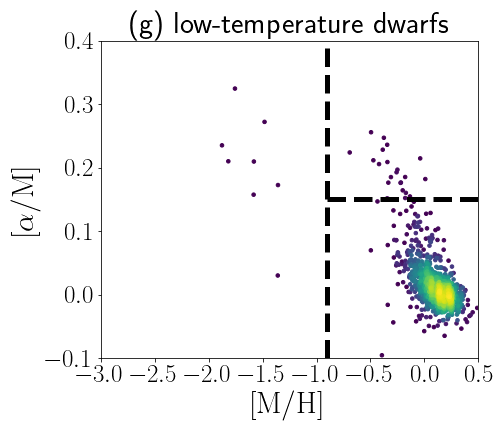} 
\includegraphics[width=0.24\textwidth ]{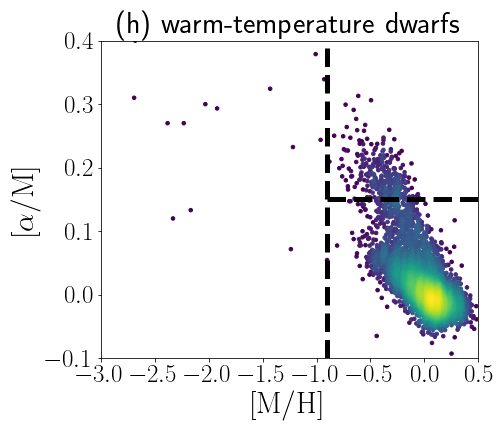} 
\caption{
The distribution of stars in our training and test data 
in the photometric space (top row) and 
in the chemical abundance space (bottom row). 
}
\label{fig:train_different_MHAM_CMD}
\end{figure*}

\subsection{Training/test data}
\label{sec:training_data}

As shown in Fig.~\ref{fig:data_and_model},
we construct a data set of stars with known chemistry 
taken from either 
APOGEE DR17 data set 
\citep{Abdurrouf2022ApJS..259...35A}  
or the data set in \cite{Li2022ApJ...931..147L}. 
We first introduce these data sets in 
Sections \ref{sec:APOGEE}-\ref{sec:Li2022}. 
Then we illustrate the properties of the combined catalog 
in Section \ref{sec:APOGEE_and_Li2022}.

\subsubsection{APOGEE DR17}
\label{sec:APOGEE}

From APOGEE DR17 \citep{Abdurrouf2022ApJS..259...35A}, 
we select 115,249 stars 
from the value-added catalog (\texttt{allStar-dr17-synspec\_rev1.fits})
that satisfy the following criteria:
(i) Both \mh\ and \am\ are available;
(ii) 0-6th, 8-10th, and 16-27th bits in the ASPCAPFLAG are zero;
(iii) Third and fourth flags of STARFLAG are zero 
(avoiding objects with a very bright neighbor and 
objects with low signal-to-noise ratio); 
(iv) Fourth flag of EXTRATARG is zero 
(adopting the spectrum with the highest signal-to-noise ratio for stars with multiple observations); 
(v) Color excess $E(B-V)<0.1$ is satisfied 
(We multiply a scaling factor 0.86 to the color excess from the \citealt{Schlegel1998ApJ...500..525S} dust map, following \citealt{Schlafly2011ApJ...737..103S}.); 
(vi) Galactic latitude satisfies $|b|>5^\circ$; and 
(vii) XP spectra coefficients are available from Gaia DR3. 
The criteria (i)-(iv) are designed to select 
clean sample of APOGEE stars with reliable 
chemical abundances, 
while maintaining the sample size. 
The criteria (v) and (vi) aim to 
exclude stars 
whose XP spectra are significantly altered by the reddening 
\citep{BailerJones2010MNRAS.403...96B}.

\subsubsection{Metal-poor star catalog from \cite{Li2022ApJ...931..147L}}
\label{sec:Li2022}

Since there are few APOGEE stars with $\mh \lesssim -2.5$, 
we also use the metal-poor stars in \cite{Li2022ApJ...931..147L} 
in addition to the APOGEE stars (see \citealt{Andrae2023ApJS..267....8A}). 
For brevity, we call this metal-poor sample as Li2022 sample. 
Among 385 stars in the Li2022 sample, 
we select 299 stars that satisfy the following criteria:
(i) Both [Fe/H] and [Mg/Fe] are determined from Subaru observation; 
(ii) Color excess $E(B-V)<0.1$ is satisfied 
(adopting the value in Table 1 of \citealt{Aoki2022ApJ...931..146A}, 
which is based on the 3D dust map of \citealt{Green2018MNRAS.478..651G}); 
(iii) XP spectra coefficients are available from Gaia DR3. 

In the following analysis, 
we regard [Fe/H] as \mh\ and [Mg/Fe] as \am, 
and merge the Li2022 sample to the APOGEE sample. 
We note that there is a star that is included 
in both APOGEE DR17 and Li2022, 
and we have confirmed for this star 
that its chemical abundances from these catalogs 
agree well with each other. 
For this duplicated star, 
we use the chemical abundances from \cite{Li2022ApJ...931..147L} 
and discard the corresponding entry in the APOGEE catalog.

\subsubsection{Combined data of APOGEE and Li2022 sample}
\label{sec:APOGEE_and_Li2022}

The combined sample of APOGEE and Li2022  
consists of 115,547 unique stars. 
The sample covers a wide range 
in metallicity ($-4.4 \leq \mh \leq + 0.6$) 
and in $\alpha$-abundance ($-0.3 \leq \am \leq + 1.1$). 
We randomly divide these stars into 
the training data set (80\%) 
and 
the test data set (20\%). 
The training data will be used in Section \ref{sec:model} 
to construct models to infer $(\mh, \am)$ from the XP spectra. 
The test data will be used to assess the performance of the models.

To understand the limitation of our models, 
it is important to understand 
the distribution of these stars 
in the color-magnitude diagram (CMD)\footnote{
The dust reddening is corrected as 
$G_0 = G - 2.497 E(B-V)$ 
and 
$(G_\mathrm{BP}-G_\mathrm{RP})_0 = (G_\mathrm{BP}-G_\mathrm{RP}) - 1.321 E(B-V)$ \citep{Wang2019ApJ...877..116W}. 
In computing the absolute magnitude $M_{G,0}$, 
we correct the parallax $\varpi$ by 
$\varpi_\mathrm{corrected} = \varpi - (-0.017 \; \mathrm{mas})$ 
with a constant zero-point offset \citep{Lindegren2021A&A...649A...4L}. 
}
and in the chemical space. 
As a basis for discussion, 
we photometrically define three types of stars 
(giants, low-temperature dwarfs, and warm-temperature dwarfs\footnote{
\label{footnote:CMD}
Before we photometrically select giants and dwarfs, 
we select stars with good parallax measurements 
(\texttt{parallax\_over\_error>5} in Gaia DR3). 
After this parallax selection, 
giants are defined by 
$1 < (G_\mathrm{BP}-G_\mathrm{RP})_0 < 3$ and 
$M_{G,0}<4$. 
Low-temperature dwarfs are defined by
$1 < (G_\mathrm{BP}-G_\mathrm{RP})_0 < 1.5$ and 
$3 \times (G_\mathrm{BP}-G_\mathrm{RP})_0 + 1 < M_{G,0}$ 
$< 3 \times (G_\mathrm{BP}-G_\mathrm{RP})_0 + 4$. 
Warm-temperature dwarfs are defined by
$0.75 < (G_\mathrm{BP}-G_\mathrm{RP})_0 < 1$ and 
$8 \times (G_\mathrm{BP}-G_\mathrm{RP})_0 - 4 < M_{G,0}$
$<3 \times (G_\mathrm{BP}-G_\mathrm{RP})_0 + 4$. 
We note that this parallax selection of the training data 
is done only for illustration in this Section. 
In other words, 
when we train our ML models, 
we use all the stars in the training data, 
including stars with poor parallax measurements. 
};
see Fig.~\ref{fig:train_different_MHAM_CMD}(a)). 
Also, we chemically define three groups of stars 
in the (\mh, \am)-space 
(low-$\alpha$, high-$\alpha$, and metal-poor stars\footnote{
Low-$\alpha$ stars (`thin' disk stars) are defined by
$-0.9 < \mh$ and $\am < 0.15$.
High-$\alpha$ stars (`thick' disk stars) are defined by
$-0.9 < \mh$ and $0.15 < \am$.
Metal-poor stars (`halo' stars) are defined by 
$\mh < -0.9$. 
}
(see Fig.~\ref{fig:train_different_MHAM_CMD}(e)).

The stars in the training data are mostly giants and dwarfs
(see Fig.~\ref{fig:train_different_MHAM_CMD}(a)). 
In particular, 
giants are the dominant members of the training data 
independent of the stellar chemistry 
(see Figs.~\ref{fig:train_different_MHAM_CMD}(b)-(d)). 
Most of the low/warm-temperature dwarfs in the training data 
belong to the low-$\alpha$ subsample 
(see Figs.~\ref{fig:train_different_MHAM_CMD}(g)(h)). 
Due to the paucity of the metal-poor dwarfs in the training data, 
we expect that it would be difficult for our models 
to infer (\mh,\am) of metal-poor dwarfs, 
which is confirmed in our later analysis.

\revise{
A careful reader may notice that the low-$\alpha$ sequence 
in Fig.~\ref{fig:train_different_MHAM_CMD}(e) appears broader than 
in Fig.~\ref{fig:train_different_MHAM_CMD}(f). 
This is due to a systematic error in the \am\ values from the APOGEE data, where the \am\ of dwarfs is slightly lower than that of giants by approximately 0.03 dex.
}
\footnote{
\revise{
This small but systematic difference in \am\ between giants and dwarfs  
propagates to our model predictions 
(e.g., see Figs.~\ref{fig:chemistry_dep}(a)(e)).
}
}

\subsection{External test data: GALAH DR3}
\label{sec:GALAH_DR3_data}

To aid the test process of our models, 
we also prepare an external test data set 
taken from GALAH DR3. 
From GALAH DR3, 
we obtain the value-added catalog 
(\texttt{GALAH\_DR3\_main\_allstar\_v2.fits}; 
\citealt{Buder2021MNRAS.506..150B}). 
We extract stars that satisfy the following criteria:
(i) Both [Fe/H] and \afe\ are determined with high reliability 
(\texttt{snr\_c3\_iraf>30}, 
\texttt{flag\_sp=0},  
\texttt{flag\_fe\_h=0}, 
\texttt{flag\_alpha\_fe=0});  
(ii) Color excess $E(B-V)<0.1$ is satisfied 
(adopting the procedure in Section \ref{sec:APOGEE}); 
(iii) XP spectra coefficients are available from Gaia DR3. 
(iv) Stars are not included in the combined training/test data of APOGEE DR17 and Li2022.

This catalog contains 178,814 stars, 
including $\sim8,000$ low-temperature dwarfs (with $\teff \lesssim 5000$ K). 
In contrast, the combined training/test data  
(from APOGEE DR17 and Li2022) 
only include $\sim 1000$ such dwarfs. 
Thus, this external test data are useful to 
test the performance of our models for low-temperature dwarfs.

\section{Construction of the model} \label{sec:model}

\subsection{Quantile Regression Forests (QRF)} 
\label{sec:QRF}

In this paper, 
we use an ML method named 
`Quantile Regression Forests (QRF)' 
to infer (\mh, \am) from the XP spectra. 
The QRF is a non-parametric tree-based ensemble method, 
which is a generalized version of the Random Forest (RF) method. 
Since the QRF is not as widely used as the RF, 
we first describe the difference between RF and QRF.

Let us denote the input data as $X$ (in our case, the information of the XP spectra) 
and the target label to be estimated as $Y$ (in our case, \mh\ or \am). 
In the RF, the algorithm uses multiple decision trees and 
tries to find the expected value of $Y$ given the data $X$, $E(Y \mid X)$. 
In the QRF, the algorithm also uses multiple decision trees, but it tries to find the probability distribution of $Y$ given the data $X$. 
Namely, for a given quantile $q$ in the range $0\leq q \leq 1$, the QRF estimates the value of $y$ such that the conditional probability 
$P(Y \leq y \mid X) = q$ is satisfied. 
In some implementations, 
the RF and QRF use the same set of decision trees. 
In such a case, while the QRF uses all the information of the decision trees to compute the probability distribution of $P(Y \mid X)$, 
the RF returns only the summary statistics of the decision trees $E(Y \mid X)$.

We adopt a publicly available QRF package 
\texttt{quantile-\hspace{0mm}forest} 
(\url{https://pypi.org/project/quantile-forest/}; 
\citealt{Johnson2024}). 
In this implementation, 
the QRF can estimate only a scalar target $Y$. 
Therefore, we construct a QRF model for \mh\ and \am\ separately. 
Namely, we only estimate 
$P(\mh \mid X)$
and 
$P(\am \mid X)$ for each star, 
and we do not estimate 
$P(\mh, \am \mid X)$ 
for each star.

\subsection{Input data for the QRF models}

Since the XP coefficients are designed to represent the observed stellar flux as a function of the (psuedo-) wavelength, 
we first normalize the coefficients by multiplying a factor 
accounting for the apparent magnitude of the star. 
For each star, 
we use its $G$-band magnitude $G$ and 
compute (i) the normalized coefficient vector 
\eq{
\vector{C} = \vector{c} \times 10^{0.4(G-13)},
}
(ii) the normalized mean BP and RP spectra ($\vector{F}_\mathrm{BP}$ and $\vector{F}_\mathrm{RP}$), 
for which the $i$th element is given by
\eq{
F_{\mathrm{BP},i} &= f_\mathrm{BP}(u_i) \times 10^{0.4(G-13)}, \\
F_{\mathrm{RP},i} &= f_\mathrm{RP}(u_i) \times 10^{0.4(G-13)}. 
}
In the main analysis of this paper, 
we use the 1310-dimensional vector  
\eq{
X = (\vector{C}, \vector{F}_\mathrm{BP}, \vector{F}_\mathrm{RP})
} 
as the input of our QRF model. 
In Section \ref{sec:interpret}, 
we use QRF models for which we only use 1200-dimensional information 
$X=(\vector{F}_\mathrm{BP}, \vector{F}_\mathrm{RP})$ 
to enhance the interpretability of the model.

\subsection{Training the QRF models}

We train the QRF model
by using the input vector $X$ and the target label $Y$ (either \mh\ or \am) 
of the APOGEE and Li2022 samples. 
We use 80\% of the data as the training data ($N_\mathrm{train} = 92,437$ stars) 
and 20\% of the data as the test data ($N_\mathrm{test} = 23,110$ stars).

We set two hyperparameters, 
\texttt{n\_estimators=100} and 
\texttt{max\_features=0.5}. 
The former hyperparameter, \texttt{n\_estimators}, 
is the number of trees in the forest, 
and our choice of 100 is the default value. 
Our choice of the latter hyperparameter, \texttt{max\_features=0.5}, 
is in between two widely adopted values of 
0.333 \citep{hastie01statisticallearning}  
and 
1.0 (default value). 
When the QRF makes the decision tree, 
the algorithm keeps splitting the feature space 
(in our case, 1310-dimensional space of $X$). 
Our choice of \texttt{max\_features=0.5} 
means that the algorithm uses randomly selected 50\% of the entire dimension (i.e., 655 dimensions) 
when looking for the best split. 
We have confirmed that the choice of the hyperparameters 
do not strongly affect the performance of the model. 
As explained in Section \ref{sec:QRF}, 
we separately build a model for estimating \mh\ (hereafter QRF-MH model)
and 
a model for estimating \am\ (hereafter QRF-AM model).

Since the QRF model can evaluate $P(Y \mid X)$, 
it can predict the label $Y$ for any specified percentile $q$. 
For brevity, we denote  
$y^{\mathrm{pred}, q}_{k}$ 
as the $q$th percentile value of the predicted value of the label $Y$ for $k$th star.\footnote{
Throughout this paper, 
we choose nine percentile points, 
$q=2.5$, 10, 16, 25, 50, 75, 84, 90, and 97.5, 
and evaluate the labels at these percentiles. 
} 
(For example, $y^{\mathrm{pred}, 50}_{k}$ corresponds to 
the predicted median value of the label $Y$ for $k$th star). 
Also, we denote 
$y_k^{\mathrm{spec}}$ to denote the spectroscopically determined (true) label 
for $k$th star. 
We separately train the QRF-MH and QRF-AM models such that the loss function 
\eq{\label{eq:loss}
loss = 
\frac{1}{N_\mathrm{train}} 
\sum_{k \in \mathrm{Training \; data}} 
{\left( y^{\mathrm{pred}, 50}_{k} - y^{\mathrm{spec}}_{k} \right)}^2
}
is minimized.

\begin{figure*}
\centering 
\includegraphics[width=0.48\textwidth ]{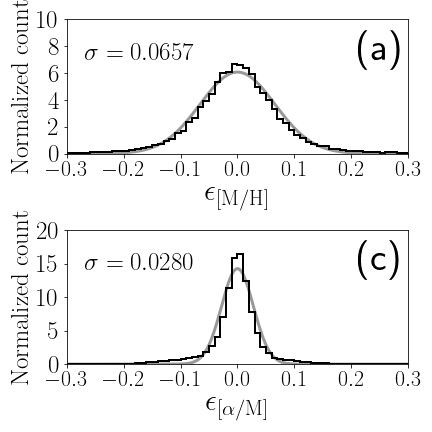} 
\includegraphics[width=0.48\textwidth ]{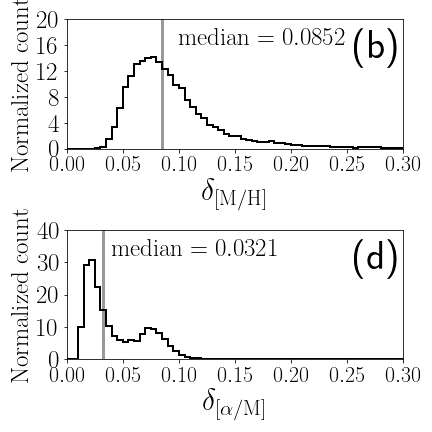}
\caption{
The performance of our models. 
(a) 
The histogram of \epsilonMH. 
We also show a fitted Gaussian distribution 
and the standard deviation $\sigma$ (shown on the panel). 
(b) 
The histogram of \deltaMH, 
which has a single-peaked distribution 
with the median value shown on the panel. 
(c)
The same as panel (a) but for the histogram of \epsilonAM. 
(d)
The histogram of \deltaAM, 
which has a double-peaked distribution 
with the median value shown on the panel. 
}
\label{fig:histogram_epsilon_delta}
\end{figure*}

\section{Validation of the model using the test data} \label{sec:validation}

\subsection{Root mean squared error}
\label{sec:rmse}

Once the QRF model is trained, 
we apply the model to the test data 
to measure the performance of the model. 
We define the root mean squared error (RMSE) for $\mh$ and $\am$ by 
\eq{\label{eq:RMSE}
\mathrm{RMSE} = 
\sqrt{
\frac{1}{N_\mathrm{test}} 
\sum_{k \in \mathrm{Test \; data}} 
{\left( y^{\mathrm{pred}, 50}_{k} - y^{\mathrm{spec}}_{k} \right)}^2
}. 
}
The results are summarized in Table~\ref{table:rmse}. 
The overall performance of our models are represented by the 
RMSE values for the entire test data, 
which is 0.0890 dex for \mh\ and 0.0436 dex for \am. 
It is reassuring that 
these numbers are comparable to 
the corresponding numbers in the literature 
which used more modern ML architecture 
(e.g., \citealt{Leung2023MNRAS.tmp.2896L,Li2023arXiv230914294L}). 
We note that we are only focusing on 
stars in low-dust extinction regions with $E(B-V)<0.1$, 
while other authors also try to estimate the chemistry 
for stars with strong dust extinction \citep{Rix2022ApJ...941...45R, Andrae2023ApJS..267....8A, Li2023arXiv230914294L}.

\revise{
As a simple benchmark, we construct a dummy regressor model 
which makes predictions using a simple rule. 
In our case, the `prediction' of the dummy regressor 
is simply a shuffled value from the test data. 
The RMSE values for this dummy regressor (shown in Table~\ref{table:rmse}) 
provide an estimate of the scatter in the labels within the test data. Therefore, the ratio between the RMSE of our models and that of the dummy regressor offers insights into the predictive power of our models. 
}

We note that the performance of the models are most reliable 
for meal-rich stars with $\mh>-1$, 
and the performance becomes worse at lower \mh. 
However, even at $\mh<-1$, 
the RMSE value for \am\ is as small as $0.0726$ dex, 
which is still useful to understand the distribution of stars 
in the (\mh, \am)-space.

We also divide our samples into giants and low/warm-temperature dwarfs 
and evaluate the RMSE values. 
We see that the performance for giants are most reliable, 
which is naturally understandable because 
our training data are dominated by giants.

We also evaluate the RMSE values 
by using the external test data taken from GALAH DR3. 
For the external test data, 
we see a similar trend to the results for the test data 
comprised of APOGEE and Li2022 sample.

\subsection{Accuracy and precision for the entire test data}
\label{sec:accuracy_precision}

To further evaluate our models, 
we introduce two types of quantities. 
First, we introduce the difference $\epsilon$ between 
our predicted chemical abundance and the `true' chemical abundance:
\eq{
\epsilon_{\mathrm{[M/H]}} &= \mh^{\mathrm{pred},50} - \mh^\mathrm{spec}, \label{eq:epsilonMH}\\
\epsilon_{[\alpha/\mathrm{M}]} &= \am^{\mathrm{pred},50} - \am^\mathrm{spec}, \label{eq:epsilonAM}
}
for each star in the test data. 
The quantity $\epsilon$ satisfies 
$-\infty < \epsilon < \infty$, 
and it reflects the accuracy of our models. 

Secondly, we introduce $\delta$ defined as
\eq{
\delta_{\mathrm{[M/H]}} &= \frac{1}{2} \left( \mh^{\mathrm{pred},84} - \mh^{\mathrm{pred},16} \right), \\
\delta_{[\alpha/\mathrm{M}]} &= \frac{1}{2} \left( \am^{\mathrm{pred},84} - \am^{\mathrm{pred},16} \right), 
}
for each star in the test data. 
The quantity $\delta$ satisfies $0 \leq \delta$, 
and it reflects the precision of the labels for each star. 
We note that 
we can evaluate $\epsilon$ only for test data, 
while 
we can evaluate $\delta$ for any stars with XP spectra. 
\revise{
Since we do not account for the uncertainties in Gaia XP spectra
(see Section \ref{sec:XPdata}),
these measurement uncertainties do not propagate into the uncertainty of our model predictions. 
In other words, the uncertainty in our models 
(represented by $\delta_{\mathrm{[M/H]}}$ or $\delta_{\mathrm{[\alpha/M]}}$)
arises solely from the randomness in the ensemble of trees.
}

\subsubsection{Accuracy and precision for the entire test data}

Figs.~\ref{fig:histogram_epsilon_delta}(a) 
and \ref{fig:histogram_epsilon_delta}(c)
show the histogram of \epsilonMH\ and \epsilonAM\ for the entire test data. 
The distributions of \epsilonMH\ or \epsilonAM\
are almost symmetric around zero, 
which indicates that there is no obvious systematic error 
in inferring (\mh,\am) in our models. 
The distributions of \epsilonMH\ or \epsilonAM\ 
can be approximated by 
Gaussian distributions, with standard deviations 
$\sigma=0.0657$ and \revise{$0.0280$}, respectively. 
These numbers are 30\%--40\% smaller than 
the corresponding RMSE values in Table.~\ref{table:rmse}, 
which indicates that the histograms in 
Figs.~\ref{fig:histogram_epsilon_delta}(a) 
and \ref{fig:histogram_epsilon_delta}(c)
have a fatter tail than a Gaussian distribution.

Figs.~\ref{fig:histogram_epsilon_delta}(b) 
and \ref{fig:histogram_epsilon_delta}(d) 
show the histogram of \deltaMH\ and \deltaAM\ for the entire test data. 
We note 
that the typical precision of our models 
can be inferred from the median 
of the distribution of \deltaMH\ or \deltaAM. 
The median value of $\deltaMH$ ($=0.0852$ dex) 
is roughly consistent with the RMSE (0.0890 dex) of our model for \mh. 
Interestingly, the histogram of $\deltaAM$ 
has two peaks, 
at 0.02 dex (first peak) and at 0.07 dex (secondary peak). 
As we will see in Fig.~\ref{fig:test_data_am}(a),
the first peak corresponds to a case when high-$\alpha$ (or low-$\alpha$) stars 
are correctly assigned high (or low) $\alpha$ abundances, 
while the second peak corresponds to a case in which high-$\alpha$ (or low-$\alpha$) stars 
are incorrectly assigned low (or high) $\alpha$ abundances.

\revise{
The performance of our models shown in Fig.~\ref{fig:histogram_epsilon_delta} 
reflects the overall performance across the entire test dataset.
However, our test data spans a wide magnitude range ($8 \lesssim G \lesssim 18$), 
and the model performance varies with $G$, 
being better for brighter stars and worse for fainter ones. 
For example, the annotated summary statistics in Fig.~\ref{fig:histogram_epsilon_delta} 
($\sigma$ in panels (a) and (c), median value in (b) and (d)) 
change by a factor of $\sim 2$ (i.e., deteriorate) if we only use stars with $15<G$ (the faintest 4\% of the test data). 
Conversely, 
these summary statistics 
change by a factor of 0.75-0.95 (i.e., improve) if we only use stars with $G<10$ (the brightest 6\% of the test data). 
}

\startlongtable
\begin{deluxetable*}{l l l l l}
\tablecaption{
Root mean squared error 
evaluated by applying our QRF-MH and QRF-AM models to the test data
\label{table:rmse}}
\tablewidth{0pt}
\tabletypesize{\scriptsize}
\tablehead{
\colhead{Sample$^\mathrm{(a)}$} &
\colhead{RMSE in \mh}  &
\colhead{RMSE in \am}  &
\colhead{RMSE in \mh}  &
\colhead{RMSE in \am} \\
\colhead{} &
\colhead{of our model} &
\colhead{of our model} &
\colhead{of a benchmark model$^\mathrm{(b)}$} &
\colhead{of a benchmark model$^\mathrm{(b)}$} 
}
\startdata
\hline
{Test data} & {} & {} & {} & {}\\
\hline
{\hspace{4mm}Entire test data}      & {0.0890} & {0.0436} & {0.547} & {0.140} \\
{\hspace{4mm}Stars with [M/H]$>-1$} & {0.0768} & {0.0416} & {0.478} & {0.138} \\
{\hspace{4mm}Stars with [M/H]$<-1$} & {0.220}  & {0.0726} & {1.30} & {0.173} \\
\hline
{\hspace{4mm}Giants}                 & {0.0767} & {0.0449} & {0.514} & {0.140} \\
{\hspace{4mm}Giants with [M/H]$>-1$} & {0.0724} & {0.0441} & {0.471} & {0.139} \\
{\hspace{4mm}Giants with [M/H]$<-1$} & {0.162}  & {0.0652} & {1.27} & {0.185}  \\
\hline
{\hspace{4mm}Low-$T$ dwarfs}                 & {0.110} & {0.0332} & {0.618} & {0.128} \\
{\hspace{4mm}Low-$T$ dwarfs with [M/H]$>-1$} & {0.0842}& {0.0332} & {0.612} & {0.127} \\
{\hspace{4mm}Low-$T$ dwarfs with [M/H]$<-1$} & {1.14}  & {0.0290} & {1.46} & {0.306} \\
\hline
{\hspace{4mm}Warm-$T$ dwarfs}                 & {0.0784} & {0.0356} & {0.525} & {0.138} \\
{\hspace{4mm}Warm-$T$ dwarfs with [M/H]$>-1$} & {0.0755} & {0.0354} & {0.517} & {0.137} \\
{\hspace{4mm}Warm-$T$ dwarfs with [M/H]$<-1$} & {0.373}  & {0.0790} & {1.65} & {0.221}  \\
\hline
{External test data (GALAH)} & {} & {} \\
\hline
{\hspace{4mm}Entire external test data} & {0.132}  & {0.0817} & {0.443} & {0.151} \\
{\hspace{4mm}Stars with [Fe/H]$>-1$}    & {0.125} & {0.0781} & {0.409} & {0.147} \\
{\hspace{4mm}Stars with [Fe/H]$<-1$}    & {0.343}  & {0.192} & {1.31} & {0.271} \\
\hline
{\hspace{4mm}Giants}                  & {0.124} & {0.0777} & {0.472} & {0.164} \\
{\hspace{4mm}Giants with [Fe/H]$>-1$} & {0.118} & {0.0754} & {0.411} & {0.162} \\
{\hspace{4mm}Giants with [Fe/H]$<-1$} & {0.230} & {0.126}  & {1.31} & {0.212} \\
\hline
{\hspace{4mm}Low-$T$ dwarfs}                  & {0.174}  & {0.0795} & {0.432} & {0.135} \\
{\hspace{4mm}Low-$T$ dwarfs with [Fe/H]$>-1$} & {0.168} & {0.0780} & {0.429} & {0.134} \\
{\hspace{4mm}Low-$T$ dwarfs with [Fe/H]$<-1$} & {0.924}  & {0.330} & {1.08} & {0.429}  \\
\hline
{\hspace{4mm}Warm-$T$ dwarfs}                  & {0.133} & {0.0812} & {0.433} & {0.141} \\
{\hspace{4mm}Warm-$T$ dwarfs with [Fe/H]$>-1$} & {0.129}& {0.0798}  & {0.429} & {0.140} \\
{\hspace{4mm}Warm-$T$ dwarfs with [Fe/H]$<-1$} & {0.602} & {0.280}  & {1.16} & {0.347} \\
\hline
\enddata
\tablecomments{
(a) Giants and dwarfs are selected from the test data 
in the same manner as in Fig.~\ref{fig:train_different_MHAM_CMD} 
(see footnote \ref{footnote:CMD}). 
\revise{
(b) We use a dummy regressor (a baseline model that randomly shuffles the test data to predict labels) as a benchmark. 
}
}
\end{deluxetable*}

\begin{figure*}
\centering 
\includegraphics[width=0.48\textwidth ]{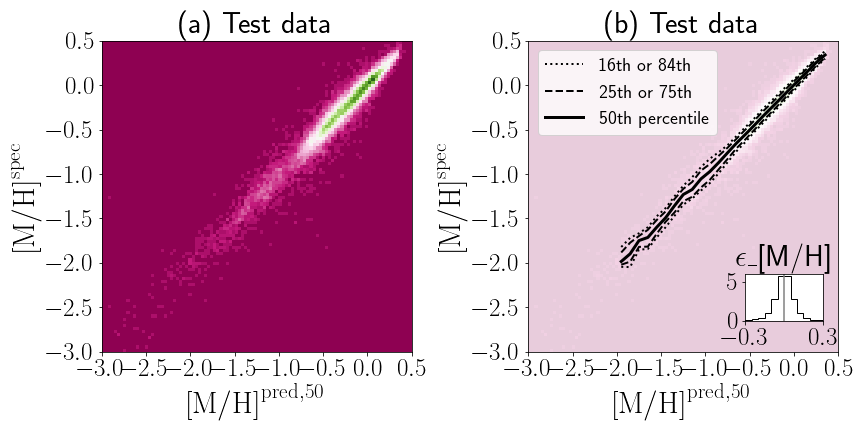} 
\includegraphics[width=0.48\textwidth ]{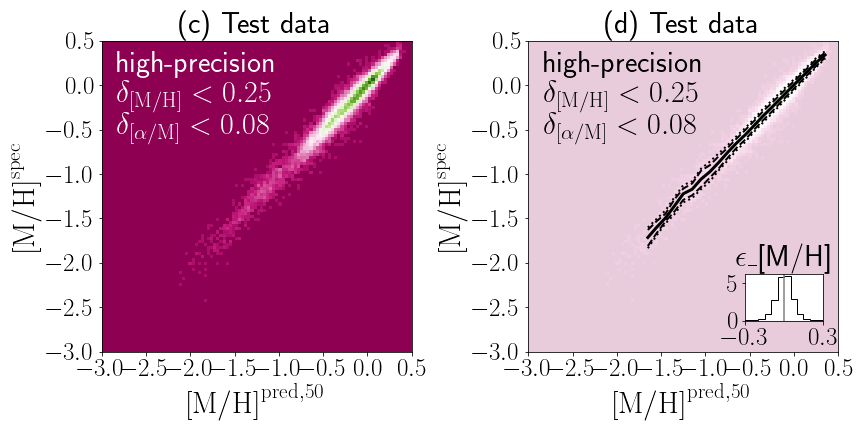} \\
\includegraphics[width=0.48\textwidth ]{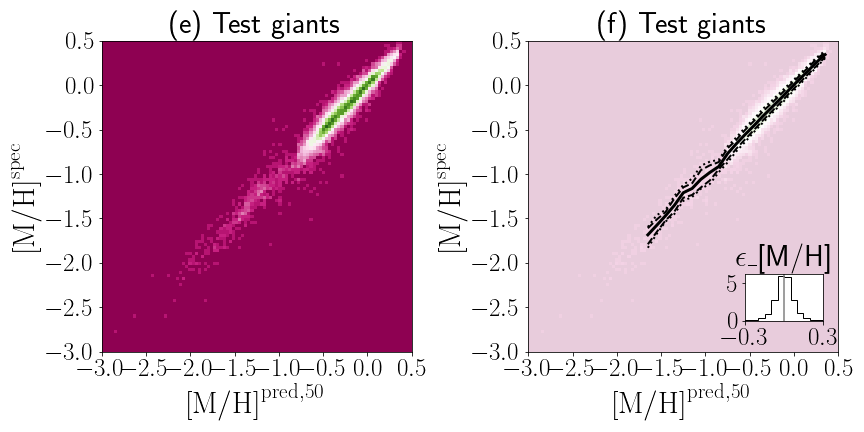} 
\includegraphics[width=0.48\textwidth ]{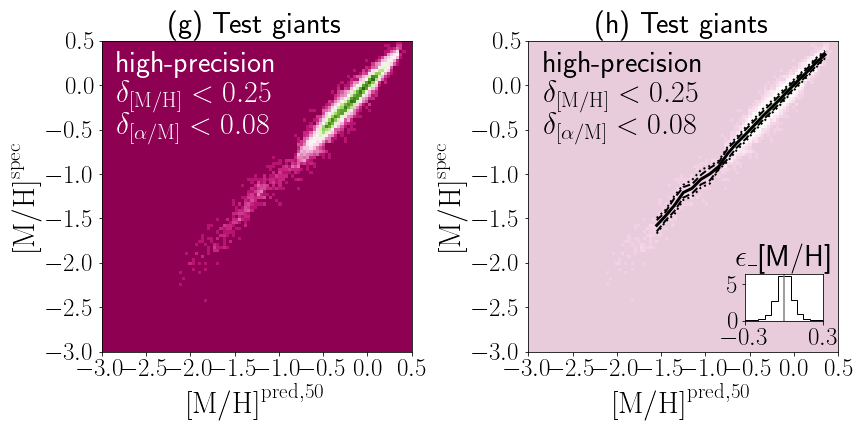} \\
\includegraphics[width=0.48\textwidth ]{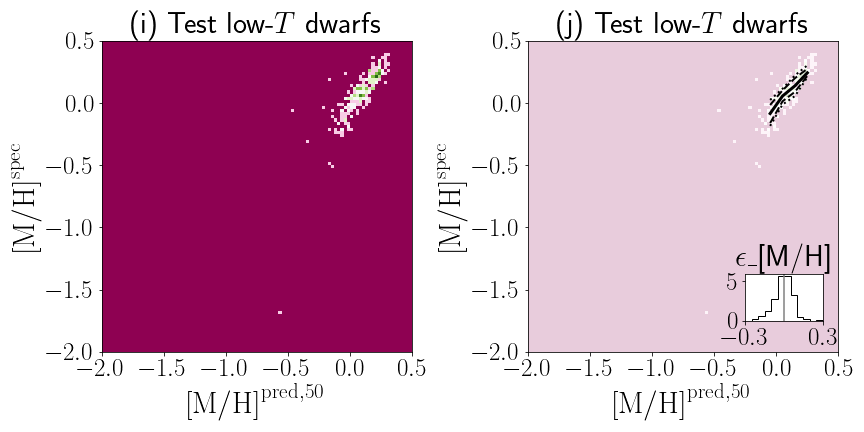} 
\includegraphics[width=0.48\textwidth ]{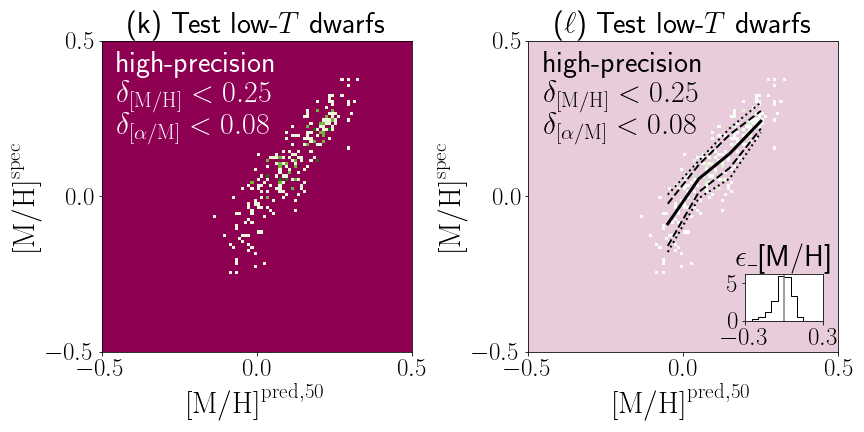} \\
\includegraphics[width=0.48\textwidth ]{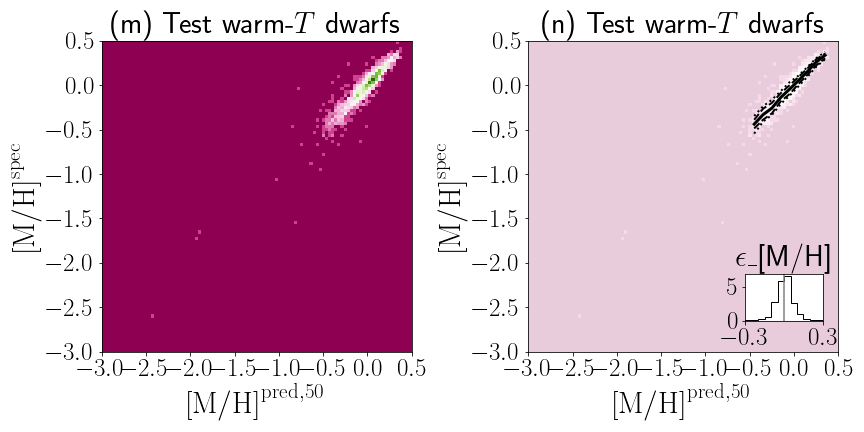} 
\includegraphics[width=0.48\textwidth ]{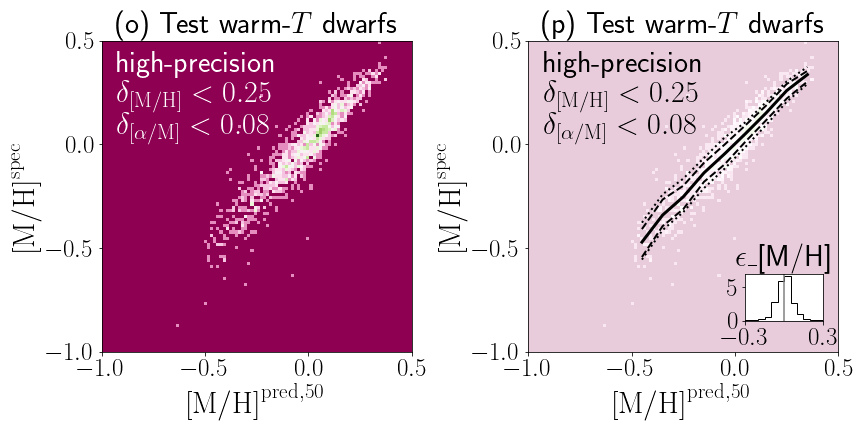} 
\caption{
The performance of the QRF-MH model (our model for estimating \mh) 
applied to the test data. 
In all panels, we show 
the predicted abundance (horizontal axis) and spectroscopic abundance (vertical axis). 
From the top to bottom rows, 
we show the distribution of 
stars in the entire test data (top row), 
giants in the test data (second row), 
low-temperature dwarfs in the test data (third row), and 
warm-temperature dwarfs in the test data (bottom row). 
The first column from the left displays the density distribution. 
The second column shows the (2.5th, 16th, 50th, 84th, 97.5th) percentile values of the spectroscopic abundance 
as a function of the predicted abundance. 
We also show the normalized histogram of 
the deviation of our predicted abundance relative to the 
spectroscopic abundance $(\epsilonMH =\mh^\mathrm{pred,50} - \mh^\mathrm{spec})$. 
The third and fourth columns are the same as the first and second columns, respectively, 
but selecting only high-precision subsample with 
$\deltaMH <0.25$ and $\deltaAM < 0.08$. 
In some panels, 
the ranges of the horizontal and vertical axes are different 
from other panels for clarity. 
}
\label{fig:test_data_mh}
\end{figure*}

\begin{figure*}
\centering 
\includegraphics[width=0.48\textwidth ]{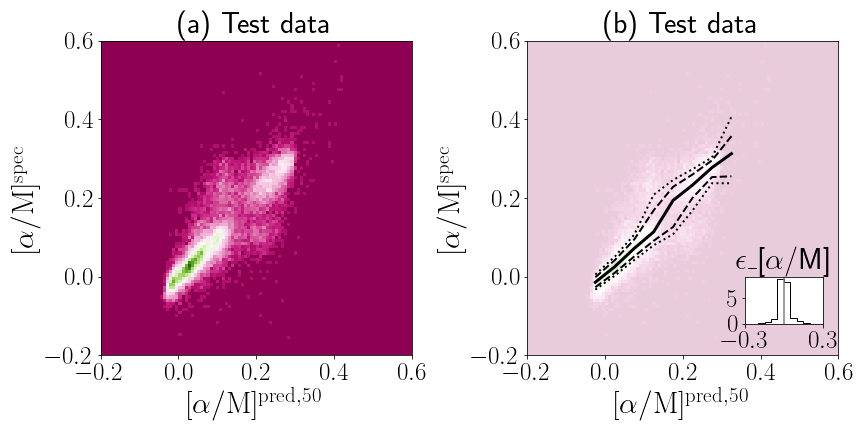} 
\includegraphics[width=0.48\textwidth ]{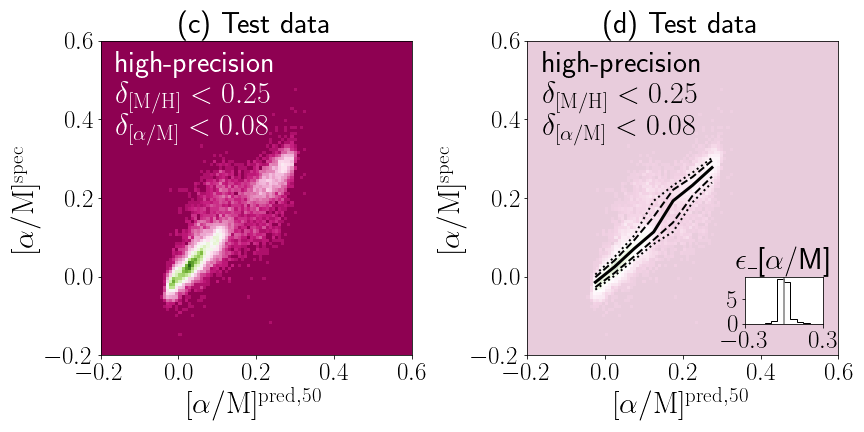} \\
\includegraphics[width=0.48\textwidth ]{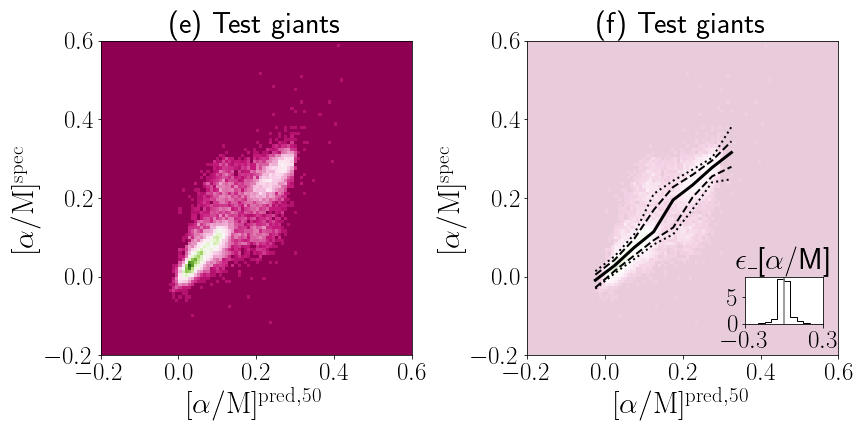} 
\includegraphics[width=0.48\textwidth ]{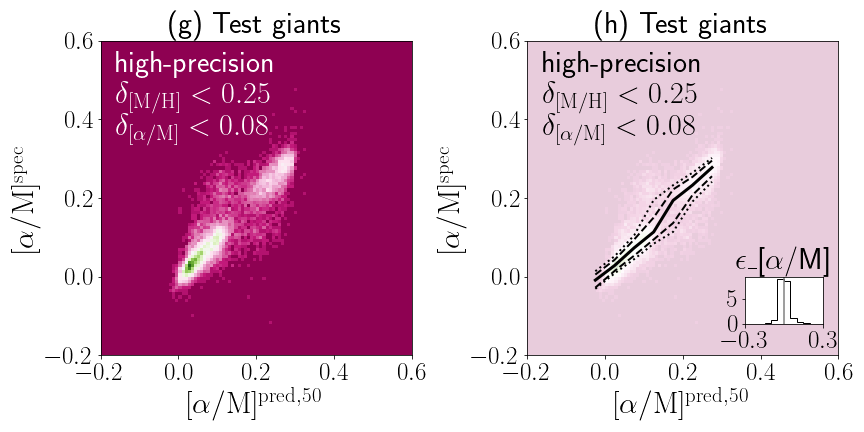} \\
\includegraphics[width=0.48\textwidth ]{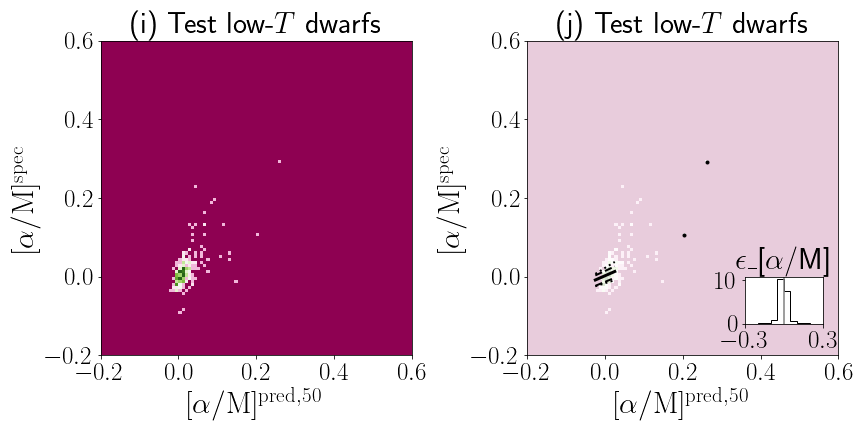} 
\includegraphics[width=0.48\textwidth ]{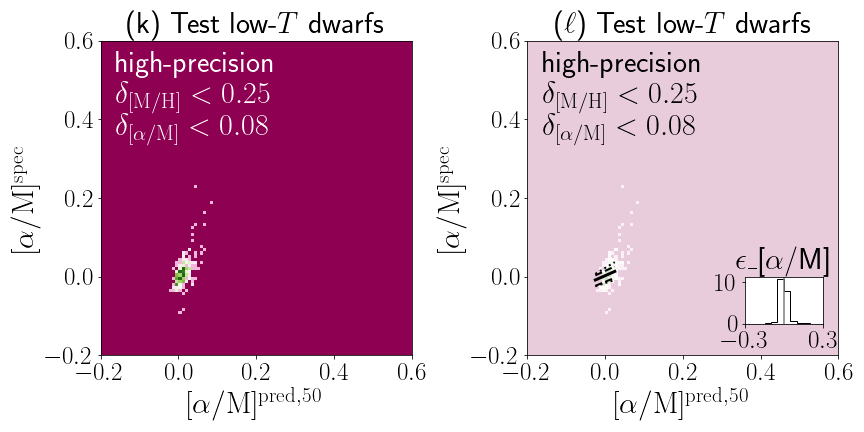} \\
\includegraphics[width=0.48\textwidth ]{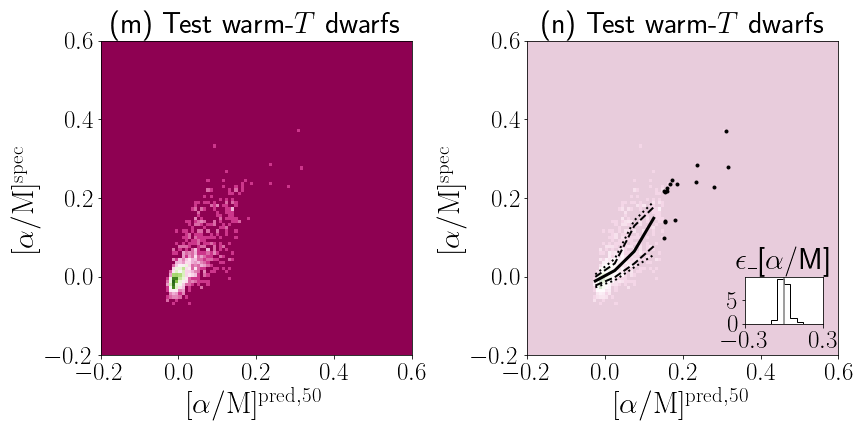} 
\includegraphics[width=0.48\textwidth ]{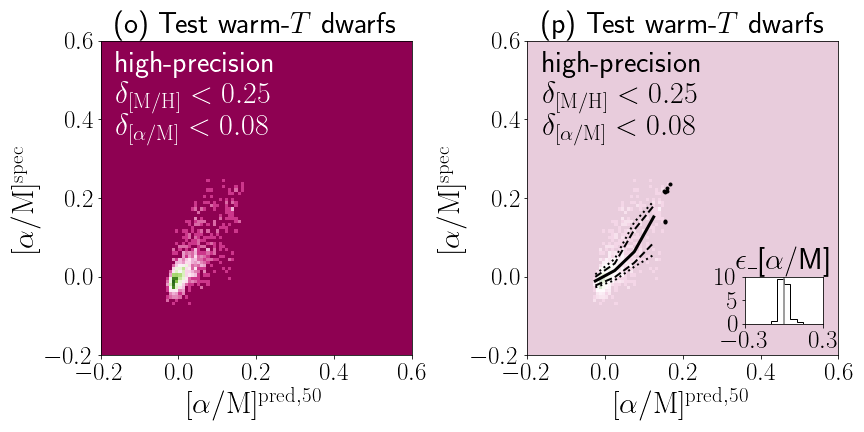} 
\caption{
The same as Fig.~\ref{fig:test_data_mh}, but showing 
the detailed performance of the QRF-AM model (our model for estimating \am) 
applied to the test data. 
On panels (j), (n), and (p), 
we plot the stars with $\am^\mathrm{pred,50}> 0.15$ with black dots 
to highlight that the majority of them  
are genuine high-$\alpha$ stars ($\am^\mathrm{spec} > 0.15$). 
}
\label{fig:test_data_am}
\end{figure*}

\subsubsection{Accuracy of \mh\ for giants and dwarfs}
\label{sec:accuracy_mh_giant_dwarf}

Fig. \ref{fig:test_data_mh} presents 
the distribution of stars in the test data 
in the $(\mh^\mathrm{pred,50}, \mh^\mathrm{spec})$-space. 
(We note that Fig. \ref{fig:GALAH_data_mh} 
is the same as Fig. \ref{fig:test_data_mh}, 
but using the external test data from GALAH DR3.)
In Fig. \ref{fig:test_data_mh}, 
the top row corresponds to the results for the entire test data. 
The second, third, and fourth rows 
correspond to the different photometric selection of stars.

\noindent (1) {\it {Giants}}

The second row of Fig. \ref{fig:test_data_mh} 
shows the performance of our model to estimate \mh, 
by using the giants in the test data. 
(Because the test data are dominated by giants, 
the second row looks similar to the top row.)

We see from the clear diagonal distribution in Fig.~\ref{fig:test_data_mh}(e)  
that our QRF-MH model can nicely predict \mh\ at $-3 \lesssim \mh$. 
Fig.~\ref{fig:test_data_mh}(f) shows the same distribution,
but oveplotting the percentile values of $\mh^\mathrm{spec}$ 
as a function of $\mh^\mathrm{pred,50}$. 
The tight alignment of these percentile value profiles 
indicates the high precision in predicting \mh\ in our model. 
The inset in Fig.~\ref{fig:test_data_mh}(f) shows the 
histogram of $\epsilonMH$. 
The symmetric distribution of this histogram 
centered around $\epsilonMH \sim 0$ 
indicates that there is no obvious systematic error in \mh\ 
for giants in the test data.

One of the advantages of our models is that we can evaluate the uncertainties \deltaMH\ and \deltaAM\ in our prediction of the labels. 
By requiring $\deltaMH<0.25$ and $\deltaAM<0.08$, we define a high-precision subsample of the stars and displayed the results in Figs.~\ref{fig:test_data_mh}(g) and (h). 
We do not see a drastic change from Fig.~\ref{fig:test_data_mh}(e) to Fig.~\ref{fig:test_data_mh}(g), because a large fraction of giants  satisfy these criteria (see Fig.~\ref{fig:histogram_epsilon_delta}(b) for a reference).

\noindent (2) {\it {Low-temperature dwarfs}}

The third row in Fig.~\ref{fig:test_data_mh} 
is the same as the second row, 
but using low-temperature dwarfs. 
Although our test data contain 
only 257 low-temperature dwarfs 
(240 of them are high-precision), 
we can recognize the diagonal feature in panels (i) and (k).  
The histograms of $\epsilonMH$ 
has a slightly fatter tail at $\epsilonMH<0$ than at $\epsilonMH>0$, 
but the estimation of \mh\ is probably acceptable for low-temperature dwarfs, 
as long as $\mh \gtrsim -0.1$. 

The above-mentioned results are supported 
by the external test data from GALAH DR3, 
which contains a larger number of low-temperature dwarfs 
(see the third row of Fig.~\ref{fig:GALAH_data_mh} in Appendix \ref{sec:test_GALAH}). 
In particular, 
there is an almost-linear trend between 
$\mh^\mathrm{pred,50}$ and $\mathrm{[Fe/H]}^\mathrm{spec}_\mathrm{GALAH}$ 
at $-0.3 \lesssim \mh^\mathrm{pred,50}$.

\noindent (3) {\it {Warm-temperature dwarfs}}

The bottom row in Fig.~\ref{fig:test_data_mh} 
is the same as the second row, 
but using warm-temperature dwarfs. 
We have 1518 low-temperature dwarfs 
(of which 1470 are high-precision), 
and we recognize a diagonal feature in panels (m) and (o) 
at $-0.5 \lesssim \mh^\mathrm{pred,50}$. 
The histograms of $\epsilonMH$ look symmetric, 
suggesting negligible systematic error in estimating \mh\ for warm-temperature dwarfs.

The above-mentioned results are confirmed 
by the external test data from GALAH DR3 
(see the bottom row of Fig.~\ref{fig:GALAH_data_mh}).

\subsubsection{Accuracy of \am\ for giants and dwarfs}
\label{sec:accuracy_am_giant_dwarf}

Fig.~\ref{fig:test_data_am} presents 
the distribution of stars in the test data 
in the $(\am^\mathrm{pred,50}, \am^\mathrm{spec})$-space. 
(We note that Fig. \ref{fig:GALAH_data_am} 
is the same as Fig. \ref{fig:test_data_am}, 
but using the GALAH data.)
Again, 
the top row of Fig.~\ref{fig:test_data_am} corresponds to the results for the entire test data. 
The other rows 
correspond to the different photometric selection of stars.

\noindent (1) {\it {Giants}}

The second row in Fig.~\ref{fig:test_data_am} 
shows the performance of our model for giants. 
In Fig.~\ref{fig:test_data_am}(e), 
the majority of stars are distributed almost diagonally, 
suggesting that our QRF-AM model can estimate realistic \am\ 
for most of the giants in the test data. 
There are some stars with off-diagonal distribution in Fig.~\ref{fig:test_data_am}(e). 
Some of these stars are recognized by our model 
as low-$\alpha$ stars but actually high-$\alpha$ stars.\footnote{
\revise{
In an attempt to understand 
the cause of these misclassifications, 
we examine the plot of $(\epsilonMH, \epsilonAM)$. 
We find that, on average, stars in our test data follow 
a smooth relationship of $\epsilonAM = -0.2 \epsilonMH$. 
This suggests that our model tends to overestimate \am\ by 0.02 dex 
when it underestimates \mh\ by 0.10 dex. 
However, this trend does not explain the misclassifications. 
(This observed relationship may offer insights for further refining our 
\am estimates, though this is beyond the scope of the current paper.)
}
} 
\revise{
(This misclassification happens to 20\% of the low-$\alpha$ giants with $\mh>-0.9$ and $\am<0.15$). 
Other stars are recognized by our model 
as high-$\alpha$ stars but actually low-$\alpha$ stars. 
(This misclassification happens to 6\% of the high-$\alpha$ giants with $\mh>-0.9$ and $\am>0.15$). 
}
The fraction of such stars is reduced 
if we select the high-precision subset of stars, 
as shown in panels (g) and (h). 
\revise{
(The misclassification fractions mentioned above become 17\% and 3\%, respectively.)
}
The symmetric shape of the histograms of $\epsilonAM$ 
suggests that we have no obvious systematic error in \am\ 
for giants.

\noindent (2) {\it {Low-temperature dwarfs}}

The third row in Fig.~\ref{fig:test_data_am} 
is the same as the second row, 
but using low-temperature dwarfs. 
In panels (j) and ($\ell$), 
we see a marginal, positive correlation between 
$\am^\mathrm{pred,50}$ and $\am^\mathrm{spec}$. 
In Fig.~\ref{fig:test_data_am}(j), 
we have 
two low-temperature dwarfs with $\am^\mathrm{pred,50}>0.15$ (shown by black dots). 
Among these two stars, 
one star (50 \%) actually have $\am^\mathrm{spec}_\mathrm{GALAH}>0.15$. 
However, due to the limited number of low-temperature dwarfs in the test data, 
the trend is not clear.

In this regard, it is worth mentioning  
the performance of our QRF-AM model for the external test data from GALAH 
(see the third row of Fig.~\ref{fig:GALAH_data_am}). 
Notably, 
as seen in Fig.~\ref{fig:GALAH_data_am}(j), 
there is a positive correlation 
between $\am^\mathrm{pred,50}$ and 
$\afe^\mathrm{spec}_\mathrm{GALAH}$ 
at $0 \lesssim \am^\mathrm{pred,50} \lesssim 0.3$. 
In Fig.~\ref{fig:GALAH_data_am}($\ell$), 
we have 
55 low-temperature dwarfs with $\am^\mathrm{pred,50}>0.15$ (shown by black dots). 
Among these 55 stars, 
41 stars (75 \%) actually have $\am^\mathrm{spec}_\mathrm{GALAH}>0.15$. 
(In Fig.~\ref{fig:GALAH_data_am}(j), this fraction is $38\% = (432/1151)$, 
which is reduced because the sample also includes low-precision stars.) 
From this exercise, we see that our QRF-AM model extract some information on \am\ from the XP spectra for low-temperature dwarfs, 
especially for high-precision sample.

\noindent (3) {\it {Warm-temperature dwarfs}}

The bottom row in Fig.~\ref{fig:test_data_am} 
is the same as the second row, 
but using warm-temperature dwarfs. 
In panels (n) and (p), 
we recognize a clear positive trend between 
$\am^\mathrm{pred,50}$ and $\am^\mathrm{spec}$ 
at $-0.05 < \am^\mathrm{pred,50} < 0.2$ 
although the number of stars with high-$\am^\mathrm{pred,50}$ is limited.

In Fig.~\ref{fig:test_data_am}(p), 
we have 7 warm-temperature dwarfs with $\am^\mathrm{pred,50}>0.15$ (shown by black dots). 
Among these 7 stars, 
5 stars (71 \%) actually have $\am^\mathrm{spec}>0.15$. 
In Fig.~\ref{fig:test_data_am}(n), 
this fraction is $75\% (=12/16)$,
which is surprisingly good given that 
the sample also include low-precision stars. 
(As a reference, 
in Fig.~\ref{fig:GALAH_data_am}(n) 
and 
Fig.~\ref{fig:GALAH_data_am}(p), 
this fraction is 
$75\% (=696/932)$ and 
$87\% (=34/39)$, respectively.)
From this exercise, 
we see that our QRF-AM model extract some information on \am\ from the XP spectra for warm-temperature dwarfs. 
However, as we will discuss 
in Section \ref{sec:chemo_dynamical_correlation} 
(see Fig.~\ref{fig:local_disc_dwarf_warmT}), 
we are not very comfortable with our estimates of (\mh, \am) 
due to an independent analysis 
on the chemo-dynamical correlation of the warm-temperature dwarfs. 

\subsection{Benchmarking the uncertainty in our model predictions}

\revise{
So far, we have mainly used the median values of 
$\mh^\mathrm{pred,50}$ and $\am^\mathrm{pred,50}$ 
to evaluate our model predictions. 
Here we use other percentile values of our predictions 
(such as $\mh^\mathrm{pred,16}$ or $\am^\mathrm{pred,16}$) 
to assess the reliability of our uncertainty estimates. 
If our uncertainty estimates are accurate, 
we expect the fraction of stars for which
$\mh^\mathrm{spec}$ falls below the predicted percentiles--such as 
$\mh^\mathrm{pred,2.5}$, 
$\mh^\mathrm{pred,16}$, 
$\mh^\mathrm{pred,25}$, 
$\mh^\mathrm{pred,50}$, 
$\mh^\mathrm{pred,75}$, 
$\mh^\mathrm{pred,84}$, and 
$\mh^\mathrm{pred,97.5}$--to 
be 2.5\%, 16\%, 25\%, 50\%, 75\%, 84\%, and 97.5\%, respectively. 
We expect the same for our \am\ predictions. 
To investigate this, 
we use the test data (not used in the training procedure). 
The corresponding fractions we observed were:
\begin{itemize}
\item For \mh: 0.6\%, 9\%, 18\%, 50\%, 69\%, 90\%, and 99.2\%.
\item For \am: 1.7\%, 13\%, 22\%, 50\%, 79\%, 88\%, and 98.6\%.
\end{itemize}
These results show that 
81\% of the stars in the test data satisfy $\mh^\mathrm{pred,16}<\mh^\mathrm{spec}<\mh^\mathrm{pred,84}$, 
and 
75\% of the stars satisfy $\am^\mathrm{pred,16}<\am^\mathrm{spec}<\am^\mathrm{pred,84}$, 
whereas we would expect 68\% in an ideal case. 
These findings suggest that our uncertainty estimates are slightly conservative, but they remain reasonable and useful 
for assessing the uncertainties in our model predictions.
}

\startlongtable
\begin{deluxetable*}{l l}
\tablecaption{Description of the output data 
\label{table:columns}}
\tablewidth{0pt}
\tabletypesize{\scriptsize}
\tablehead{
\colhead{Quantity} &
\colhead{Description} 
}
\startdata
{source\_id} & {source\_id in Gaia DR3 } \\
{phot\_g\_mean\_mag} & {$G$-band magnitude (Gaia DR3)} \\
{phot\_bp\_mean\_mag} & {$G_\mathrm{BP}$-band magnitude (Gaia DR3)} \\
{phot\_rp\_mean\_mag} & {$G_\mathrm{RP}$-band magnitude (Gaia DR3)} \\
{l} & {Galactic longitude (Gaia DR3)} \\
{b} & {Galactic latitude (Gaia DR3)} \\
{ra} & {Right ascension (Gaia DR3)} \\
{dec} & {Declination (Gaia DR3)} \\
{parallax} & {Parallax without zero-point correction (Gaia DR3)} \\
{parallax\_error} & {Parallax error (Gaia DR3)} \\
{ebv\_dustmaps\_086} & {$E(B-V)$ estimated from \cite{Schlegel1998ApJ...500..525S} dust map multiplied by a factor 0.86} \\
{bool\_in\_training\_sample} & {Boolean data indicating whether the star is included in the training data} \\
{bool\_flag\_cmd\_good} & {Boolean data indicating whether the star satisfies the simple CMD cut} \\
{mh\_2p5\_qrf} & {2.5th percentile value of the predicted \mh} \\
{mh\_10\_qrf} & {10th percentile value of the predicted \mh} \\
{mh\_16\_qrf} & {16th percentile value of the predicted \mh} \\
{mh\_25\_qrf} & {25th percentile value of the predicted \mh} \\
{mh\_50\_qrf} & {50th percentile value of the predicted \mh} \\
{mh\_75\_qrf} & {75th percentile value of the predicted \mh} \\
{mh\_84\_qrf} & {84th percentile value of the predicted \mh} \\
{mh\_90\_qrf} & {90th percentile value of the predicted \mh} \\
{mh\_97p5\_qrf} & {97.5th percentile value of the predicted \mh} \\
{alpham\_2p5\_qrf} & {2.5th percentile value of the predicted \am} \\
{alpham\_10\_qrf} & {10th percentile value of the predicted \am} \\
{alpham\_16\_qrf} & {16th percentile value of the predicted \am} \\
{alpham\_25\_qrf} & {25th percentile value of the predicted \am} \\
{alpham\_50\_qrf} & {50th percentile value of the predicted \am} \\
{alpham\_75\_qrf} & {75th percentile value of the predicted \am} \\
{alpham\_84\_qrf} & {84th percentile value of the predicted \am} \\
{alpham\_90\_qrf} & {90th percentile value of the predicted \am} \\
{alpham\_97p5\_qrf} & {97.5th percentile value of the predicted \am} \\
\hline
\enddata
\end{deluxetable*}

\section{Results}
\label{sec:results}

\subsection{The catalog of \mh\ and \am} 
\label{sec:catalog}

After training the QRF-MH and QRF-AM models, we first apply our models for the entire catalog of stars with Gaia XP spectra (219 million stars). 
Among these stars, we publish the labels for 48 million stars with $E(B-V) < 0.1$ at the Zenodo database (\url{https://zenodo.org/records/10902172}).\footnote{
We also publish 134 million stars with 
$0.1 < E(B-V) < 1$ at the same location as a reference for future studies. 
That is, we publish in total 182 million stars 
with $E(B-V) < 1$. 
}
We describe the columns of the published catalog 
in Table~\ref{table:columns}.

\begin{figure}
\centering 
\includegraphics[width=0.4\textwidth ]{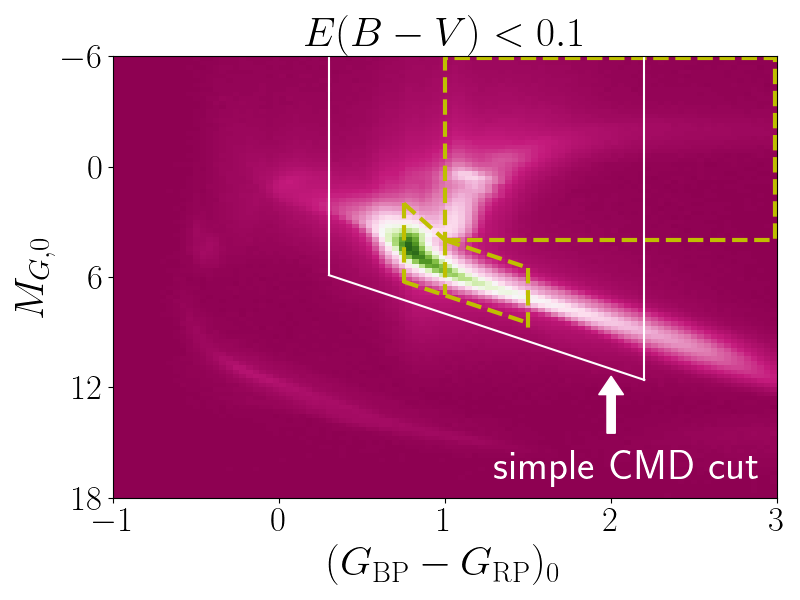} \\
\caption{
CMD of the stars in our catalog. 
The region above the white solid line satisfies  
\texttt{{bool\_flag\_cmd\_good}=True}, 
which roughly corresponds to the region covered by our training data. 
The regions surrounded by yellow dashed lines 
are the same as in Fig.~\ref{fig:train_different_MHAM_CMD}. 
We note that the stars in the catalog satisfies 
$E(B-V)<0.1$. 
}
\label{fig:CMD_ebv_dependence}
\end{figure}

Fig.~\ref{fig:CMD_ebv_dependence} 
shows the CMD of our published catalog. 
We see that some fraction of stars in the catalog 
are hot dwarfs and white dwarfs, 
which are not represented by our training data 
(see also Fig.~\ref{fig:train_different_MHAM_CMD}). 
The users of the catalog need to avoid these stars. 
As a simple way to omit these stars, 
we select stars that satisfy
\eq{
0.3 &< (G_\mathrm{BP}-G_\mathrm{RP})_0 < 2.2 \label{eq:0p3bprp2p2}\\
M_{G,0} &< 3 \times (G_\mathrm{BP}-G_\mathrm{RP})_0 + 5 ,
}
and flag these stars as \texttt{{bool\_flag\_cmd\_good}=True} in our catalog. 
The corresponding region is shown by the white solid line 
in Fig.~\ref{fig:CMD_ebv_dependence}. 
Among 48 million stars with $E(B-V)<0.1$, 
about 46 million stars satisfy this simple CMD cut. 
\revise{
This flag can be used to 
avoid stars for which we have less confidence on 
the estimated chemistry. 
}
However, 
we do not attempt to state that all the stars with 
\texttt{{bool\_flag\_cmd\_good}=True} 
have reliable estimates of their chemistry. 
The reliability of the estimated values of (\mh, \am)
should be carefully analyzed by using other information, 
such as the stellar evolutionary stage (e.g., giants or dwarfs), parallax,\footnote{
For stars with poor parallax data 
(e.g., \texttt{parallax\_over\_error <5}), 
\revise{
the point-estimate of $M_\mathrm{G,0}$ is associated with a large uncertainty. 
However, our flag \texttt{{bool\_flag\_cmd\_good}} 
remains useful for these stars as well, 
because both dwarfs and giants in our catalog almost always satisfy
\texttt{{bool\_flag\_cmd\_good}=True}. 
See Appendix \ref{appendix:flag} for some discussion. 
Those stars with negative parallax are flagged as 
\texttt{{bool\_flag\_cmd\_good}=False} 
because the point-estimate of $M_{G,0}$ is not well-defined. 
However, such stars are more likely to be distant objects rather than foreground white dwarfs, and as long as the color condition in equation (\ref{eq:0p3bprp2p2}) is satisfied, our model predictions should remain reliable.
}
}
$\deltaMH$ or $\deltaAM$.

\begin{figure*}
\centering 
\includegraphics[width=0.40\textwidth ]{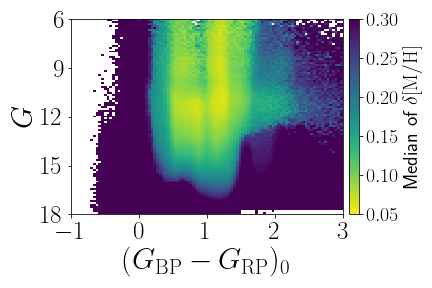}
\includegraphics[width=0.40\textwidth ]{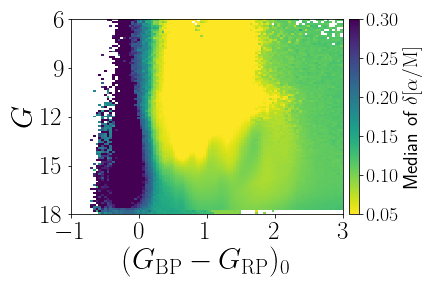}
\caption{
\revise{
The dependence on $(G, (G_\mathrm{BP}-G_\mathrm{RP})_0)$ 
of $\deltaMH$ (left) and $\deltaAM$ (right). 
Here, we use stars with low dust extinction ($E(B-V)<0.1$) 
but we do not apply any cuts such as the parallax cut. 
}
}
\label{fig:bprp_G_color_coded_by_delta}
\end{figure*}

\begin{figure*}
\centering 
\includegraphics[width=0.24\textwidth ]{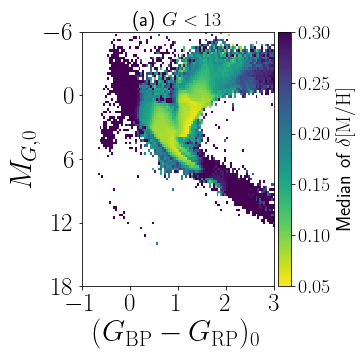}
\includegraphics[width=0.24\textwidth ]{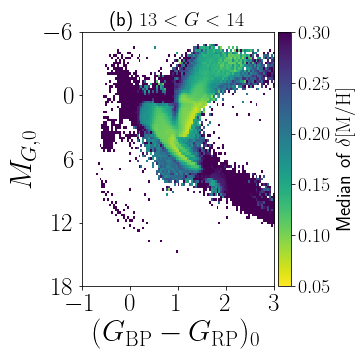}
\includegraphics[width=0.24\textwidth ]{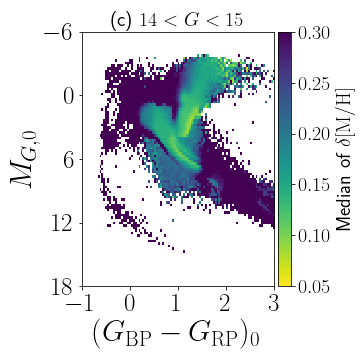}
\includegraphics[width=0.24\textwidth ]{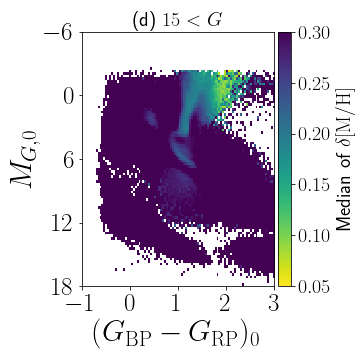}\\
\includegraphics[width=0.24\textwidth ]{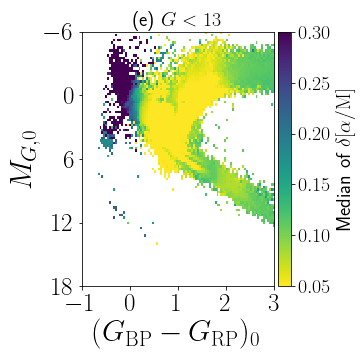}
\includegraphics[width=0.24\textwidth ]{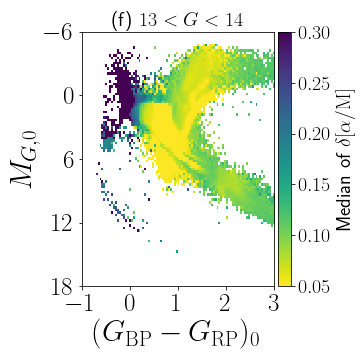}
\includegraphics[width=0.24\textwidth ]{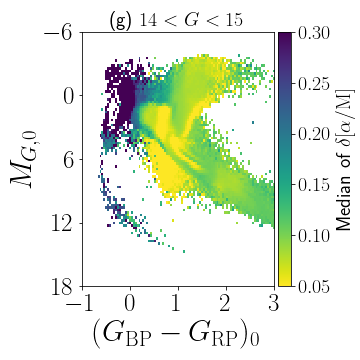}
\includegraphics[width=0.24\textwidth ]{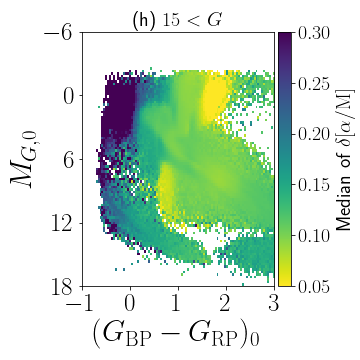}
\caption{
\revise{
The color-magnitude diagram of stars in our catalog, color-coded by 
the uncertainties $\deltaMH$ and $\deltaAM$. 
In all plots, 
we use stars with low dust extinction ($E(B-V)<0.1$) 
and marginally acceptable parallax measurements 
(fractional parallax error smaller than 100\% or \texttt{parallax\_over\_error}$>1$). 
(a)-(d) The map of $\deltaMH$ with different magnitude ranges. 
(e)-(h) The map of $\deltaAM$ with different magnitude ranges. 
}
}
\label{fig:CMD_color_coded_by_delta}
\end{figure*}

\begin{figure*}
\centering 
\includegraphics[width=0.24\textwidth ]{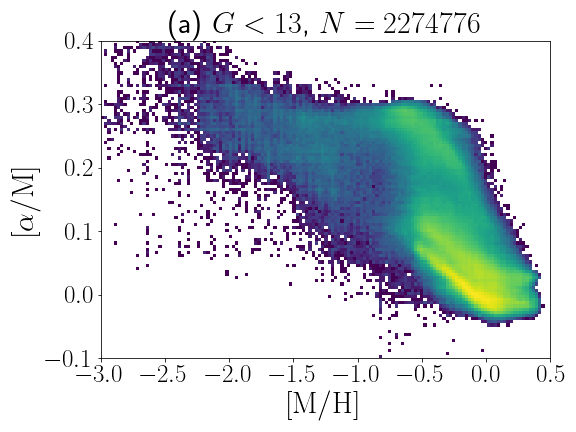}
\includegraphics[width=0.24\textwidth ]{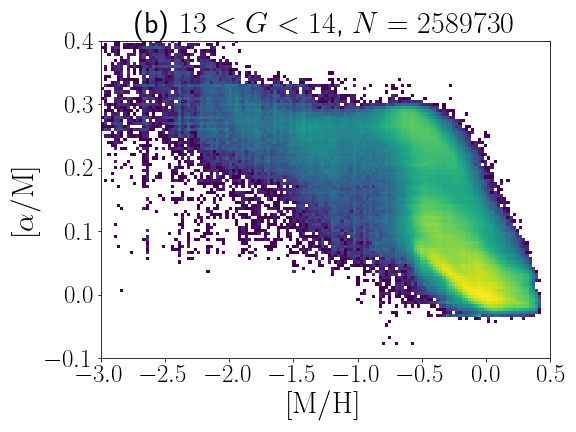}
\includegraphics[width=0.24\textwidth ]{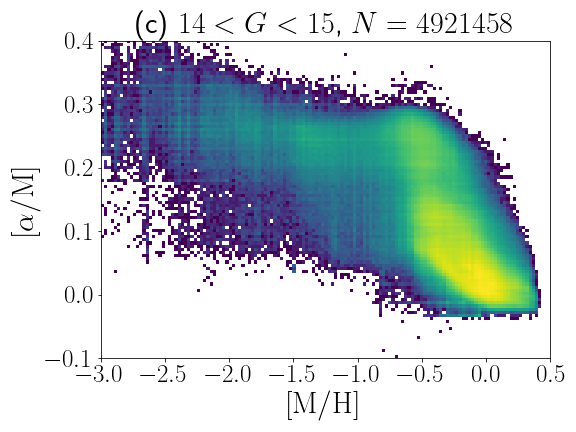}
\includegraphics[width=0.24\textwidth ]{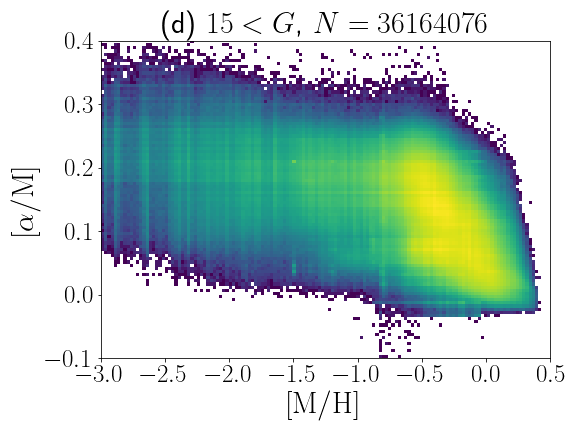}\\
\includegraphics[width=0.24\textwidth ]{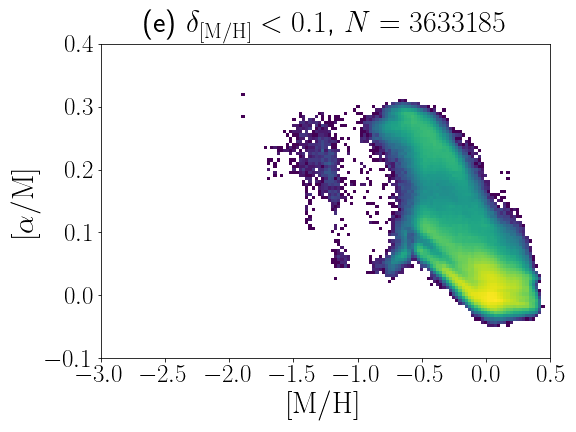}
\includegraphics[width=0.24\textwidth ]{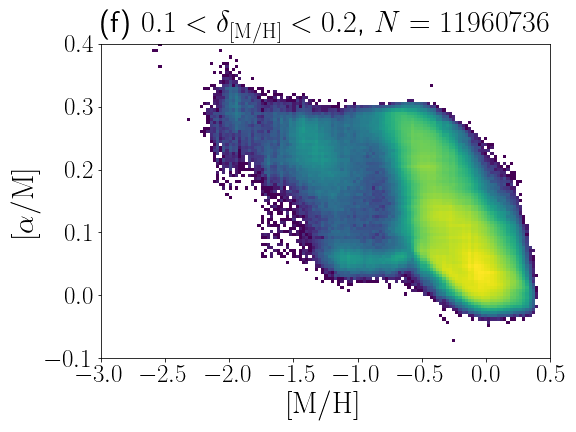}
\includegraphics[width=0.24\textwidth ]{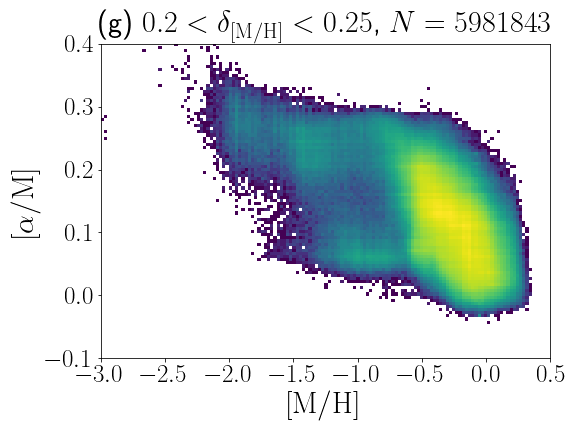}
\includegraphics[width=0.24\textwidth ]{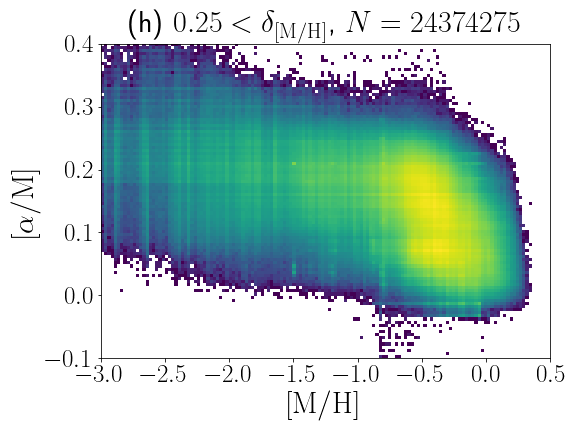} \\
\includegraphics[width=0.24\textwidth ]{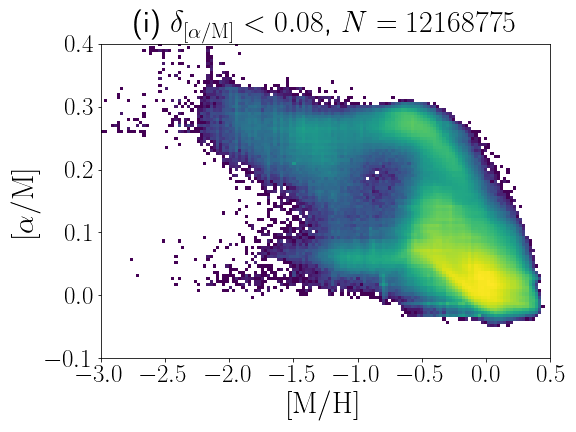}
\includegraphics[width=0.24\textwidth ]{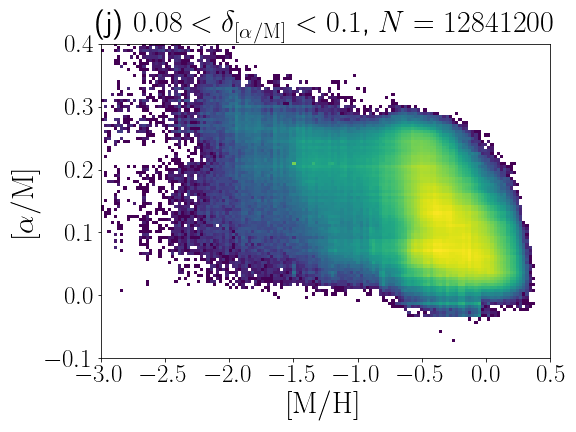}
\includegraphics[width=0.24\textwidth ]{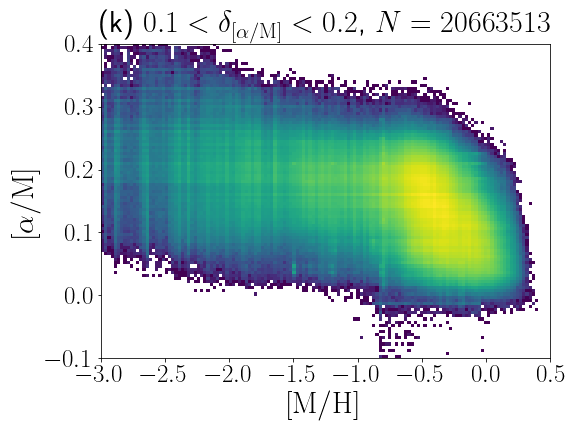}
\includegraphics[width=0.24\textwidth ]{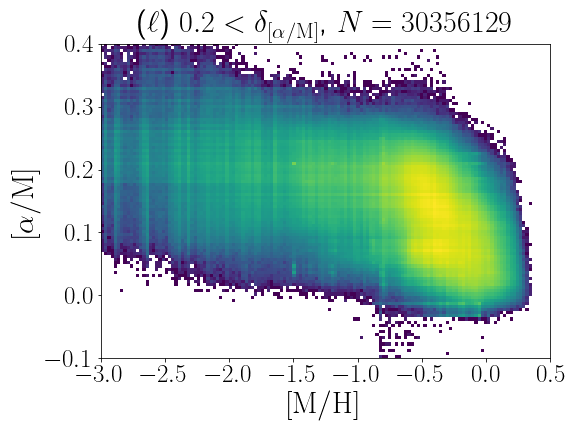}
\caption{
The (\mh,\am)-distribution of the cataloged stars with various cuts. 
(We have excluded stars in the training data.) 
(Top row) 
The distribution of stars 
with different $G$-band magnitude ranges. 
(Middle row) 
The distribution of stars 
with different $\deltaMH$ ranges. 
(Bottom row) 
The distribution of stars 
with different $\deltaAM$ ranges. 
}
\label{fig:chemistry_dep}
\end{figure*}

\begin{figure*}
\centering 
\includegraphics[width=0.24\textwidth ]{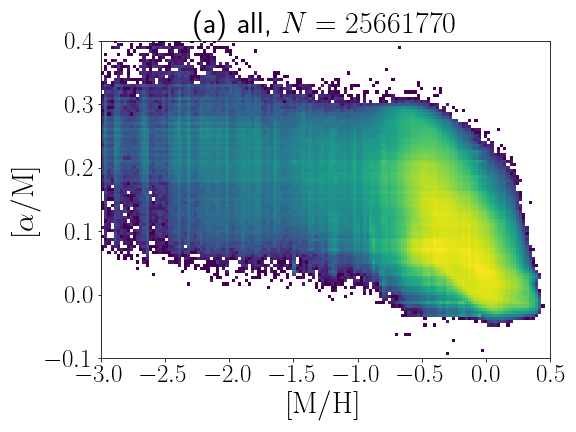}
\includegraphics[width=0.24\textwidth ]{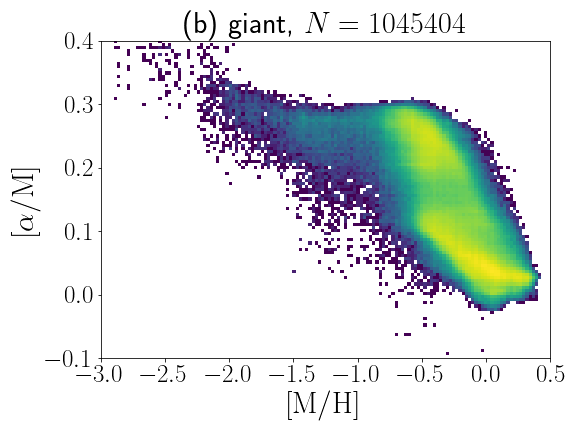}
\includegraphics[width=0.24\textwidth ]{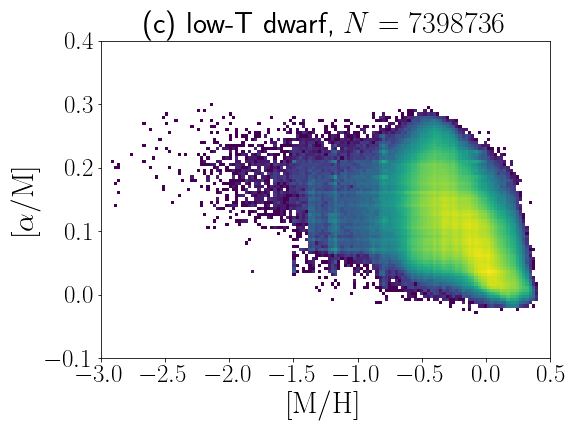}
\includegraphics[width=0.24\textwidth ]{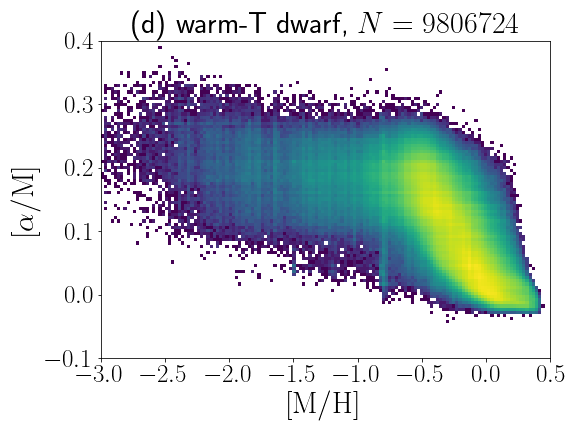}\\
\includegraphics[width=0.24\textwidth ]{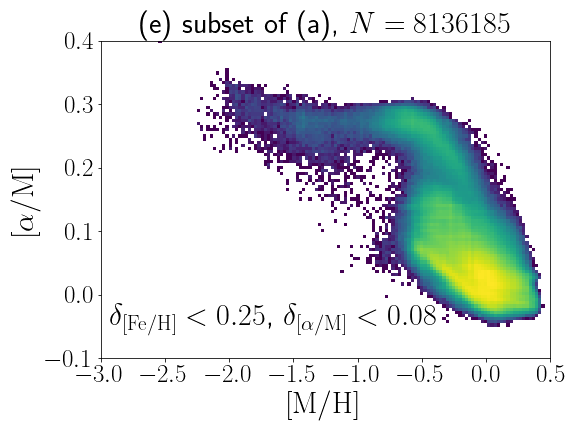}
\includegraphics[width=0.24\textwidth ]{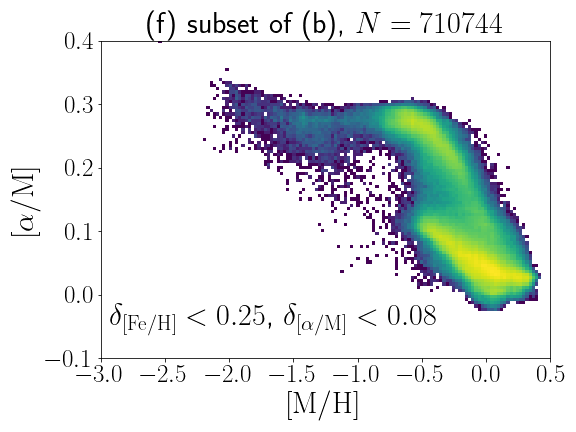}
\includegraphics[width=0.24\textwidth ]{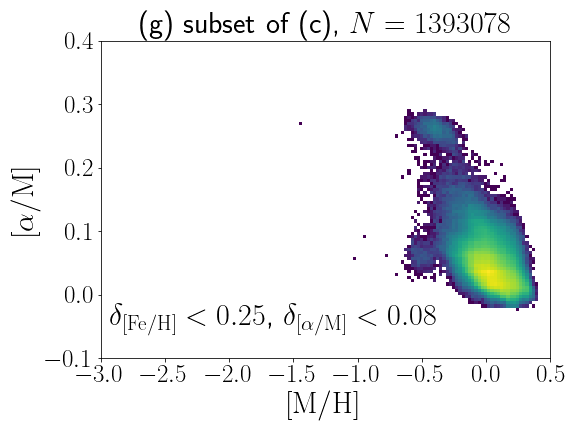}
\includegraphics[width=0.24\textwidth ]{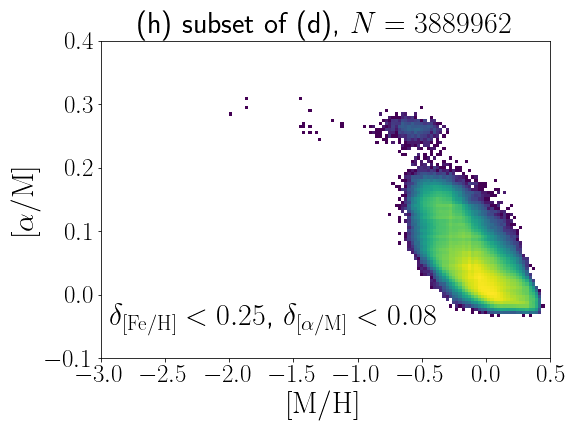}
\caption{
The distribution of stars in the 
$(\mh, \am)$-space estimated from our models. 
In all of the plots, 
we show Solar-neighbor stars ($7\kpc<R<9 \kpc$, $|z| < 3 \kpc$) 
with low dust extinction ($E(B-V)<0.1$) 
and good parallax measurements 
(fractional parallax error smaller than 20\%).  
We also note that 
we have excluded stars in the training data. 
(a) Stars without further cuts. 
(b) Giant stars. 
(c) Low-temperature dwarf stars.  
(d) Warm-temperature dwarf stars. 
Panels (e)-(h) 
are the same as panels 
(a)-(d), respectively, 
but with further constraints of 
$\deltaMH <0.25$ and $\deltaAM < 0.08$. 
}
\label{fig:all_clean}
\end{figure*}

\subsection{Chemical distribution of stars} \label{sec:chemical_distribution}

Here we investigate the distribution of 
stars with Gaia XP spectra in the chemistry space. 
In Sections \ref{sec:chemistry_G_dep}-\ref{sec:chemistry_delta_dep}, 
we restrict ourselves to low-extinction stars 
($E(B-V)<0.1$) 
that are not in the training data. 
In Section \ref{sec:chemistry_CMD}, 
we further restrict ourselves to stars 
that are in the Solar cylinder ($7 < R/\kpc< 9$, $|z| < 3 \kpc$) 
and have good parallax measurements 
(\texttt{parallax\_over\_error>5} in Gaia DR3).

In the following, 
when we refer to our estimated values of \mh\ or \am, 
we simply use $\mh = \mh^\mathrm{pred,50}$ 
or $\am = \am^\mathrm{pred,50}$ for brevity.

\subsubsection{Dependence on $G$ and \revise{$(G_\mathrm{BP}-G_\mathrm{RP})$}} \label{sec:chemistry_G_dep}

\revise{
Fig.~\ref{fig:bprp_G_color_coded_by_delta} shows how the median values of $\deltaMH$ and $\deltaAM$ 
vary as a function of $(G, (G_\mathrm{BP}-G_\mathrm{RP})_0)$. 
We see that our estimates for \mh\ and \am\ become unreliable for blue stars with $(G_\mathrm{BP}-G_\mathrm{RP})_0 \lesssim 0.3$. 
Also, both $\deltaMH$ and $\deltaAM$ deteriorate for stars with $G \gtrsim 15$. 
}

\revise{
To further explore how the color and magnitude affect the performance of our models, 
we focus on stars with marginally good parallax data (with \texttt{parallax\_over\_error}$>1$) 
and construct the CMD of the stars in our catalog. 
Fig.~\ref{fig:CMD_color_coded_by_delta} shows the CMDs of the stars 
in our catalog, 
grouped by different magnitude ranges and color-coded 
by the median values of $\deltaMH$ and $\deltaAM$. 
We see smaller uncertainties in \mh\ and \am\ 
in regions corresponding to the main-sequence and red giant branch, 
as these areas are well represented in our training data. 
In contrast, stars with colors outside the range $0.3<(G_\mathrm{BP}-G_\mathrm{RP})_0<2.2$ 
exhibit larger uncertainties in \mh\ and \am, 
due to the limited color coverage in our training dataset.
}

The top row in Fig.~\ref{fig:chemistry_dep} shows 
the $(\mh, \am)$-distribution for different ranges of $G$. 
We see a clear bimodality of low-$\alpha$ and high-$\alpha$ sequences 
at $G<13$ (Fig.~\ref{fig:chemistry_dep}(a)), 
and the bimodality is marginally recognized at $13<G<15$ 
(Figs.~\ref{fig:chemistry_dep}(b) and \ref{fig:chemistry_dep}(c)). 
In contrast, 
we do not see the bimodal structure at $15<G$. 
This result indicates that 
our estimates of \mh\ and \am\ are reasonable for bright stars 
and deteriorate with increasing $G$. 
The difficulty in estimating \mh\ or \am\ for faint stars is consistent with \cite{Witten2022MNRAS.516.3254W}.

\revise{
In general, fainter stars have larger uncertainties in Gaia XP spectra. 
Consequently, the fact that we ignore measurement uncertainties in Gaia XP spectra 
(see Section \ref{sec:XPdata}) may negatively affect our model predictions, particularly for faint stars.
While addressing how to properly account for these uncertainties in Gaia XP spectra for fainter stars is beyond the scope of this paper, it is an important topic for future investigation. 
}

\subsubsection{Dependence on \deltaMH\ and \deltaAM}
\label{sec:chemistry_delta_dep}

The middle and bottom rows in Fig.~\ref{fig:chemistry_dep} 
show the chemical distribution of stars 
for different ranges of the precision of our estimates
(\deltaMH\ and \deltaAM). 
We see a bimodality of low-$\alpha$ and high-$\alpha$ sequences 
for stars with $\deltaMH < 0.1$ (Fig.~\ref{fig:chemistry_dep}(e)) 
and for stars with $\deltaAM < 0.08$ (Fig.~\ref{fig:chemistry_dep}(i)), 
but the bimodality is not as clearly seen for stars with 
other ranges of \deltaMH\ or \deltaAM. 
This result indicates that 
the values of \deltaMH\ and \deltaAM\ 
can be used to select stars 
for which we have reliable estimates of (\mh,\am).

\subsubsection{Dependence on the stellar evolutionary stage}
\label{sec:chemistry_CMD}

In Fig.~\ref{fig:all_clean}, 
we investigate the chemical distribution 
of stars with different evolutionary stages. 
For Solar-neighbor stars with good parallax measurements 
(see the first paragraph of Section \ref{sec:chemical_distribution}), 
we select giants and low/warm-temperature dwarfs 
based on the location in the CMD 
in the same manner as in Section \ref{sec:APOGEE_and_Li2022}.

We note that \cite{Witten2022MNRAS.516.3254W} 
pointed out that estimating \am\ is difficult (for faint stars with $G=16$) at $\teff > 5000$ K, 
which roughly corresponds to $(G_\mathrm{BP}-G_\mathrm{RP})_0 \lesssim 1.1$ 
(see also \citealt{Gavel2021A&A...656A..93G}). 
The warm-temperature main-sequence stars 
(with $(G_\mathrm{BP}-G_\mathrm{RP})_0 < 1$)
are selected to investigate this issue.

Panels (a)-(d) in Fig.~\ref{fig:all_clean} show the chemical distribution 
of stars 
without any CMD selections (panel (a)), 
of giants (panel (b)), 
of low-temperature dwarfs (panel (c)), 
and 
of warm-temperature dwarfs (panel (d)). 
Panels (e)-(h) are the same as panels (a)-(d), respectively, 
but with further constraints of 
$\deltaMH < 0.25$ and $\deltaAM<0.08$ (i.e., high-precision subsample).

It is intriguing to note that the clear bimodality in the (\mh, \am)-space 
is visible for giants, even before applying the high-precision selection. 
This result indicates 
that our estimates of (\mh, \am) for giants have higher reliability than dwarfs. 
We also note that 
the fraction of stars with high-precision estimates of (\mh, \am) is as high as $\sim$68\% for giants. 
(As shown in Fig.~\ref{fig:all_clean}, 
we have 1,054,004 and 710,744 giants in panels (b) and (f), respectively.)

For warm- and low-temperature dwarfs, 
we do not see the clear bimodality in panels (c) or (d), 
but the bimodality can be seen in their high-precision subsample (in panels (g) and (h)). 
These results indicate 
that most stars with $\am > 0.17$ in panel (c) 
have large uncertainties in (\mh, \am) 
and may not be {\it true} high-$\alpha$ stars.

\subsection{Chemo-dynamical correlation}
\label{sec:chemo_dynamical_correlation}

Here we investigate 
how the kinematics of low-$\alpha$ and high-$\alpha$ 
disk stars in the Solar neighborhood 
change as a function of their chemistry 
(see also \citealt{Li2023arXiv230914294L}). 
Here, we select 
giants and low/warm-temperature dwarfs 
with high-precision (\mh, \am) from our analysis 
(as selected in Section \ref{sec:chemical_distribution})
and 
with radial velocity 
\revise{
and reliable parallax (\texttt{parallax\_over\_error>5}) 
available from Gaia DR3. 
Appendix \ref{appendix:coordinate} describes our procedure to 
derive the position, velocity, and orbital eccentricity of the stars. }

\begin{figure*}
\raggedright
\includegraphics[width=0.3\textwidth ]{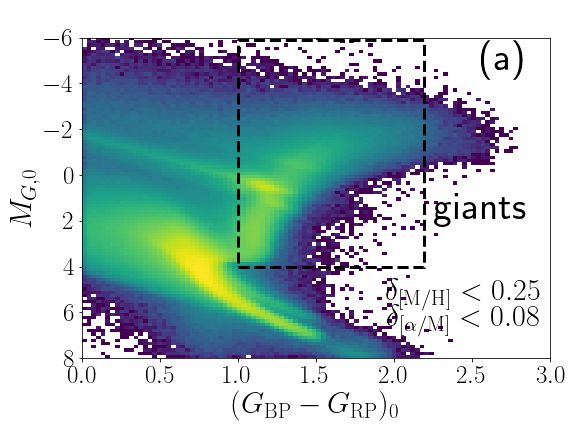}
\includegraphics[width=0.3\textwidth ]{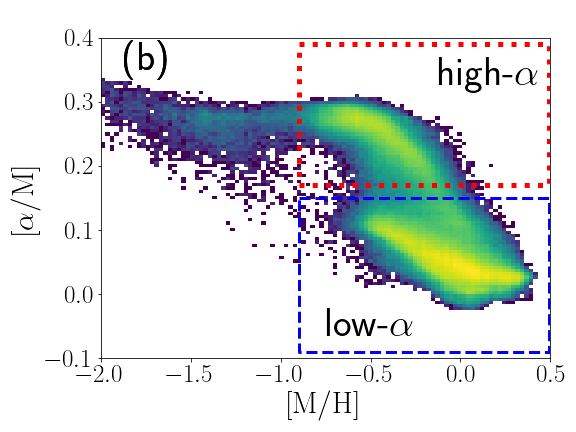}
\includegraphics[width=0.3\textwidth ]{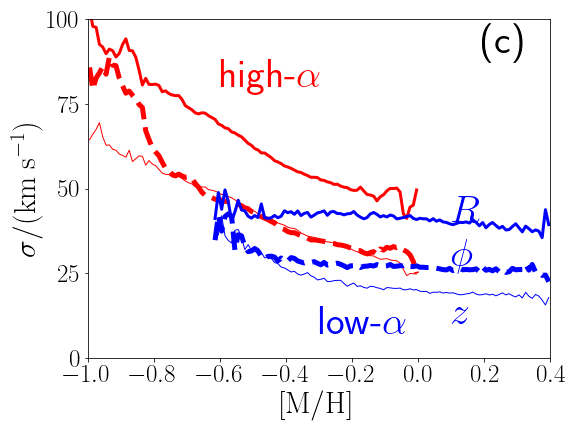}
\includegraphics[width=0.3\textwidth ]{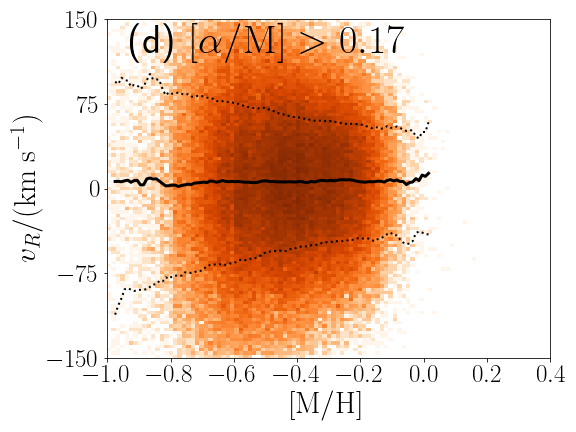}
\includegraphics[width=0.3\textwidth ]{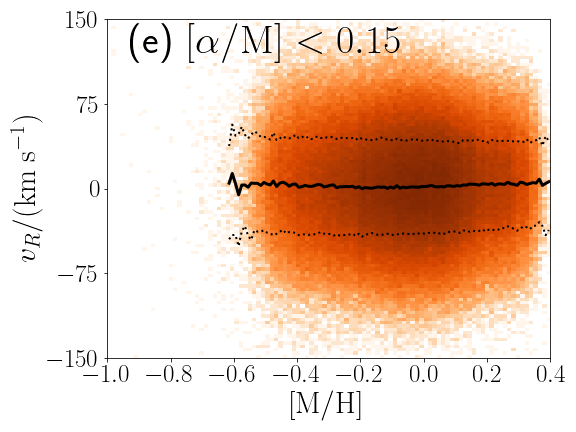}
\includegraphics[width=0.3\textwidth ]{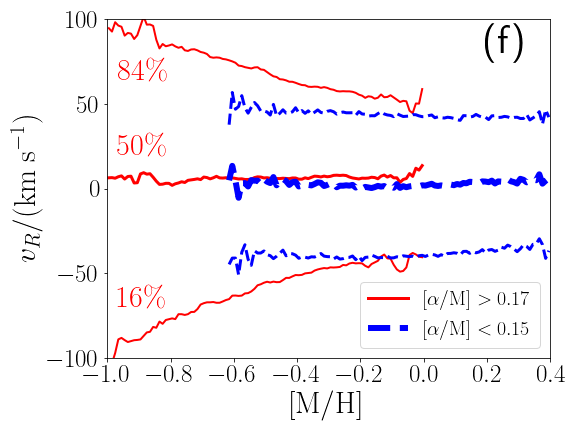} \\
\includegraphics[width=0.3\textwidth ]{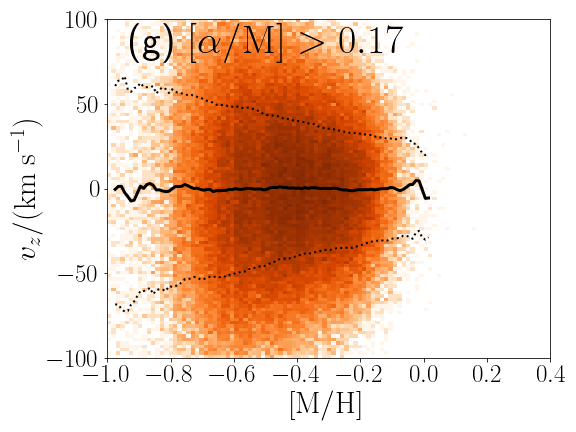}
\includegraphics[width=0.3\textwidth ]{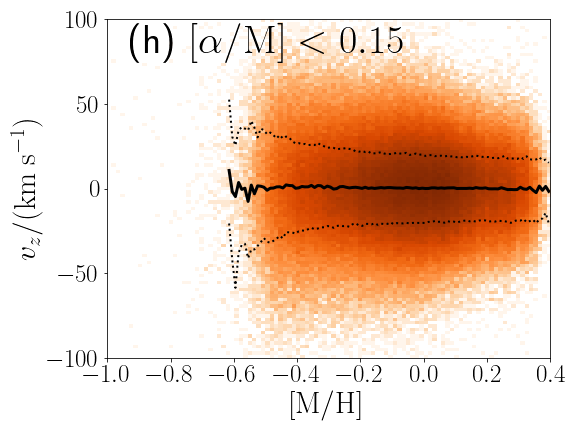}
\includegraphics[width=0.3\textwidth ]{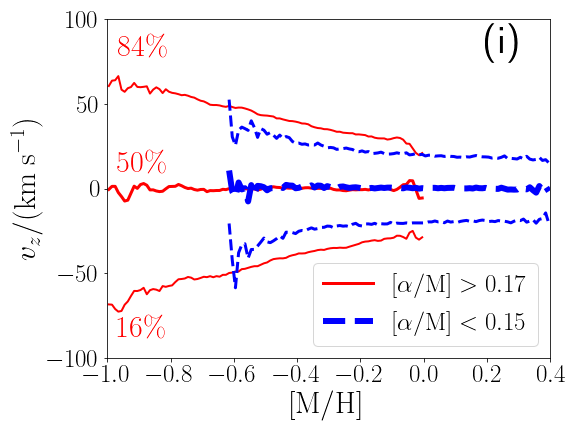} \\
\includegraphics[width=0.3\textwidth ]{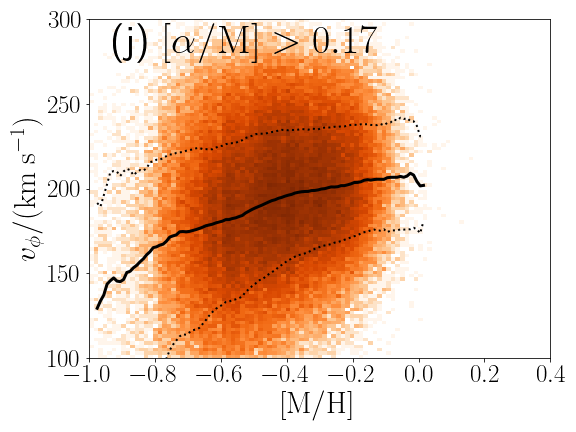}
\includegraphics[width=0.3\textwidth ]{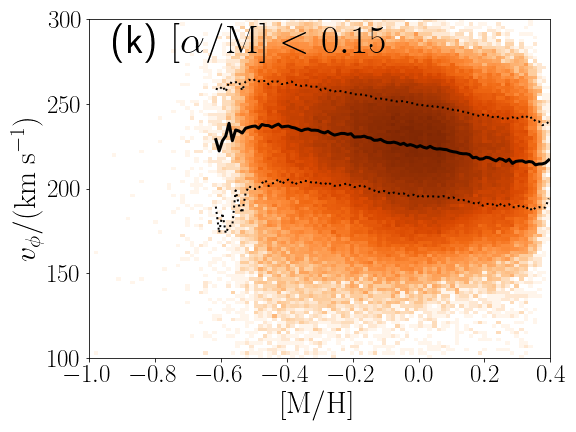}
\includegraphics[width=0.3\textwidth ]{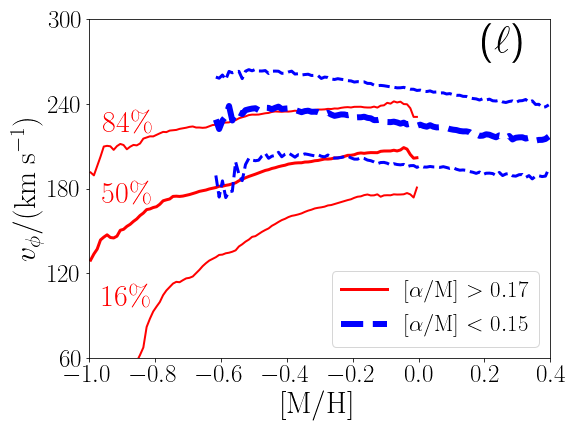} \\
\includegraphics[width=0.3\textwidth ]{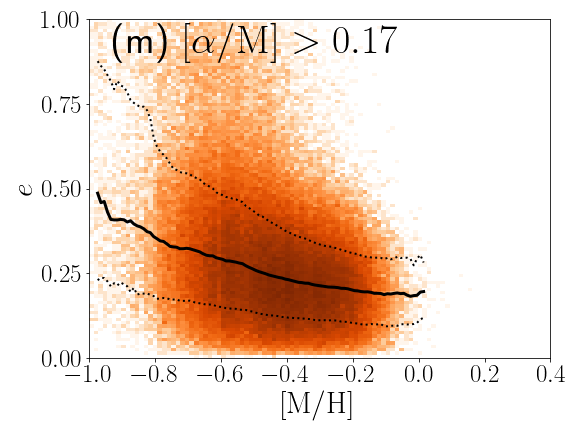}
\includegraphics[width=0.3\textwidth ]{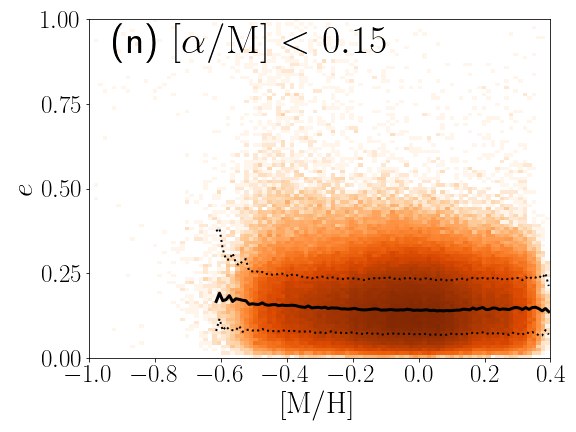}
\includegraphics[width=0.3\textwidth ]{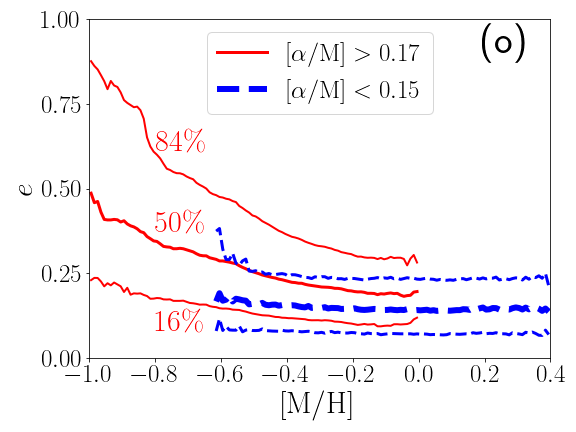}
\caption{
Chemo-dynamics of local giants with high-precision $(\mh, \am)$. 
(a) CMD selection of giants. 
(b) Selection box for high-$\alpha$ and low-$\alpha$ stars. 
(c) Velocity dispersion of high-$\alpha$ and low-$\alpha$ stars 
as a function of \mh\ in our catalog. 
(d)-(f) The distribution of the Galactocentric cylindrical radial velocity $v_R$ as a function of \mh. 
(g)-(i) The distribution of the Galactocentric vertical velocity $v_z$. 
(j)-($\ell$) The distribution of the Galactocentric azimuthal velocity $v_\phi$. 
(m)-(o) The distribution of the orbital eccentricity $e$. 
In panels (d)-(o), the three lines correspond to the 16th, 50th, and 84th percentile values. 
}
\label{fig:local_disc_giants_1bprp}
\end{figure*}

As shown in Table~\ref{table:rmse},
our models can infer most reliable estimates of (\mh, \am) for giants. 
Fig.~\ref{fig:local_disc_giants_1bprp} 
shows the chemo-dynamical correlation for local giants. 
As a basis for our discussion, 
we divide metal-rich stars at $-0.9 < \mh$ 
into low-$\alpha$ stars with $\am<0.15$ 
and high-$\alpha$ stars with $\am>0.17$. 
As we see from Fig.~\ref{fig:local_disc_giants_1bprp}(b), 
the \mh-distributions of these two populations 
overlap at $-0.6 < \mh < 0$. 
Notably, 
the low-$\alpha$ and high-$\alpha$ stars 
are dynamically distinct from each other 
in terms of the velocity distribution (panels (c), (f), (i), and ($\ell$)) 
and the eccentricity distribution (panel (o)). 
For example, 
as seen in Fig.~\ref{fig:local_disc_giants_1bprp}($\ell$), 
the median velocity $v_\phi$ for low-$\alpha$ stars 
declines as a function of \mh, 
while that for high-$\alpha$ stars 
increases as a function of \mh. 
This trend in $v_\phi$ is consistent 
with previous analysis of local disk stars 
\citep{Lee2011ApJ...738..187L}. 
Also, 
the velocity dispersion 
\revise{(panel (c))}\footnote{
\revise{
The velocity dispersion $\sigma_i$ ($i=R,\phi,z$) 
is defined such that 
the width between the 16th and 84th percentile values of $v_i$ 
is equal to $2\sigma_i$.
}
} 
and the median eccentricity 
\revise{(panels (m),(n), and (o))}
are almost constant as a function of \mh\ for low-$\alpha$ stars, 
but decline steadily for high-$\alpha$ stars. 
These results indicate that 
our model can tell the difference 
between high-\am\ stars and low-\am\ stars 
even if their \mh\ are similar.

\begin{figure*}
\raggedright
\includegraphics[width=0.3\textwidth ]{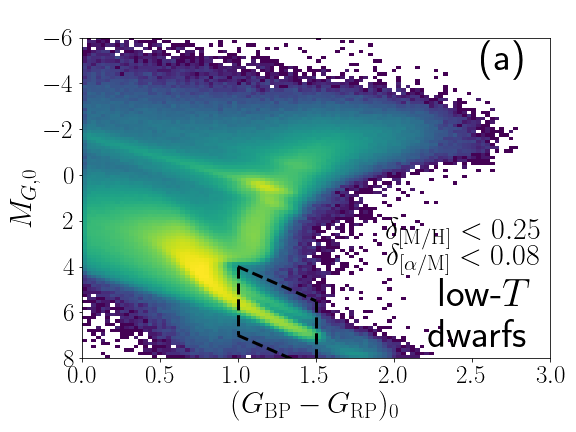}
\includegraphics[width=0.3\textwidth ]{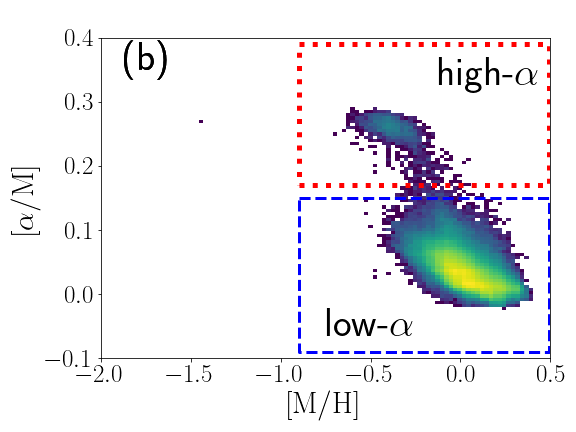}
\includegraphics[width=0.3\textwidth ]{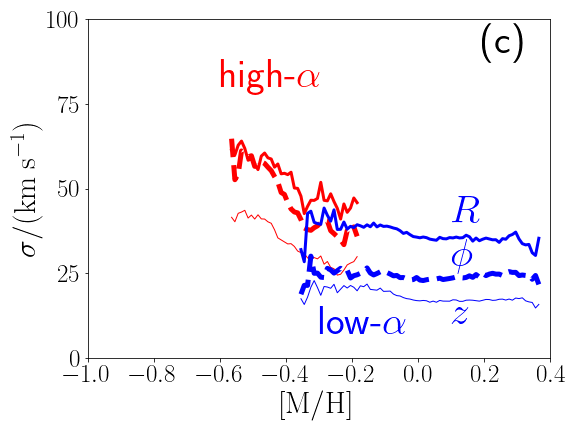}
\includegraphics[width=0.3\textwidth ]{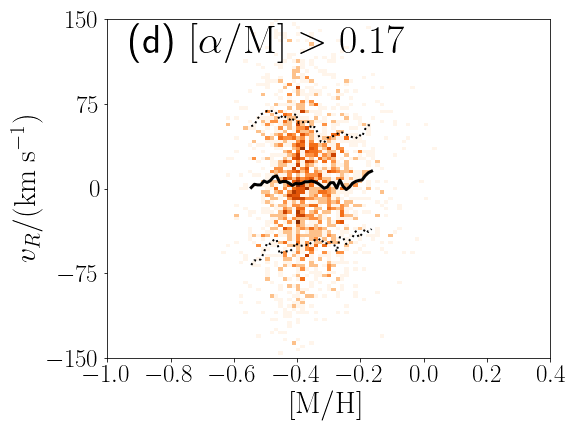}
\includegraphics[width=0.3\textwidth ]{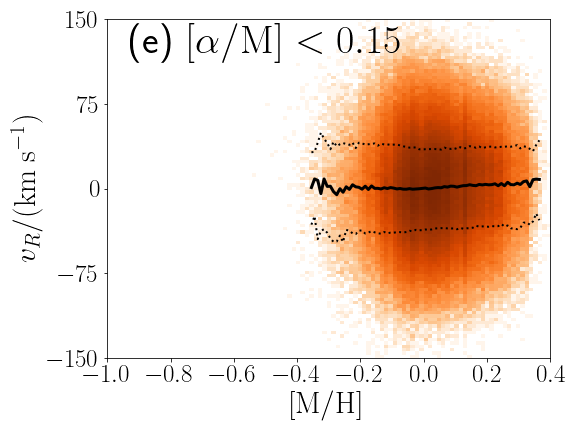}
\includegraphics[width=0.3\textwidth ]{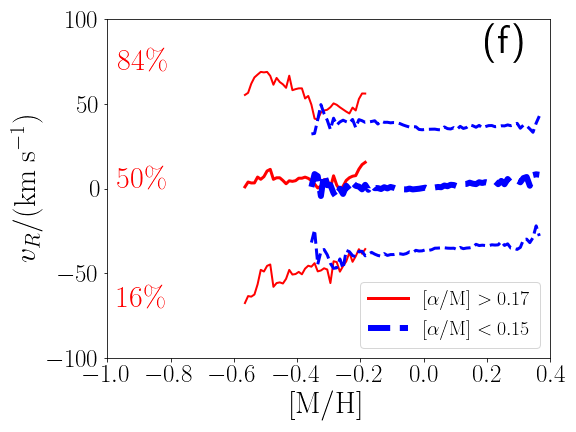}
\includegraphics[width=0.3\textwidth ]{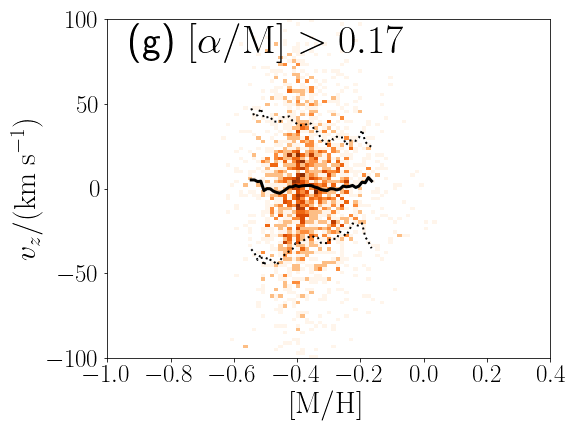}
\includegraphics[width=0.3\textwidth ]{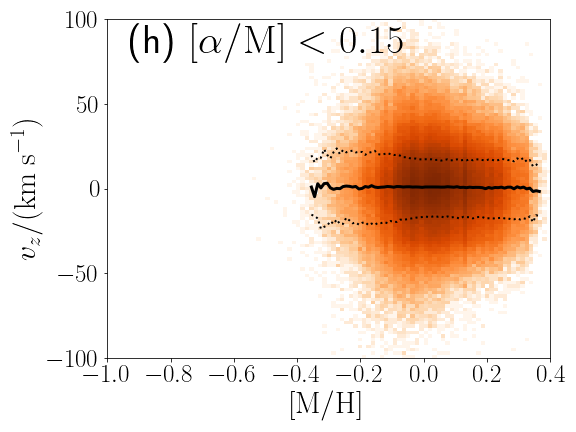}
\includegraphics[width=0.3\textwidth ]{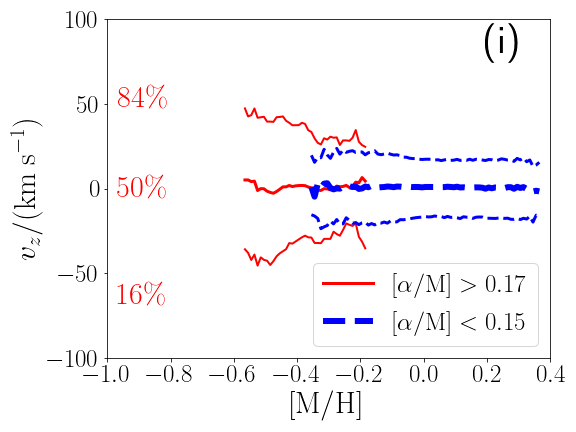}
\includegraphics[width=0.3\textwidth ]{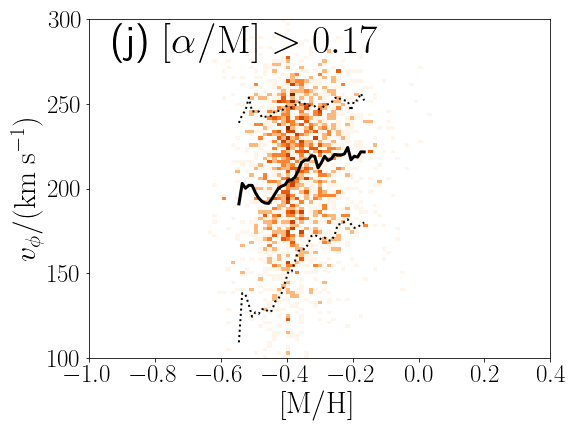}
\includegraphics[width=0.3\textwidth ]{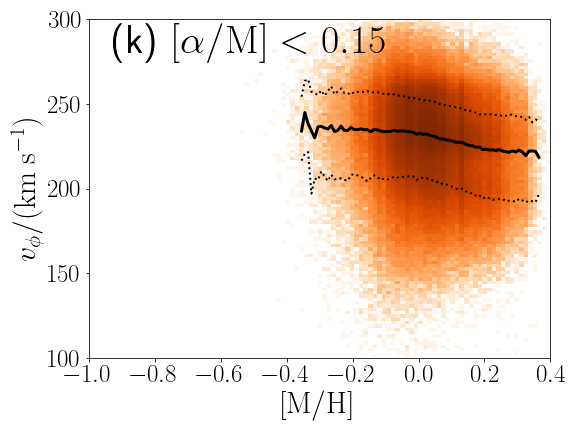}
\includegraphics[width=0.3\textwidth ]{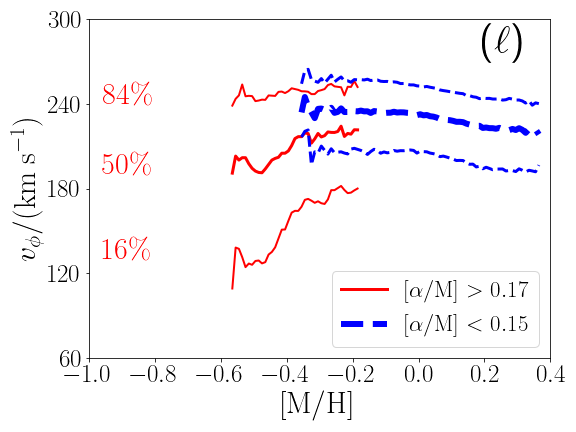}
\includegraphics[width=0.3\textwidth ]{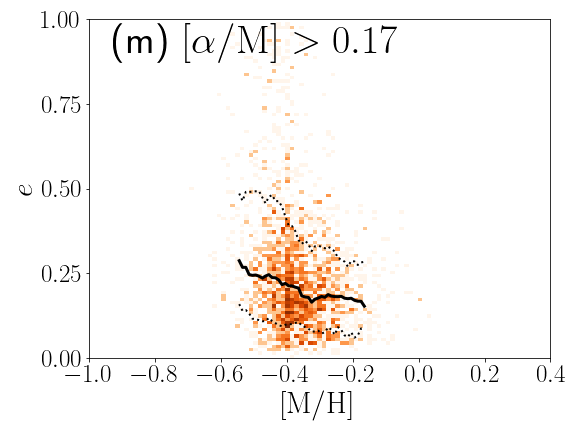}
\includegraphics[width=0.3\textwidth ]{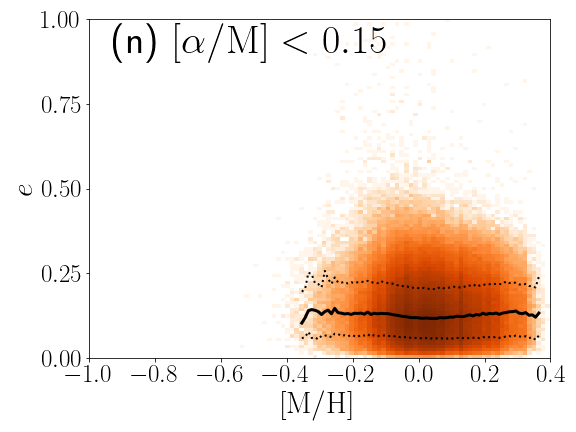}
\includegraphics[width=0.3\textwidth ]{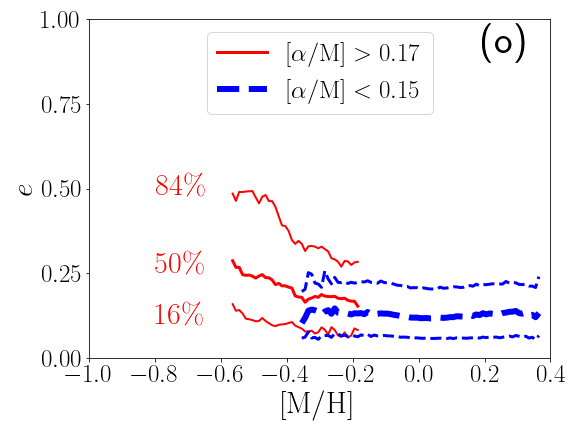}
\caption{
The same as Fig.~\ref{fig:local_disc_giants_1bprp}, 
but for local low-temperature dwarfs in our catalog. 
We note that the chemo-dynamical correlations for the low-temperature dwarfs 
(right-most column of this figure) 
are similar to those for giants 
(right-most column of Fig.~\ref{fig:local_disc_giants_1bprp}). 
This result is a supporting evidence that, 
at high-metallicity region (e.g., $\mh > -0.6$)
our estimates of (\mh, \am) 
for the high-precision subsample of low-temperature dwarfs are 
reliable. 
}
\label{fig:local_disc_dwarf_lowT}
\end{figure*}

Fig.~\ref{fig:local_disc_dwarf_lowT} 
is the same as Fig.~\ref{fig:local_disc_giants_1bprp}, 
but for local low-temperature dwarfs. 
Due to the limited sample size, 
the \mh-distributions of high-$\alpha$ and low-$\alpha$ subsamples 
only overlap at $-0.35 < \mh < -0.2$. 
However, we can still see a different trends 
for high-$\alpha$ and low-$\alpha$ stars 
at this overlapping metallicity region. 
As seen in Fig.~\ref{fig:local_disc_dwarf_lowT}($\ell$), 
we see a different trend in $v_\phi$ 
for low-$\alpha$ and high-$\alpha$ stars. 
Other panels in Fig.~\ref{fig:local_disc_dwarf_lowT} 
also show similar trends to the corresponding panels 
in Fig.~\ref{fig:local_disc_giants_1bprp}. 
These results provide a supporting evidence that 
our models can infer \am\ 
for low-temperature dwarfs.

\begin{figure*}
\raggedright
\includegraphics[width=0.3\textwidth ]{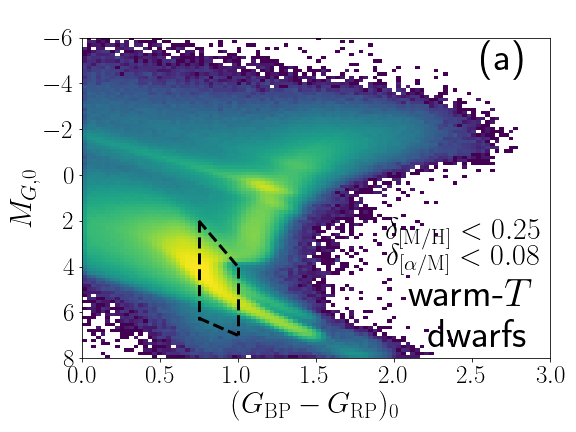}
\includegraphics[width=0.3\textwidth ]{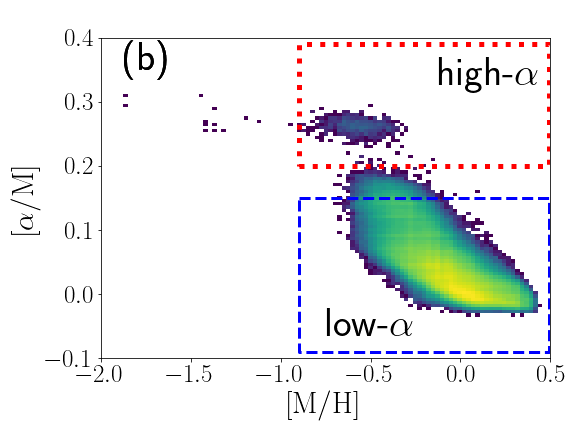}
\includegraphics[width=0.3\textwidth ]{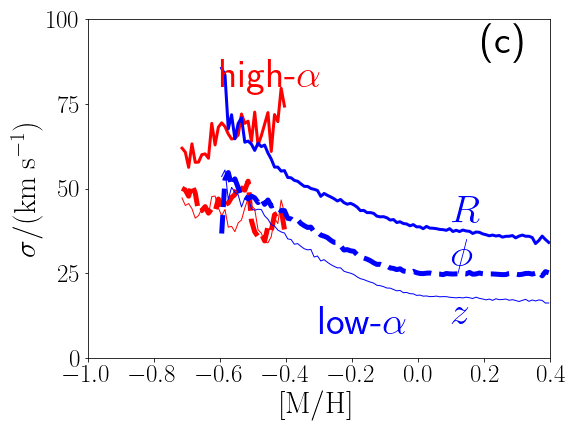}
\includegraphics[width=0.3\textwidth ]{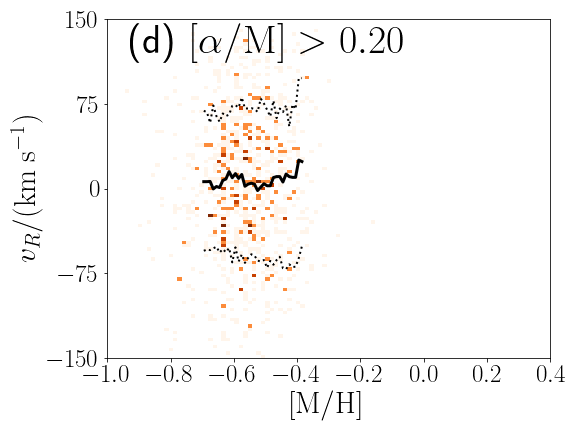}
\includegraphics[width=0.3\textwidth ]{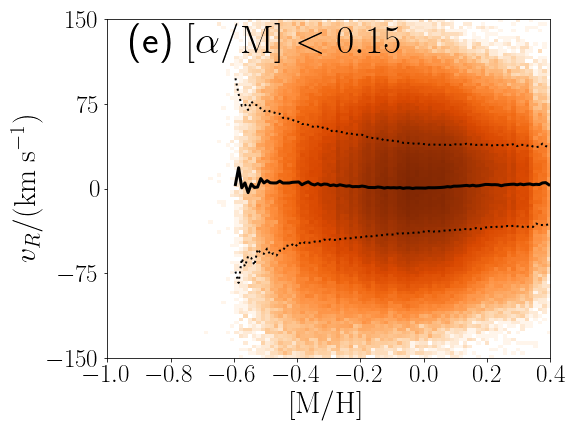}
\includegraphics[width=0.3\textwidth ]{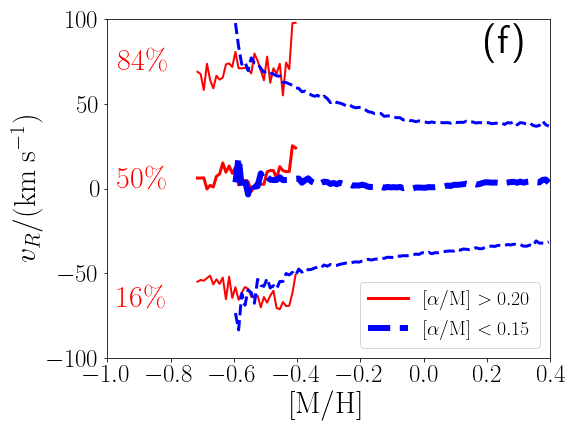}
\includegraphics[width=0.3\textwidth ]{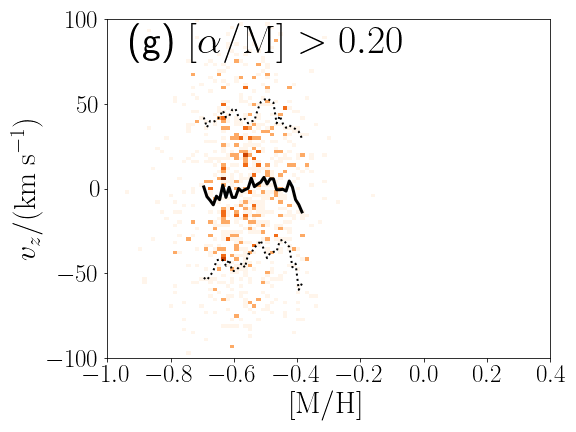}
\includegraphics[width=0.3\textwidth ]{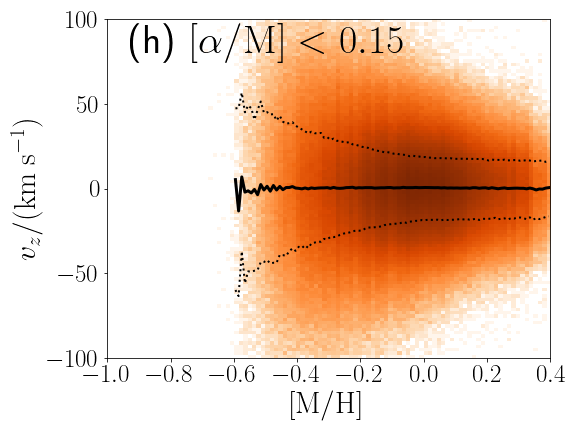}
\includegraphics[width=0.3\textwidth ]{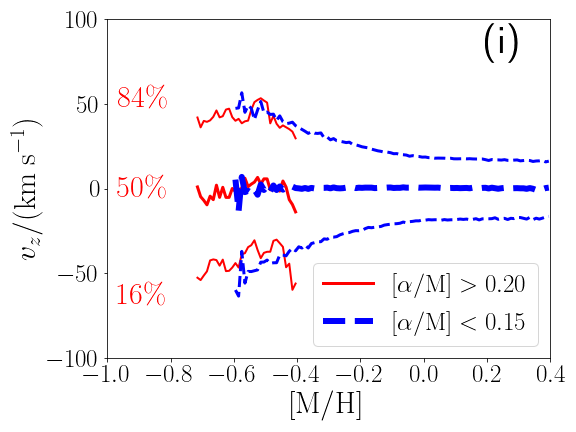}
\includegraphics[width=0.3\textwidth ]{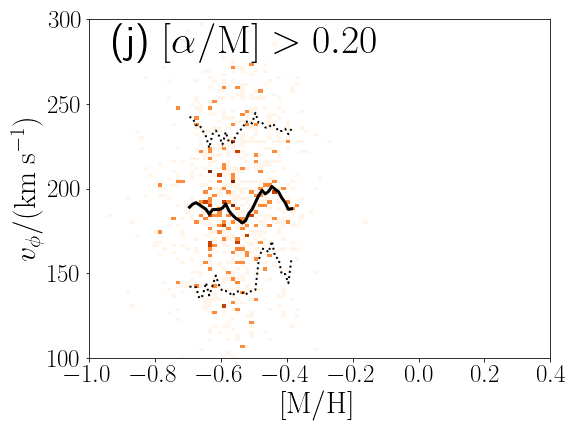}
\includegraphics[width=0.3\textwidth ]{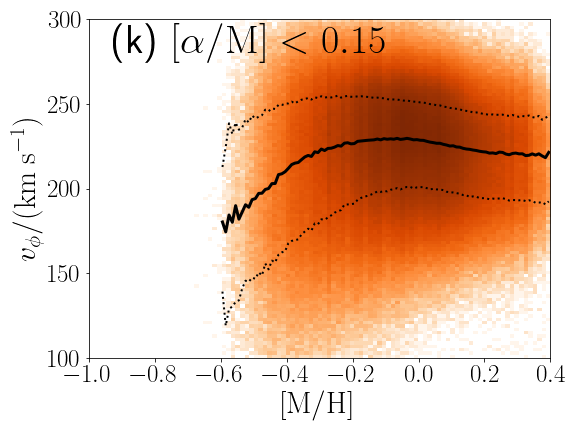}
\includegraphics[width=0.3\textwidth ]{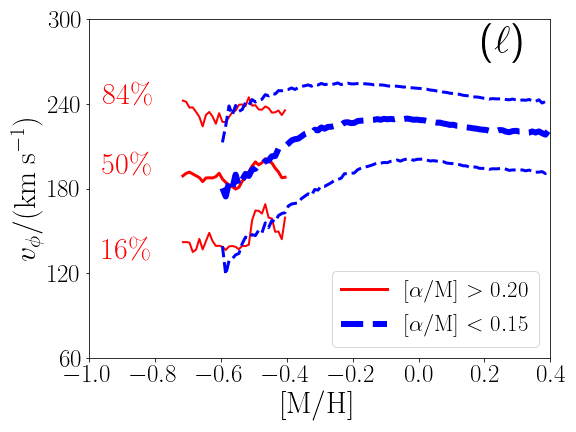}
\includegraphics[width=0.3\textwidth ]{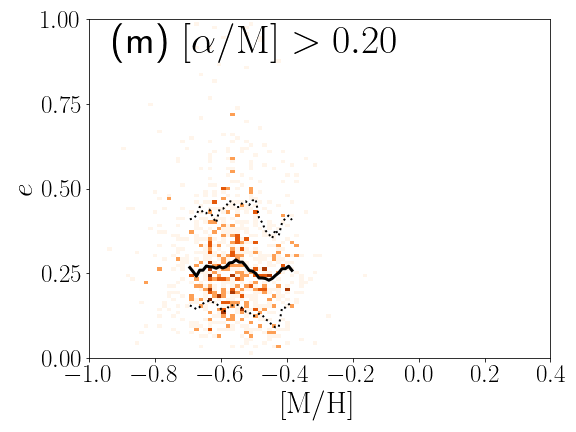}
\includegraphics[width=0.3\textwidth ]{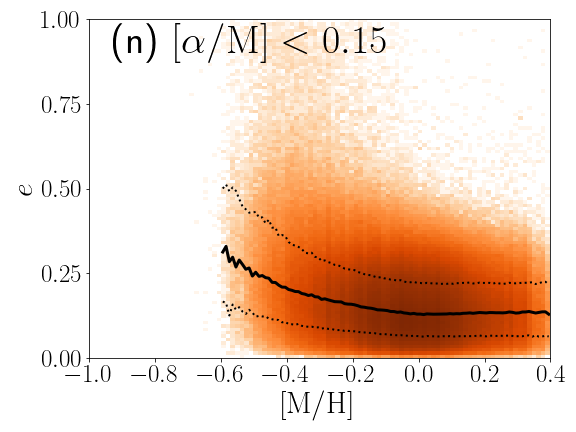}
\includegraphics[width=0.3\textwidth ]{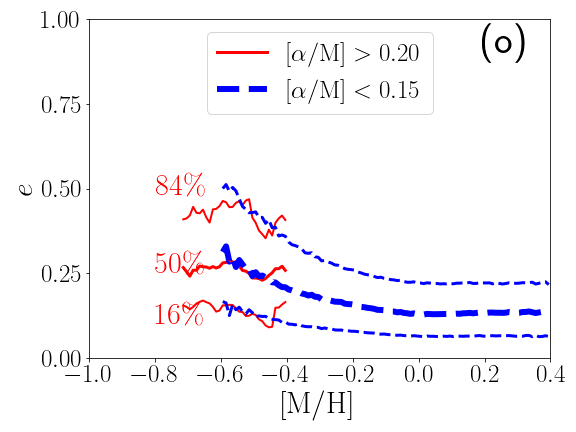}
\caption{
The same as Fig.~\ref{fig:local_disc_giants_1bprp}, 
but for local warm-temperature dwarfs in our catalog. 
We note that, for the warm-temperature dwarfs, 
the dynamical properties of high-$\alpha$ stars and low-$\alpha$ stars 
are similar to each other at their overlapping \mh\ (i.e., $\mh \simeq -0.5$). 
This result is in contrast to the result for giants and low-temperature dwarfs. 
This result implies that our estimates of \am\ 
for warm-temperature dwarfs 
may be less reliable than those for giants and low-temperature dwarfs. 
}
\label{fig:local_disc_dwarf_warmT}
\end{figure*}

Fig.~\ref{fig:local_disc_dwarf_warmT} 
is the same as Fig.~\ref{fig:local_disc_giants_1bprp}, 
but for local warm-temperature dwarfs. 
At first glance, 
the distinct bimodal distribution of warm-temperature dwarfs in the (\mh, \am)-space (Fig.~\ref{fig:local_disc_dwarf_warmT}(b)) 
gives an impression that our models can 
infer realistic estimates of \am. 
However, this may not be the case, 
based on the other panels in this figure. 
For the warm-temperature dwarf sample, 
we do not see distinct kinematical trends for 
low-$\alpha$ and high-$\alpha$ stars, 
unlike the samples of giants and low-temperature dwarfs. 
This result indicates that, 
our models can not tell the difference 
between high-\am\ and low-\am\ warm-temperature dwarfs 
for a given \mh, 
and therefore \am\ for warm-temperature dwarfs are 
(much) less reliable than \am\ for giants or low-temperature dwarfs, 
supporting the claims by \cite{Gavel2021A&A...656A..93G} and \cite{Witten2022MNRAS.516.3254W}. 
However, this result is 
slightly at odds with the results in 
Figs.~\ref{fig:test_data_am}(n)(p) and Figs.~\ref{fig:GALAH_data_am}(n)(p), 
where we see that those stars 
for which we predict to be high-$\alpha$ stars 
are more likely to be true high-$\alpha$ stars. 
We do not have a clear understanding on the reliability of (\mh, \am) for warm-temperature dwarfs. 
In any case, given the results in \cite{Witten2022MNRAS.516.3254W}, 
we need to be careful about using our (\mh, \am) for warm-temperature dwarfs.

\subsection{Chemistry of stars with Gaia-Sausage-Enceladus-like orbits}
\label{sec:GSE}

\begin{figure*}
\raggedright
\includegraphics[width=0.45\textwidth ]{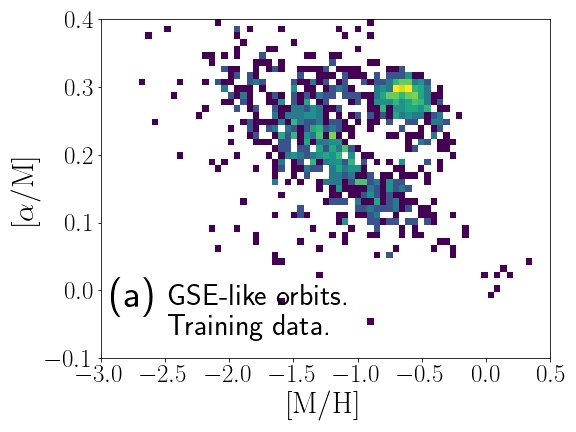}
\includegraphics[width=0.45\textwidth ]{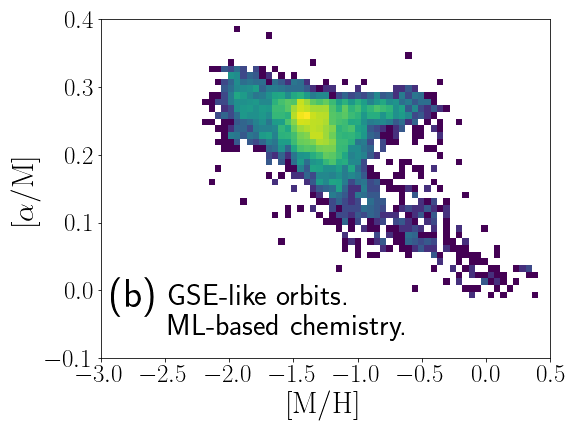}\\
\includegraphics[width=0.45\textwidth ]{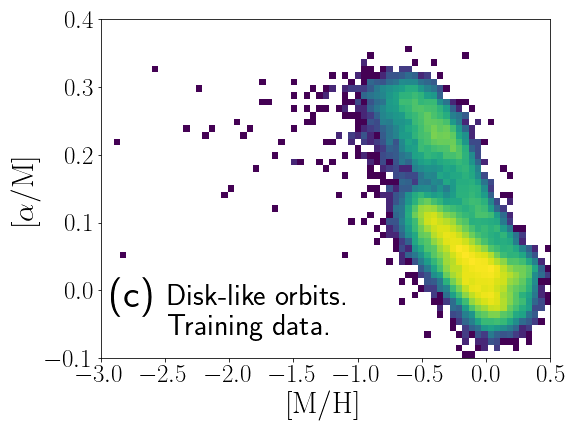}
\includegraphics[width=0.45\textwidth ]{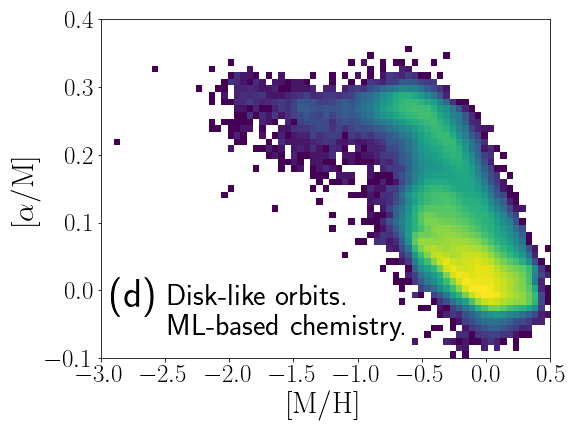}
\caption{
\revise{
The distribution of stars in chemical space 
for the kinematically selected sample. 
(a) Spectroscopically determined chemical abundances for 
stars in the training data with GSE-like orbits. 
The diagonal feature corresponds to the GSE component. 
(b) ML-based chemical abundances derived from Gaia XP spectra with GSE-like orbits. 
A clear GSE component is visible in the ML-based abundances.
(c) The same as (a), but for stars with disk-like orbits. 
The low-$\alpha$ and high-$\alpha$ components are clearly distinguishable. 
(d) The same as (b), but for stars with disk-like orbits. 
The chemical distributions in panels (c) and (d) are similar to each other. 
}
}
\label{fig:GSE}
\end{figure*}

\revise{
The stellar halo of the Milky Way is believed to have formed through mergers with dwarf galaxies and other stellar systems that accreted onto the proto-Milky Way. 
One of the largest merger remnants is known as the Gaia Sausage Enceladus 
(GSE; \citealt{Belokurov2018MNRAS.478..611B, Helmi2018Natur.563...85H}), 
which is characterized by nearly radial orbits. 
The GSE stars exhibit a diagonal trend in the (\mh,\am)-space 
(see Fig.~\ref{fig:GSE}(a) and description below), 
which is reminiscent of the chemical patterns 
seen in classical dwarf galaxies 
\citep{Tolstoy2009ARA&A..47..371T}. 
Given these kinematic and chemical properties, it is widely accepted that GSE is the remnant of the last major merger event experienced by the Milky Way.
Here we use the GSE stars to test the reliability of our ML models 
in inferring (\mh, \am), inspired by the approach of 
\cite{Gavel2021A&A...656A..93G}. 
}

\revise{
We define stars with GSE-like orbits by using 
the following criteria:
(i) The orbital energy $E$ satisfies $-1.5 \times 10^5 \;(\kms)^2 < E< 0$;
(ii) The azimuthal angular momentum $J_\phi$ satisfies $|J_\phi|<500 \kpc \kms$; 
(iii) Gaia DR3 provides radial velocity and reliable parallax ($\texttt{parallax\_over\_error}>5$). 
Criteria (i) and (ii) are motivated by recent studies on GSE properties (e.g., \citealt{Belokurov2023MNRAS.518.6200B}). 
When evaluating $(E,J_\phi)$, 
we assume a Galactic potential and Solar motion, 
as detailed in Appendix \ref{appendix:coordinate}. 
Criterion (iii) ensures that the uncertainties in $(E,J_\phi)$ are minimal. 
}

\revise{
Fig.~\ref{fig:GSE}(a) shows the distribution of stars 
in $(\mh,\am)$-space for our training data, 
using abundances derived from APOGEE or Li2022. 
We observe a diagonal feature in $(\mh,\am)$-space, 
spanning from $(\mh,\am)\simeq (-2, 0.3)$ to 
$(\mh,\am)\simeq (-0.5, 0.1)$. 
This diagonal feature is identified in the literature as the GSE component  
(e.g., Fig.5 of \citealt{Mackereth2019MNRAS.482.3426M}). 
Additionally, we note a blob at $(\mh,\am)\simeq (-0.7, 0.3)$, 
which likely corresponds to the high-eccentricity tail of the thick-disk component.
Fig.~\ref{fig:GSE}(b) shows the distribution of stars 
in $(\mh,\am)$-space from our catalog, 
where we use abundances predicted by our ML models, 
$(\mh^\mathrm{pred,50},\am^\mathrm{pred,50})$. 
In this case, 
we restrict the sample to the high-precision subsample defined by
$\deltaMH<0.25$, $\deltaAM<0.08$, and \texttt{{bool\_flag\_cmd\_good}=True}. 
Intriguingly, 
we can confirm the diagonal feature 
representing the GSE component 
and the high-$\alpha$ blob 
corresponding to the thick-disk component. 
While the high-$\alpha$ blob is 
horizontally connected to the GSE component, 
our result is reassuring 
because it demonstrates that our ML models are able to chemically identify GSE stars. 
In contrast to the models in \cite{Gavel2021A&A...656A..93G}, 
which were unable to chemically identify GSE stars, 
our models show a more reasonable behavior. 
}

\revise{
For completeness, 
we also examine the chemical properties of 
stars with disk-like orbits. 
We perform the same analysis, but with the selection criteria for
$(E,J_\phi)$ adjusted such that 
$E<1\times 10^5 (\kms)^2$ and $1500 \kpc\kms < J_\phi < 2500 \kpc\kms$ are satisfied. 
(Here, positive $J_\phi$ corresponds to prograde orbits.) 
As before, we find that the chemical distributions of the stars 
in the training data (Fig.~\ref{fig:GSE}(c)) 
and those of our catalog data (Fig.~\ref{fig:GSE}(d)) 
are similar to each other. 
}

\revise{
We emphasize that our models rely solely on Gaia XP spectra and 
$G$-band photometry to derive $(\mh,\am)$, and do not incorporate any positional or kinematic information about the stars. 
Consequently, our models do not leverage any correlation between chemistry and kinematics. 
The fact that our catalog displays a chemo-dynamical correlation similar to that in the training data suggests that our models effectively extract meaningful chemical information from Gaia XP spectra.
}

\section{Interpretation of the QRF models} \label{sec:interpret}

To better understand/interpret how our models work, 
we conduct an additional analysis 
using the grouped Permutation Feature Importance (gPFI) method. 
We note that the analysis in this Section 
is independent from the main analysis 
in this paper.

\subsection{Simple QRF models}
\label{sec:simpleQRF}

In the main analysis of this paper, 
we construct our models 
that use the 1310-dimensional data vector 
$X = (\vector{F}_\mathrm{BP}, \vector{F}_\mathrm{RP}, \vector{C})$ 
as their inputs. 
In this Section, 
we construct simpler models 
in the same manner as in Section \ref{sec:model} 
except that we use a 1200-dimensional data vector 
\eq{
X = (\vector{F}_\mathrm{BP}, \vector{F}_\mathrm{RP}) . 
}
Because 
the information of the normalized spectral coefficients $\vector{C}$ 
and 
that of the normalized mean BP and RP spectra $(\vector{F}_\mathrm{BP}, \vector{F}_\mathrm{RP})$ 
are redundant (see Fig.~\ref{fig:XPschematic}),  
removing $\vector{C}$ from $X$ 
makes the interpretation of the model more transparent. 
Following Section \ref{sec:model}, 
we separately train  
a model to estimate \mh\ (Simple-QRF-MH model)
and 
another model to estimate \am\ (Simple-QRF-AM model).

As in Section \ref{sec:rmse}, 
we apply the Simple-QRF-MH model and Simple-QRF-AM model 
to the test data and evaluate the RMSE values. 
We find that the RMSE values are 
$0.119$ dex for \mh\ 
and 
$0.0556$ dex for \am, 
when we use the entire test data.\footnote{
Given that 
we use the same set of training/test data as in Section \ref{sec:model}, 
it is intriguing to note that 
these RMSE values are 
$\sim$30\% larger (worse) than those in the main analysis 
($0.0890$ dex for \mh\ and $0.0436$ dex for \am; 
see Table~\ref{table:rmse}). 
Actually, this is the reason why we choose 
$X = (\vector{F}_\mathrm{BP}, \vector{F}_\mathrm{RP}, \vector{C})$ 
as the input vector in the main part of this paper rather than 
$X = (\vector{F}_\mathrm{BP}, \vector{F}_\mathrm{RP})$. 
\revise{See Appendix \ref{appendix:dimension} for some discussion.}
}
In the following, 
we will analyze the Simple-QRF-MH and Simple-QRF-AM models.

\begin{figure*}
\centering 
\includegraphics[width=0.9\textwidth ]{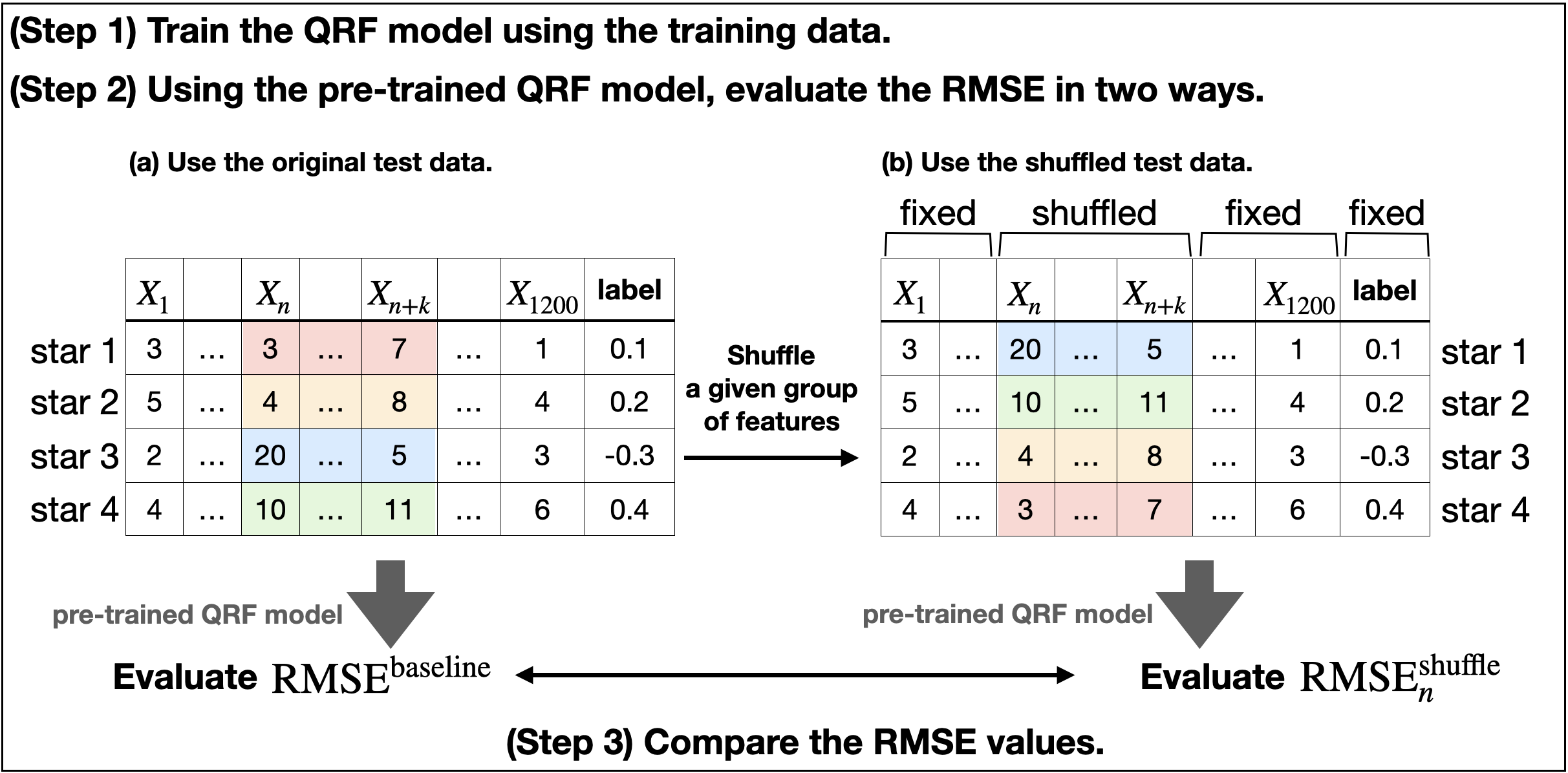} 
\caption{
The schematic explanation of gPFI analysis. 
We make a shuffled test data 
in which the grouped features $(X_n, \cdots, X_{n+k})$ 
are shuffled randomly. 
(We do not shuffle the labels in the test data.) 
If the grouped features are important, shuffling them would make the prediction worse.
Thus, by investigating the ratio of the RMSE in (b) to the RMSE in (a), we can evaluate the importance of the grouped features.
}
\label{fig:gPFI_schematic}
\end{figure*}

\begin{figure*}
\centering 
\includegraphics[width=0.47\textwidth ]{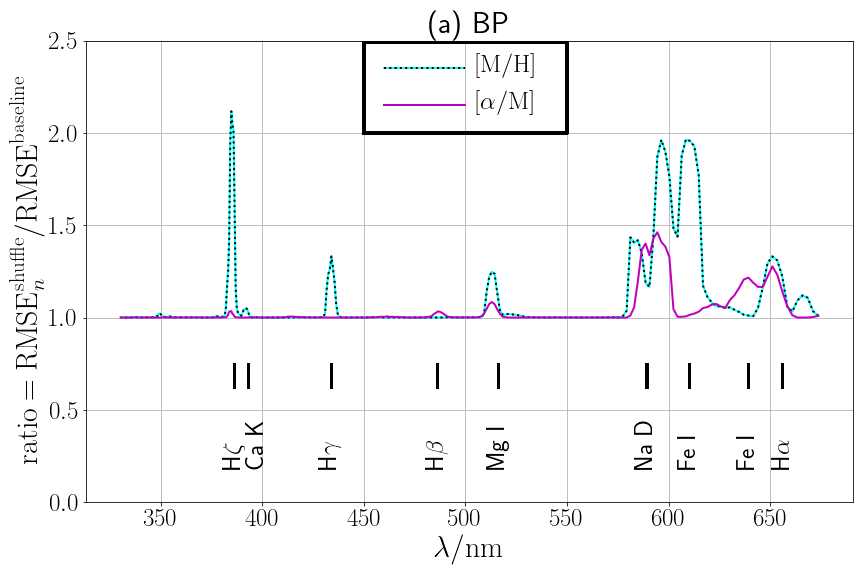} 
\includegraphics[width=0.47\textwidth ]{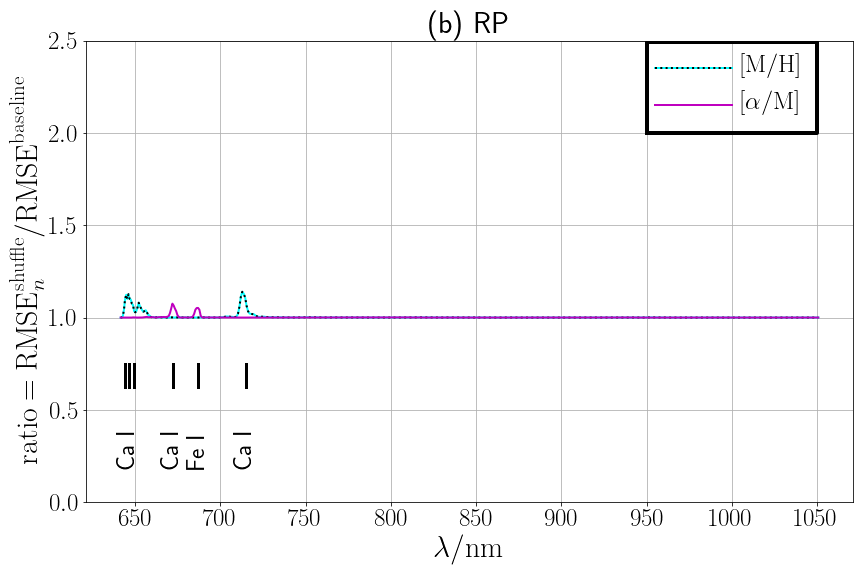} 
\caption{
The importance of each wavelength range in the XP spectra 
in inferring the labels (\mh, \am) 
estimated from the gPFI analysis. 
The vertical axis ($\mathrm{ratio} = \mathrm{RMSE}^\mathrm{shuffle}_n/\mathrm{RMSE}^\mathrm{baseline}$) 
indicates how the performance of the ML models 
deteriorate 
if the flux data near a given wavelength region are shuffled. 
In panels (a) 
The results for BP and RP spectra are shown in panels (a) and (b), respectively. 
We also plot the locations of some important absorption lines. 
We note that the BP spectra contain much more information than RP spectra. 
Intriguingly, our model seem to extract information on \am\ 
from the spectral shape near Na D lines (at 589.0 nm and 589.6 nm). 
}
\label{fig:gPFI_result_bp_rp}
\end{figure*}

\subsection{gPFI analysis}
\label{sec:gPFI}

The procedure of the gPFI analysis 
is schematically described in Fig.~\ref{fig:gPFI_schematic}. 
For simplicity, let us focus on the procedure 
for the Simple-QRF-MH model.  
As we have mentioned in the previous subsection, 
this model has a RMSE in \mh\ 
of 0.119 dex 
when applied to the test data set. 
We use this value as the baseline value, 
$\mathrm{RMSE}^\mathrm{baseline}$. 
In the gPFI analysis, 
we apply this pre-trained Simple-QRF-MH model 
to a `shuffled' test data set. 
To be specific,  
among the 1200-dimensional information of $X$, 
we combine $k$ consecutive features $\{ X_n, \cdots, X_{n+k}\}$ as a group,\footnote{
In general, the grouped features 
are not necessarily consecutive. 
We group consecutive features 
because the flux values of neighboring 
wavelengths are correlated. 
} 
and shuffle this group of features 
in the test data to create 
a shuffled test data set. 
In the shuffled test data, 
the stellar labels are not shuffled. 
We then compute the RMSE value for the shuffled test data 
(denoted as $\mathrm{RMSE}^\mathrm{shuffled}_n$)
and compare the value with the baseline value $\mathrm{RMSE}^\mathrm{baseline}$.

The interpretation of the gPFI method is simple. 
On the one hand, 
suppose that the shuffled features do not contain any information on \mh. 
In this case, 
shuffling them would result in almost identical RMSE value ($\mathrm{RMSE}^\mathrm{shuffled}_n/\mathrm{RMSE}^\mathrm{baseline}\simeq1$).\footnote{
\revise{
By construction, QRF models (and other RF-based models) 
naturally ignore input features that do not provide useful information for predicting the label $Y$. 
When a QRF model is trained, decision trees are constructed by randomly splitting the feature space $X$,
with each split chosen to maximize the prediction quality. 
and constructs a decision tree 
such that the split is beneficial to predict the label. 
If the $n$th dimension in the input vector ($X_n$) contains no information that helps predict $Y$, 
then $X_n$ will not contribute to effective splits, 
and will therefore tend to be ignored in the tree-building process. 
In such a case, a trained QRF model will predict $Y$ without using $X_n$. 
As a result, if $X_n$ is shuffled (i.e., its values are randomly permuted), it would have little or no impact on the predicted label $Y$, since the model does not rely on $X_n$ in the first place. 
However, if $X_n$ does have some weak or indirect predictive power, shuffling it might cause small changes in the prediction, though the effect would be minimal. 
A unique characteristic of our QRF models, 
is that they focus on the most important part of the spectrum. 
This approach contrasts with forward-modeling methods, such as those used in \cite{Zhang2023MNRAS.524.1855Z} 
where the full coefficient vector and its covariance are used to maximize the likelihood or posterior of the data.
}
} 
On the other hand, 
suppose that the shuffled features contain some useful information on \mh. 
In this case, 
shuffling them would increase the RMSE value (i.e., deteriorate the prediction of \mh), 
so that the ratio 
$\mathrm{RMSE}^\mathrm{shuffled}_n/\mathrm{RMSE}^\mathrm{baseline}$ becomes notably larger than 1. 
The gPFI analysis measures the importance of the grouped features  
by measuring how the RMSE increases due to the shuffling.

We remind that $X$ is the normalized flux value 
sampled at 1200 points in the pseudo-wavelength range. 
Thus, $\{ X_n, \cdots, X_{n+k}\}$ 
corresponds to the normalized flux at a certain wavelength region 
within the BP band if $n+k \leq 600$, 
and 
within the RP band if $601 \leq n$. 
In our analysis, 
we choose integer values at 
$132 \leq n \leq 510$ 
(BP domain; corresponding to $680 \geq \lambda/\mathrm{nm} \geq 330$)
and 
$733 \leq n \leq 1107$ 
(RP domain; corresponding to $640 \leq \lambda/\mathrm{nm} \leq 1050$). 
We also set $k=5$, but the choice of $k$ 
only has a small effect on the result.

\subsection{Results of gPFI analysis} \label{sec:gPFI_results}

We conduct the gPFI analysis for 
Simple-QRF-MH model and Simple-QRF-AM model, 
as shown in Fig.~\ref{fig:gPFI_result_bp_rp}. 
In Fig.~\ref{fig:gPFI_result_bp_rp}, 
the horizontal axis is the physical wavelength 
that is converted from the pseudo wavelength 
(see Fig.~\ref{fig:XPschematic}). 
(To be specific, 
we compute the $n$th and $(n+k)$th 
wavelengths and use the mean value as the horizontal axis.) 
The vertical axis is evaluated by 
\eq{
\mathrm{(ratio)} = 
\mathrm{RMSE}^\mathrm{shuffled}_n / 
\mathrm{RMSE}^\mathrm{baseline}. 
}

From Fig.~\ref{fig:gPFI_result_bp_rp}, 
we see that our models extract information on (\mh, \am) 
from several narrow wavelength regions, mostly in the BP spectra.\footnote{ 
\revise{
In the following, 
we assign line labels based on the spectrum of Arcturus, as provided in  \cite{Hinkle2000vnia.book.....H}. 
}
}
Importantly, a large portion of the information on \am\ 
seem to arise from the spectral feature near the Na D lines 
(at 589.0 nm and 589.6 nm). 
As demonstrated in Appendix \ref{sec:Na_Mg}, 
there is an observational trend such that 
stars with higher [Na/Fe] tend to have higher [Mg/Fe] 
for a given metallicity [Fe/H]. 
Thus, we interpret that 
our Simple-QRF-AM model probably extracts information on \am\ 
from the Na D lines through this correlation. 
Fig.~\ref{fig:gPFI_result_bp_rp}(a) indicates that 
the Mg I line at 516 nm is also used to infer \am.
Therefore, reassuringly, our Simple-QRF-AM model is not entirely dependent on the data near the Na D lines.

Interestingly, the same Mg I line 
is also used by our Simple-QRF-MH model to infer \mh. 
This result means that the Simple-QRF-MH model 
is using the correlation between the Mg abundance and the overall metal abundance. 
Also, Fig.~\ref{fig:gPFI_result_bp_rp}(a) indicates 
that some 
\revise{Balmer lines} 
(e.g., H$\zeta$ and H$\gamma$) 
are also used to extract information on \mh. 
These lines are sensitive to the surface temperature (color), 
and thus they are probably informative to infer \mh\ 
through the color-metallicity relationship \citep{Yang2022A&A...659A.181Y}.

The gPFI analysis is helpful for humans to understand 
how the ML models extract information on (\mh, \am). 
However, we have some wavelength regions 
for which we do not fully understand why the models find useful (or useless). 
For example, 
we see notable features at 
639 nm in Fig.~\ref{fig:gPFI_result_bp_rp}(a)
or at 687 nm in Fig.~\ref{fig:gPFI_result_bp_rp}(b), 
but we do not understand why these wavelength regions are
more important than other wavelength regions.\footnote{
\revise{
According to the NIST Atomic Spectra Database \citep{NIST2024} 
available at \url{https://www.nist.gov/pml/atomic-spectra-database}, 
there are several strong Fe I lines, 
including those at 
639.4,
640.0,
641.2,
642.1,  
643.1, 
690.6, and
692.2 nm. 
These Fe I lines may potentially explain the features 
at 639 nm and 687 nm in Fig.~\ref{fig:gPFI_result_bp_rp}. 
However, since these lines are from Fe I, it remains uncertain whether they are useful for extracting \am. 
}
} 
As another example, 
we do not see any features in Fig.~\ref{fig:gPFI_result_bp_rp}(b) 
near the Ca triplet lines 
(at 849.8 nm, 854.2 nm, and 866.2 nm), 
although we know that Ca triplet is a strong absorption feature 
which has long been used to infer (\mh, \am) 
from high-resolution spectroscopy.

We note that the gPFI analysis in this Section is based on 
Simple-QRF-MH and Simple-QRF-AM models, 
which are trained on 1200-dimensional data.  
These models are technically different from 
QRF-MH and QRF-AM models in the main analysis of this paper, 
which are trained on 1310-dimensional data
(see Fig.~\ref{fig:XPschematic}). 
However, we believe that the interpretation of the models 
obtained in this Section is applicable to the QRF-MH and QRF-AM models 
as well, 
because the models are quite similar to each other. 
(For example, we can naturally guess that QRF-AM model also uses 
the Na D lines to infer \am.) 
We note that gPFI analysis is just one of many tools to 
interpret the ML models, 
and we should try other tools to try to understand the ML models. 
In the future, 
we hope that our gPFI analysis may serve as a useful starting point 
to understand the blurred information encoded in Gaia XP spectra.

\section{Conclusion} \label{sec:conclusion}

In this paper, we use a tree-based ML algorithm 
to infer (\mh, \am) from the low-resolution Gaia XP spectra. 
The estimation of \mh\ from \revise{the} XP spectra have been tried by various groups 
(e.g., \citealt{Rix2022ApJ...941...45R, Andrae2023ApJS..267....8A, Leung2023MNRAS.tmp.2896L}), 
but the estimation of \am\ from \revise{the} XP spectra 
has been recognized as a difficult task, 
based on theoretical arguments 
\citep{Gavel2021A&A...656A..93G, Witten2022MNRAS.516.3254W}. 
Prior to our work, 
only one group tackled this task (\citealt{Li2023arXiv230914294L}; 
but see also 
\citealt{Guiglion2024A&A...682A...9G}), 
who used a modern ML architecture. 
The uniqueness of this paper is that 
we tackled this problem 
by using a classical, tree-based ML algorithm, 
which allows us to interpret our models 
in a straightforward manner. 
This paper is summarized as follows.

\begin{enumerate}
\item 
We separately construct a model 
to infer \mh\ (QRF-MH model) 
and a model to infer \am\ (QRF-AM model), 
by using the training data 
of giants and dwarfs 
with known chemistry \citep{Abdurrouf2022ApJS..259...35A, Li2022ApJ...931..147L} 
located in low-dust extinction region with 
$E(B-V)<0.1$. 
In the main analysis, we use 
1310-dimensional information of Gaia XP spectra 
(shown in Fig.~\ref{fig:XPschematic}), 
consisting of 
the 110-dimensional coefficient data 
and the 1200-dimensional flux data. 
The derived catalog of (\mh, \am) is publicly available 
(\url{https://zenodo.org/records/10902172}).

\item  
We investigate the performance of our models 
by using the test data.  
The RMSE values for our models 
(which indicate the typical accuracy of our models)
are 
0.0890 dex for \mh\ and 0.0436 dex for \am. 
The accuracy in \mh\ is comparable 
to that in previous works 
(e.g., \citealt{Rix2022ApJ...941...45R, 
Andrae2023ApJS..267....8A,Leung2023MNRAS.tmp.2896L}). 
The accuracy in \am\ is also comparable 
to that in \cite{Li2022ApJ...931..147L}, 
who used a modern ML architecture. 
We note that 
our models are only applicable to stars with 
low dust extinction, 
while other previous models 
(e.g., \citealt{Rix2022ApJ...941...45R, Zhang2023MNRAS.524.1855Z, Li2023arXiv230914294L}) 
are applicable to stars with high dust extinction as well. 
However, we note that these previous studies 
used not only Gaia XP spectra and Gaia's photometric information 
but also the parallax information or photometric data from external catalogs. 
In contrast, 
we only use Gaia XP spectra and $G$-band magnitude, 
without using other information. 
Improving the classical models' performance for stars with high dust extinction 
by using external data is the scope for future studies.

\item 
Our models are more reliable for 
metal-rich stars ($\mh>-1$) 
than for metal-poor stars ($\mh<-1$; see Table~\ref{table:rmse}). 
This is mainly because we have less training stars with low \mh.
In the future, 
it would be important to increase the number of low-\mh\ stars 
(with known \am) in the training sample.

\item  
Our estimates of (\mh, \am) 
are most reliable for giants 
(see Table~\ref{table:rmse} and the second row in Figs.~\ref{fig:test_data_mh}, 
\ref{fig:test_data_am}, 
\ref{fig:GALAH_data_mh}, and 
\ref{fig:GALAH_data_am}). 
This is mainly because our training data are dominated by giants. 
The (\mh, \am)-distribution of giants obtained from our analysis 
shows a clear bimodal feature of high-$\alpha$ 
and low-$\alpha$ sequence. 
The bimodality is more prominent 
if we select the high-precision subsample 
with smaller uncertainty in (\mh, \am). 
Intriguingly, as seen in Fig.~\ref{fig:local_disc_giants_1bprp}, 
the low-\am\ and high-\am\ giants show distinct kinematics, 
consistent with the results in the literature 
(e.g., \citealt{Lee2011ApJ...738..187L}). 
This chemo-dynamical correlation is 
a supporting evidence that 
our estimates of (\mh, \am) are reliable
(see also \cite{Li2023arXiv230914294L}, 
who also used this argument 
to validate their estimates of (\mh, \am) from the XP spectra). 

\item  
Our estimates of (\mh, \am) are less reliable for dwarfs than for giants (see Table~\ref{table:rmse}). 
However, we see a positive correlation 
between our prediction of \mh\ (or \am) 
and the spectroscopically determined \mh\ (or \am) 
for dwarfs (see third and fourth rows in 
Figs.~\ref{fig:test_data_mh} and \ref{fig:test_data_am}). 
Because we have small number of dwarfs in the APOGEE DR17 catalog, 
we used GALAH DR3 data as the `external' test data 
to evaluate the performance of dwarfs. 
As seen in the third and fourth row in 
Figs.~\ref{fig:GALAH_data_mh} and \ref{fig:GALAH_data_am}, 
our models predict reasonable (\mh, \am) for dwarfs as well. 
In the future, 
it would be interesting to include the GALAH data 
in the training sample so that the model performance 
can be improved for dwarfs.

\item  
The chemo-dynamical correlation seen for giants 
is also confirmed for low-temperature dwarfs 
(see Fig.~\ref{fig:local_disc_dwarf_lowT}). 
This chemo-dynamical correlation serves as 
a supporting evidence that 
our estimates of (\mh, \am) are informative 
for low-temperature dwarfs. 
In contrast, 
the chemo-dynamical correlation is not seen for 
warm-temperature dwarfs 
(see Fig.~\ref{fig:local_disc_dwarf_warmT}). 
This result is at odds with 
the apparently good performance of our models 
for warm dwarfs that we see in 
Figs.~\ref{fig:test_data_mh}, \ref{fig:test_data_am}, 
\ref{fig:GALAH_data_mh} and \ref{fig:GALAH_data_am}. 
Given the theoretical arguments 
on the difficulty of inferring \am\ 
for warm-temperature dwarfs \citep{Witten2022MNRAS.516.3254W}, 
we need to be careful in using our (\mh, \am) for warm dwarfs. 

\item 
To understand how our models infer (\mh, \am), 
we quantify which part of the input XP spectra are important 
by using a so-called gPFI method (Section \ref{sec:interpret}). 
In this analysis, 
we separately construct a model 
to infer \mh\ (Simple-QRF-MH model) 
and a model to infer \am\ (Simple-QRF-AM model), 
by using the 1200-dimensional flux information of Gaia XP spectra 
(see Fig.~\ref{fig:XPschematic}). 
We find that the information on (\mh, \am) is contained in 
several narrow wavelength regions, 
many of which are located within the blue part of the XP spectra. 
Importantly, 
the Na D lines (589 nm) and the Mg I line (516 nm) 
seem to be important to estimate \am. 
This finding is intriguing because 
the correlation between Na and Mg abundances is known from literature 
(see Fig.~\ref{fig:Na_and_Mg}) 
and Mg is a typical $\alpha$ element. 
Identifying the wavelength ranges that are useful 
to infer chemistry may be useful in the future 
to derive chemical abundances from narrow-band photometric data 
from large surveys, 
such as Pristine survey \citep{Martin2023arXiv230801344M} 
or J-PLUS/S-PLUS survey \citep{Yang2022A&A...659A.181Y}.

\item 
Various medium/high-resolution 
spectroscopic surveys are ongoing or forthcoming, 
including but not limited to 
WEAVE \citep{Jin2023MNRAS.tmp..715J}, 
4MOST \citep{deJong2019Msngr.175....3D}, and 
PFS \citep{Takada2014PASJ...66R...1T}. 
These datasets would provide useful training data to further refine our ML models, which would be beneficial in understanding the chemical distribution within the Milky Way.

\end{enumerate}

\acknowledgments

\revise{K.H. thanks the anonymous referee for constructive and helpful comments.}
K.H. thanks Ian U. Roederer, Monica Valluri, Eric Bell, Leandro Beraldo e Silva, Akifumi Okuno, Daisuke Taniguchi, Eugene Vasiliev, 
Vasily Belokurov, \revise{and Yuan-Sen Ting} for discussion. 
K.H. thanks Tadafumi Matsuno for his insightful comments on the correlation between Na and Mg abundances. 
This work is partly conducted while the author was hospitalized due to an injury, and the author thank the kindness of the people involved.  
K.H. is supported by JSPS KAKENHI Grant Numbers JP24K07101, JP21K13965 and JP21H00053.

This work has made use of data from the European Space Agency (ESA) mission
{\it Gaia} (\url{https://www.cosmos.esa.int/gaia}), processed by the {\it Gaia}
Data Processing and Analysis Consortium (DPAC,
\url{https://www.cosmos.esa.int/web/gaia/dpac/consortium}). Funding for the DPAC
has been provided by national institutions, in particular the institutions
participating in the {\it Gaia} Multilateral Agreement.

Funding for the Sloan Digital Sky Survey IV has been provided by the Alfred P. Sloan Foundation, the U.S. Department of Energy Office of Science, and the Participating Institutions. SDSS acknowledges support and resources from the Center for High-Performance Computing at the University of Utah. The SDSS web site is \url{www.sdss4.org}.

\facility{Gaia}

\software{
AGAMA \citep{Vasiliev2019_AGAMA},\;
dustmaps \citep{Green2018JOSS....3..695M}, 
\revise{
GaiaXPy \citep{Montegriffo2023A&A...674A...3M}
}, 
matplotlib \citep{Hunter2007},
numpy \citep{vanderWalt2011},
scipy \citep{Jones2001}}
quantile-forest \citep{Johnson2024}. 

\bibliographystyle{aasjournal}
\bibliography{mybibtexfile}

\begin{thebibliography}{}
\expandafter\ifx\csname natexlab\endcsname\relax\def\natexlab#1{#1}\fi
\providecommand{\url}[1]{\href{#1}{#1}}
\providecommand{\dodoi}[1]{doi:~\href{http://doi.org/#1}{\nolinkurl{#1}}}
\providecommand{\doeprint}[1]{\href{http://ascl.net/#1}{\nolinkurl{http://ascl.net/#1}}}
\providecommand{\doarXiv}[1]{\href{https://arxiv.org/abs/#1}{\nolinkurl{https://arxiv.org/abs/#1}}}

\bibitem[{{Abdurro'uf} {et~al.}(2022){Abdurro'uf}, {Accetta}, {Aerts}, {Silva
  Aguirre}, {Ahumada}, {Ajgaonkar}, {Filiz Ak}, {Alam}, {Allende Prieto},
  {Almeida}, {Anders}, {Anderson}, {Andrews}, {Anguiano}, {Aquino-Ort{\'\i}z},
  {Arag{\'o}n-Salamanca}, {Argudo-Fern{\'a}ndez}, {Ata}, {Aubert},
  {Avila-Reese}, {Badenes}, {Barb{\'a}}, {Barger}, {Barrera-Ballesteros},
  {Beaton}, {Beers}, {Belfiore}, {Bender}, {Bernardi}, {Bershady}, {Beutler},
  {Bidin}, {Bird}, {Bizyaev}, {Blanc}, {Blanton}, {Boardman}, {Bolton},
  {Boquien}, {Borissova}, {Bovy}, {Brandt}, {Brown}, {Brownstein}, {Brusa},
  {Buchner}, {Bundy}, {Burchett}, {Bureau}, {Burgasser}, {Cabang}, {Campbell},
  {Cappellari}, {Carlberg}, {Wanderley}, {Carrera}, {Cash}, {Chen}, {Chen},
  {Cherinka}, {Chiappini}, {Choi}, {Chojnowski}, {Chung}, {Clerc}, {Cohen},
  {Comerford}, {Comparat}, {da Costa}, {Covey}, {Crane}, {Cruz-Gonzalez},
  {Culhane}, {Cunha}, {Dai}, {Damke}, {Darling}, {Davidson}, {Davies},
  {Dawson}, {De Lee}, {Diamond-Stanic}, {Cano-D{\'\i}az}, {S{\'a}nchez},
  {Donor}, {Duckworth}, {Dwelly}, {Eisenstein}, {Elsworth}, {Emsellem},
  {Eracleous}, {Escoffier}, {Fan}, {Farr}, {Feng}, {Fern{\'a}ndez-Trincado},
  {Feuillet}, {Filipp}, {Fillingham}, {Frinchaboy}, {Fromenteau}, {Galbany},
  {Garc{\'\i}a}, {Garc{\'\i}a-Hern{\'a}ndez}, {Ge}, {Geisler}, {Gelfand},
  {G{\'e}ron}, {Gibson}, {Goddy}, {Godoy-Rivera}, {Grabowski}, {Green},
  {Greener}, {Grier}, {Griffith}, {Guo}, {Guy}, {Hadjara}, {Harding},
  {Hasselquist}, {Hayes}, {Hearty}, {Hern{\'a}ndez}, {Hill}, {Hogg},
  {Holtzman}, {Horta}, {Hsieh}, {Hsu}, {Hsu}, {Huber}, {Huertas-Company},
  {Hutchinson}, {Hwang}, {Ibarra-Medel}, {Chitham}, {Ilha}, {Imig}, {Jaekle},
  {Jayasinghe}, {Ji}, {Johnson}, {Jones}, {J{\"o}nsson}, {Katkov}, {Khalatyan},
  {Kinemuchi}, {Kisku}, {Knapen}, {Kneib}, {Kollmeier}, {Kong}, {Kounkel},
  {Kreckel}, {Krishnarao}, {Lacerna}, {Lane}, {Langgin}, {Lavender}, {Law},
  {Lazarz}, {Leung}, {Leung}, {Lewis}, {Li}, {Li}, {Lian}, {Liang}, {Lin},
  {Lin}, {Lin}, {Lintott}, {Long}, {Longa-Pe{\~n}a}, {L{\'o}pez-Cob{\'a}},
  {Lu}, {Lundgren}, {Luo}, {Mackereth}, {de la Macorra}, {Mahadevan},
  {Majewski}, {Manchado}, {Mandeville}, {Maraston}, {Margalef-Bentabol},
  {Masseron}, {Masters}, {Mathur}, {McDermid}, {Mckay}, {Merloni},
  {Merrifield}, {Meszaros}, {Miglio}, {Di Mille}, {Minniti}, {Minsley},
  {Monachesi}, {Moon}, {Mosser}, {Mulchaey}, {Muna}, {Mu{\~n}oz}, {Myers},
  {Myers}, {Nadathur}, {Nair}, {Nandra}, {Neumann}, {Newman}, {Nidever},
  {Nikakhtar}, {Nitschelm}, {O'Connell}, {Garma-Oehmichen}, {Luan Souza de
  Oliveira}, {Olney}, {Oravetz}, {Ortigoza-Urdaneta}, {Osorio}, {Otter},
  {Pace}, {Padilla}, {Pan}, {Pan}, {Parikh}, {Parker}, {Peirani}, {Pe{\~n}a
  Ram{\'\i}rez}, {Penny}, {Percival}, {Perez-Fournon}, {Pinsonneault},
  {Poidevin}, {Poovelil}, {Price-Whelan}, {B{\'a}rbara de Andrade Queiroz},
  {Raddick}, {Ray}, {Rembold}, {Riddle}, {Riffel}, {Riffel}, {Rix}, {Robin},
  {Rodr{\'\i}guez-Puebla}, {Roman-Lopes}, {Rom{\'a}n-Z{\'u}{\~n}iga}, {Rose},
  {Ross}, {Rossi}, {Rubin}, {Salvato}, {S{\'a}nchez}, {S{\'a}nchez-Gallego},
  {Sanderson}, {Santana Rojas}, {Sarceno}, {Sarmiento}, {Sayres}, {Sazonova},
  {Schaefer}, {Schiavon}, {Schlegel}, {Schneider}, {Schultheis}, {Schwope},
  {Serenelli}, {Serna}, {Shao}, {Shapiro}, {Sharma}, {Shen}, {Shetrone}, {Shu},
  {Simon}, {Skrutskie}, {Smethurst}, {Smith}, {Sobeck}, {Spoo}, {Sprague},
  {Stark}, {Stassun}, {Steinmetz}, {Stello}, {Stone-Martinez},
  {Storchi-Bergmann}, {Stringfellow}, {Stutz}, {Su}, {Taghizadeh-Popp},
  {Talbot}, {Tayar}, {Telles}, {Teske}, {Thakar}, {Theissen}, {Tkachenko},
  {Thomas}, {Tojeiro}, {Hernandez Toledo}, {Troup}, {Trump}, {Trussler},
  {Turner}, {Tuttle}, {Unda-Sanzana}, {V{\'a}zquez-Mata}, {Valentini},
  {Valenzuela}, {Vargas-Gonz{\'a}lez}, {Vargas-Maga{\~n}a}, {Alfaro},
  {Villanova}, {Vincenzo}, {Wake}, {Warfield}, {Washington}, {Weaver},
  {Weijmans}, {Weinberg}, {Weiss}, {Westfall}, {Wild}, {Wilde}, {Wilson},
  {Wilson}, {Wilson}, {Wolf}, {Wood-Vasey}, {Yan}, {Zamora}, {Zasowski},
  {Zhang}, {Zhao}, {Zheng}, {Zheng}, \& {Zhu}}]{Abdurrouf2022ApJS..259...35A}
{Abdurro'uf}, {Accetta}, K., {Aerts}, C., {et~al.} 2022, \apjs, 259, 35,
  \dodoi{10.3847/1538-4365/ac4414}

\bibitem[{{Andrae} {et~al.}(2023){Andrae}, {Rix}, \&
  {Chandra}}]{Andrae2023ApJS..267....8A}
{Andrae}, R., {Rix}, H.-W., \& {Chandra}, V. 2023, \apjs, 267, 8,
  \dodoi{10.3847/1538-4365/acd53e}

\bibitem[{{Aoki} {et~al.}(2022){Aoki}, {Li}, {Matsuno}, {Xing}, {Chen},
  {Christlieb}, {Honda}, {Ishigaki}, {Shi}, {Suda}, {Tominaga}, {Yan}, {Zhao},
  \& {Zhao}}]{Aoki2022ApJ...931..146A}
{Aoki}, W., {Li}, H., {Matsuno}, T., {et~al.} 2022, \apj, 931, 146,
  \dodoi{10.3847/1538-4357/ac6515}

\bibitem[{{Bailer-Jones}(2010)}]{BailerJones2010MNRAS.403...96B}
{Bailer-Jones}, C.~A.~L. 2010, \mnras, 403, 96,
  \dodoi{10.1111/j.1365-2966.2009.16125.x}

\bibitem[{{Bellazzini} {et~al.}(2023){Bellazzini}, {Massari}, {De Angeli},
  {Mucciarelli}, {Bragaglia}, {Riello}, \&
  {Montegriffo}}]{Bellazzini2023A&A...674A.194B}
{Bellazzini}, M., {Massari}, D., {De Angeli}, F., {et~al.} 2023, \aap, 674,
  A194, \dodoi{10.1051/0004-6361/202345921}

\bibitem[{{Belokurov} {et~al.}(2018){Belokurov}, {Erkal}, {Evans}, {Koposov},
  \& {Deason}}]{Belokurov2018MNRAS.478..611B}
{Belokurov}, V., {Erkal}, D., {Evans}, N.~W., {Koposov}, S.~E., \& {Deason},
  A.~J. 2018, \mnras, 478, 611, \dodoi{10.1093/mnras/sty982}

\bibitem[{{Belokurov} {et~al.}(2023){Belokurov}, {Vasiliev}, {Deason},
  {Koposov}, {Fattahi}, {Dillamore}, {Davies}, \&
  {Grand}}]{Belokurov2023MNRAS.518.6200B}
{Belokurov}, V., {Vasiliev}, E., {Deason}, A.~J., {et~al.} 2023, \mnras, 518,
  6200, \dodoi{10.1093/mnras/stac3436}

\bibitem[{{Bennett} \& {Bovy}(2019)}]{Bennett2019MNRAS.482.1417B}
{Bennett}, M., \& {Bovy}, J. 2019, \mnras, 482, 1417,
  \dodoi{10.1093/mnras/sty2813}

\bibitem[{{Buder} {et~al.}(2021){Buder}, {Sharma}, {Kos}, {Amarsi},
  {Nordlander}, {Lind}, {Martell}, {Asplund}, {Bland-Hawthorn}, {Casey}, {de
  Silva}, {D'Orazi}, {Freeman}, {Hayden}, {Lewis}, {Lin}, {Schlesinger},
  {Simpson}, {Stello}, {Zucker}, {Zwitter}, {Beeson}, {Buck}, {Casagrande},
  {Clark}, {{\v{C}}otar}, {da Costa}, {de Grijs}, {Feuillet}, {Horner},
  {Kafle}, {Khanna}, {Kobayashi}, {Liu}, {Montet}, {Nandakumar}, {Nataf},
  {Ness}, {Spina}, {Tepper-Garc{\'\i}a}, {Ting}, {Traven},
  {Vogrin{\v{c}}i{\v{c}}}, {Wittenmyer}, {Wyse}, {{\v{Z}}erjal}, \& {GALAH
  Collaboration}}]{Buder2021MNRAS.506..150B}
{Buder}, S., {Sharma}, S., {Kos}, J., {et~al.} 2021, \mnras, 506, 150,
  \dodoi{10.1093/mnras/stab1242}

\bibitem[{{Chandra} {et~al.}(2023){Chandra}, {Naidu}, {Conroy}, {Ji}, {Rix},
  {Bonaca}, {Cargile}, {Han}, {Johnson}, {Ting}, {Woody}, \&
  {Zaritsky}}]{Chandra2023ApJ...951...26C}
{Chandra}, V., {Naidu}, R.~P., {Conroy}, C., {et~al.} 2023, \apj, 951, 26,
  \dodoi{10.3847/1538-4357/accf13}

\bibitem[{{Cutri} {et~al.}(2021){Cutri}, {Wright}, {Conrow}, {Fowler},
  {Eisenhardt}, {Grillmair}, {Kirkpatrick}, {Masci}, {McCallon}, {Wheelock},
  {Fajardo-Acosta}, {Yan}, {Benford}, {Harbut}, {Jarrett}, {Lake}, {Leisawitz},
  {Ressler}, {Stanford}, {Tsai}, {Liu}, {Helou}, {Mainzer}, {Gettngs},
  {Gonzalez}, {Hoffman}, {Marsh}, {Padgett}, {Skrutskie}, {Beck}, {Papin}, \&
  {Wittman}}]{Cutri2014yCat.2328....0C}
{Cutri}, R.~M., {Wright}, E.~L., {Conrow}, T., {et~al.} 2021, VizieR Online
  Data Catalog, II/328

\bibitem[{{De Angeli} {et~al.}(2023){De Angeli}, {Weiler}, {Montegriffo},
  {Evans}, {Riello}, {Andrae}, {Carrasco}, {Busso}, {Burgess}, {Cacciari},
  {Davidson}, {Harrison}, {Hodgkin}, {Jordi}, {Osborne}, {Pancino},
  {Altavilla}, {Barstow}, {Bailer-Jones}, {Bellazzini}, {Brown}, {Castellani},
  {Cowell}, {Delchambre}, {De Luise}, {Diener}, {Fabricius}, {Fouesneau},
  {Fr{\'e}mat}, {Gilmore}, {Giuffrida}, {Hambly}, {Hidalgo}, {Holland},
  {Kostrzewa-Rutkowska}, {van Leeuwen}, {Lobel}, {Marinoni}, {Miller},
  {Pagani}, {Palaversa}, {Piersimoni}, {Pulone}, {Ragaini}, {Rainer},
  {Richards}, {Rixon}, {Ruz-Mieres}, {Sanna}, {Sarro}, {Rowell}, {Sordo},
  {Walton}, \& {Yoldas}}]{DeAngeli2023A&A...674A...2D}
{De Angeli}, F., {Weiler}, M., {Montegriffo}, P., {et~al.} 2023, \aap, 674, A2,
  \dodoi{10.1051/0004-6361/202243680}

\bibitem[{{de Jong} {et~al.}(2019){de Jong}, {Agertz}, {Berbel}, {Aird},
  {Alexander}, {Amarsi}, {Anders}, {Andrae}, {Ansarinejad}, {Ansorge},
  {Antilogus}, {Anwand-Heerwart}, {Arentsen}, {Arnadottir}, {Asplund}, {Auger},
  {Azais}, {Baade}, {Baker}, {Baker}, {Balbinot}, {Baldry}, {Banerji},
  {Barden}, {Barklem}, {Barth{\'e}l{\'e}my-Mazot}, {Battistini}, {Bauer},
  {Bell}, {Bellido-Tirado}, {Bellstedt}, {Belokurov}, {Bensby}, {Bergemann},
  {Bestenlehner}, {Bielby}, {Bilicki}, {Blake}, {Bland-Hawthorn}, {Boeche},
  {Boland}, {Boller}, {Bongard}, {Bongiorno}, {Bonifacio}, {Boudon}, {Brooks},
  {Brown}, {Brown}, {Br{\"u}ggen}, {Brynnel}, {Brzeski}, {Buchert},
  {Buschkamp}, {Caffau}, {Caillier}, {Carrick}, {Casagrande}, {Case}, {Casey},
  {Cesarini}, {Cescutti}, {Chapuis}, {Chiappini}, {Childress}, {Christlieb},
  {Church}, {Cioni}, {Cluver}, {Colless}, {Collett}, {Comparat}, {Cooper},
  {Couch}, {Courbin}, {Croom}, {Croton}, {Daguis{\'e}}, {Dalton}, {Davies},
  {Davis}, {de Laverny}, {Deason}, {Dionies}, {Disseau}, {Doel}, {D{\"o}scher},
  {Driver}, {Dwelly}, {Eckert}, {Edge}, {Edvardsson}, {Youssoufi}, {Elhaddad},
  {Enke}, {Erfanianfar}, {Farrell}, {Fechner}, {Feiz}, {Feltzing}, {Ferreras},
  {Feuerstein}, {Feuillet}, {Finoguenov}, {Ford}, {Fotopoulou}, {Fouesneau},
  {Frenk}, {Frey}, {Gaessler}, {Geier}, {Gentile Fusillo}, {Gerhard},
  {Giannantonio}, {Giannone}, {Gibson}, {Gillingham},
  {Gonz{\'a}lez-Fern{\'a}ndez}, {Gonzalez-Solares}, {Gottloeber}, {Gould},
  {Grebel}, {Gueguen}, {Guiglion}, {Haehnelt}, {Hahn}, {Hansen}, {Hartman},
  {Hauptner}, {Hawkins}, {Haynes}, {Haynes}, {Heiter}, {Helmi}, {Aguayo},
  {Hewett}, {Hinton}, {Hobbs}, {Hoenig}, {Hofman}, {Hook}, {Hopgood},
  {Hopkins}, {Hourihane}, {Howes}, {Howlett}, {Huet}, {Irwin}, {Iwert},
  {Jablonka}, {Jahn}, {Jahnke}, {Jarno}, {Jin}, {Jofre}, {Johl}, {Jones},
  {J{\"o}nsson}, {Jordan}, {Karovicova}, {Khalatyan}, {Kelz}, {Kennicutt},
  {King}, {Kitaura}, {Klar}, {Klauser}, {Kneib}, {Koch}, {Koposov},
  {Kordopatis}, {Korn}, {Kosmalski}, {Kotak}, {Kovalev}, {Kreckel}, {Kripak},
  {Krumpe}, {Kuijken}, {Kunder}, {Kushniruk}, {Lam}, {Lamer}, {Laurent},
  {Lawrence}, {Lehmitz}, {Lemasle}, {Lewis}, {Li}, {Lidman}, {Lind}, {Liske},
  {Lizon}, {Loveday}, {Ludwig}, {McDermid}, {Maguire}, {Mainieri}, {Mali},
  {Mandel}, {Mandel}, {Mannering}, {Martell}, {Martinez Delgado}, {Matijevic},
  {McGregor}, {McMahon}, {McMillan}, {Mena}, {Merloni}, {Meyer}, {Michel},
  {Micheva}, {Migniau}, {Minchev}, {Monari}, {Muller}, {Murphy},
  {Muthukrishna}, {Nandra}, {Navarro}, {Ness}, {Nichani}, {Nichol}, {Nicklas},
  {Niederhofer}, {Norberg}, {Obreschkow}, {Oliver}, {Owers}, {Pai},
  {Pankratow}, {Parkinson}, {Paschke}, {Paterson}, {Pecontal}, {Parry},
  {Phillips}, {Pillepich}, {Pinard}, {Pirard}, {Piskunov}, {Plank},
  {Pl{\"u}schke}, {Pons}, {Popesso}, {Power}, {Pragt}, {Pramskiy}, {Pryer},
  {Quattri}, {Queiroz}, {Quirrenbach}, {Rahurkar}, {Raichoor}, {Ramstedt},
  {Rau}, {Recio-Blanco}, {Reiss}, {Renaud}, {Revaz}, {Rhode}, {Richard},
  {Richter}, {Rix}, {Robotham}, {Roelfsema}, {Romaniello}, {Rosario},
  {Rothmaier}, {Roukema}, {Ruchti}, {Rupprecht}, {Rybizki}, {Ryde}, {Saar},
  {Sadler}, {Sahl{\'e}n}, {Salvato}, {Sassolas}, {Saunders}, {Saviauk},
  {Sbordone}, {Schmidt}, {Schnurr}, {Scholz}, {Schwope}, {Seifert}, {Shanks},
  {Sheinis}, {Sivov}, {Sk{\'u}lad{\'o}ttir}, {Smartt}, {Smedley}, {Smith},
  {Smith}, {Sorce}, {Spitler}, {Starkenburg}, {Steinmetz}, {Stilz}, {Storm},
  {Sullivan}, {Sutherland}, {Swann}, {Tamone}, {Taylor}, {Teillon}, {Tempel},
  {ter Horst}, {Thi}, {Tolstoy}, {Trager}, {Traven}, {Tremblay}, {Tresse},
  {Valentini}, {van de Weygaert}, {van den Ancker}, {Veljanoski}, {Venkatesan},
  {Wagner}, {Wagner}, {Walcher}, {Waller}, {Walton}, {Wang}, {Winkler},
  {Wisotzki}, {Worley}, {Worseck}, {Xiang}, {Xu}, {Yong}, {Zhao}, {Zheng},
  {Zscheyge}, \& {Zucker}}]{deJong2019Msngr.175....3D}
{de Jong}, R.~S., {Agertz}, O., {Berbel}, A.~A., {et~al.} 2019, The Messenger,
  175, 3, \dodoi{10.18727/0722-6691/5117}

\bibitem[{{De Silva} {et~al.}(2015){De Silva}, {Freeman}, {Bland-Hawthorn},
  {Martell}, {de Boer}, {Asplund}, {Keller}, {Sharma}, {Zucker}, {Zwitter},
  {Anguiano}, {Bacigalupo}, {Bayliss}, {Beavis}, {Bergemann}, {Campbell},
  {Cannon}, {Carollo}, {Casagrande}, {Casey}, {Da Costa}, {D'Orazi}, {Dotter},
  {Duong}, {Heger}, {Ireland}, {Kafle}, {Kos}, {Lattanzio}, {Lewis}, {Lin},
  {Lind}, {Munari}, {Nataf}, {O'Toole}, {Parker}, {Reid}, {Schlesinger},
  {Sheinis}, {Simpson}, {Stello}, {Ting}, {Traven}, {Watson}, {Wittenmyer},
  {Yong}, \& {{\v{Z}}erjal}}]{DeSilva2015MNRAS.449.2604D}
{De Silva}, G.~M., {Freeman}, K.~C., {Bland-Hawthorn}, J., {et~al.} 2015,
  \mnras, 449, 2604, \dodoi{10.1093/mnras/stv327}

\bibitem[{{Gaia Collaboration} {et~al.}(2016){Gaia Collaboration}, {Prusti},
  {de Bruijne}, {Brown}, {Vallenari}, {Babusiaux}, {Bailer-Jones}, {Bastian},
  {Biermann}, {Evans}, {Eyer}, {Jansen}, {Jordi}, {Klioner}, {Lammers},
  {Lindegren}, {Luri}, {Mignard}, {Milligan}, {Panem}, {Poinsignon},
  {Pourbaix}, {Randich}, {Sarri}, {Sartoretti}, {Siddiqui}, {Soubiran},
  {Valette}, {van Leeuwen}, {Walton}, {Aerts}, {Arenou}, {Cropper}, {Drimmel},
  {H{\o}g}, {Katz}, {Lattanzi}, {O'Mullane}, {Grebel}, {Holland}, {Huc},
  {Passot}, {Bramante}, {Cacciari}, {Casta{\~n}eda}, {Chaoul}, {Cheek}, {De
  Angeli}, {Fabricius}, {Guerra}, {Hern{\'a}ndez}, {Jean-Antoine-Piccolo},
  {Masana}, {Messineo}, {Mowlavi}, {Nienartowicz}, {Ord{\'o}{\~n}ez-Blanco},
  {Panuzzo}, {Portell}, {Richards}, {Riello}, {Seabroke}, {Tanga},
  {Th{\'e}venin}, {Torra}, {Els}, {Gracia-Abril}, {Comoretto},
  {Garcia-Reinaldos}, {Lock}, {Mercier}, {Altmann}, {Andrae}, {Astraatmadja},
  {Bellas-Velidis}, {Benson}, {Berthier}, {Blomme}, {Busso}, {Carry},
  {Cellino}, {Clementini}, {Cowell}, {Creevey}, {Cuypers}, {Davidson}, {De
  Ridder}, {de Torres}, {Delchambre}, {Dell'Oro}, {Ducourant}, {Fr{\'e}mat},
  {Garc{\'\i}a-Torres}, {Gosset}, {Halbwachs}, {Hambly}, {Harrison}, {Hauser},
  {Hestroffer}, {Hodgkin}, {Huckle}, {Hutton}, {Jasniewicz}, {Jordan},
  {Kontizas}, {Korn}, {Lanzafame}, {Manteiga}, {Moitinho}, {Muinonen},
  {Osinde}, {Pancino}, {Pauwels}, {Petit}, {Recio-Blanco}, {Robin}, {Sarro},
  {Siopis}, {Smith}, {Smith}, {Sozzetti}, {Thuillot}, {van Reeven}, {Viala},
  {Abbas}, {Abreu Aramburu}, {Accart}, {Aguado}, {Allan}, {Allasia},
  {Altavilla}, {{\'A}lvarez}, {Alves}, {Anderson}, {Andrei}, {Anglada Varela},
  {Antiche}, {Antoja}, {Ant{\'o}n}, {Arcay}, {Atzei}, {Ayache}, {Bach},
  {Baker}, {Balaguer-N{\'u}{\~n}ez}, {Barache}, {Barata}, {Barbier}, {Barblan},
  {Baroni}, {Barrado y Navascu{\'e}s}, {Barros}, {Barstow}, {Becciani},
  {Bellazzini}, {Bellei}, {Bello Garc{\'\i}a}, {Belokurov}, {Bendjoya},
  {Berihuete}, {Bianchi}, {Bienaym{\'e}}, {Billebaud}, {Blagorodnova},
  {Blanco-Cuaresma}, {Boch}, {Bombrun}, {Borrachero}, {Bouquillon}, {Bourda},
  {Bouy}, {Bragaglia}, {Breddels}, {Brouillet}, {Br{\"u}semeister},
  {Bucciarelli}, {Budnik}, {Burgess}, {Burgon}, {Burlacu}, {Busonero}, {Buzzi},
  {Caffau}, {Cambras}, {Campbell}, {Cancelliere}, {Cantat-Gaudin}, {Carlucci},
  {Carrasco}, {Castellani}, {Charlot}, {Charnas}, {Charvet}, {Chassat},
  {Chiavassa}, {Clotet}, {Cocozza}, {Collins}, {Collins}, {Costigan}, {Crifo},
  {Cross}, {Crosta}, {Crowley}, {Dafonte}, {Damerdji}, {Dapergolas}, {David},
  {David}, {De Cat}, {de Felice}, {de Laverny}, {De Luise}, {De March}, {de
  Martino}, {de Souza}, {Debosscher}, {del Pozo}, {Delbo}, {Delgado},
  {Delgado}, {di Marco}, {Di Matteo}, {Diakite}, {Distefano}, {Dolding}, {Dos
  Anjos}, {Drazinos}, {Dur{\'a}n}, {Dzigan}, {Ecale}, {Edvardsson}, {Enke},
  {Erdmann}, {Escolar}, {Espina}, {Evans}, {Eynard Bontemps}, {Fabre},
  {Fabrizio}, {Faigler}, {Falc{\~a}o}, {Farr{\`a}s Casas}, {Faye}, {Federici},
  {Fedorets}, {Fern{\'a}ndez-Hern{\'a}ndez}, {Fernique}, {Fienga}, {Figueras},
  {Filippi}, {Findeisen}, {Fonti}, {Fouesneau}, {Fraile}, {Fraser}, {Fuchs},
  {Furnell}, {Gai}, {Galleti}, {Galluccio}, {Garabato}, {Garc{\'\i}a-Sedano},
  {Gar{\'e}}, {Garofalo}, {Garralda}, {Gavras}, {Gerssen}, {Geyer}, {Gilmore},
  {Girona}, {Giuffrida}, {Gomes}, {Gonz{\'a}lez-Marcos},
  {Gonz{\'a}lez-N{\'u}{\~n}ez}, {Gonz{\'a}lez-Vidal}, {Granvik}, {Guerrier},
  {Guillout}, {Guiraud}, {G{\'u}rpide}, {Guti{\'e}rrez-S{\'a}nchez}, {Guy},
  {Haigron}, {Hatzidimitriou}, {Haywood}, {Heiter}, {Helmi}, {Hobbs},
  {Hofmann}, {Holl}, {Holland}, {Hunt}, {Hypki}, {Icardi}, {Irwin}, {Jevardat
  de Fombelle}, {Jofr{\'e}}, {Jonker}, {Jorissen}, {Julbe}, {Karampelas},
  {Kochoska}, {Kohley}, {Kolenberg}, {Kontizas}, {Koposov}, {Kordopatis},
  {Koubsky}, {Kowalczyk}, {Krone-Martins}, {Kudryashova}, {Kull}, {Bachchan},
  {Lacoste-Seris}, {Lanza}, {Lavigne}, {Le Poncin-Lafitte}, {Lebreton},
  {Lebzelter}, {Leccia}, {Leclerc}, {Lecoeur-Taibi}, {Lemaitre}, {Lenhardt},
  {Leroux}, {Liao}, {Licata}, {Lindstr{\o}m}, {Lister}, {Livanou}, {Lobel},
  {L{\"o}ffler}, {L{\'o}pez}, {Lopez-Lozano}, {Lorenz}, {Loureiro},
  {MacDonald}, {Magalh{\~a}es Fernandes}, {Managau}, {Mann}, {Mantelet},
  {Marchal}, {Marchant}, {Marconi}, {Marie}, {Marinoni}, {Marrese},
  {Marschalk{\'o}}, {Marshall}, {Mart{\'\i}n-Fleitas}, {Martino}, {Mary},
  {Matijevi{\v{c}}}, {Mazeh}, {McMillan}, {Messina}, {Mestre}, {Michalik},
  {Millar}, {Miranda}, {Molina}, {Molinaro}, {Molinaro}, {Moln{\'a}r},
  {Moniez}, {Montegriffo}, {Monteiro}, {Mor}, {Mora}, {Morbidelli}, {Morel},
  {Morgenthaler}, {Morley}, {Morris}, {Mulone}, {Muraveva}, {Musella},
  {Narbonne}, {Nelemans}, {Nicastro}, {Noval}, {Ord{\'e}novic},
  {Ordieres-Mer{\'e}}, {Osborne}, {Pagani}, {Pagano}, {Pailler}, {Palacin},
  {Palaversa}, {Parsons}, {Paulsen}, {Pecoraro}, {Pedrosa}, {Pentik{\"a}inen},
  {Pereira}, {Pichon}, {Piersimoni}, {Pineau}, {Plachy}, {Plum}, {Poujoulet},
  {Pr{\v{s}}a}, {Pulone}, {Ragaini}, {Rago}, {Rambaux}, {Ramos-Lerate},
  {Ranalli}, {Rauw}, {Read}, {Regibo}, {Renk}, {Reyl{\'e}}, {Ribeiro},
  {Rimoldini}, {Ripepi}, {Riva}, {Rixon}, {Roelens}, {Romero-G{\'o}mez},
  {Rowell}, {Royer}, {Rudolph}, {Ruiz-Dern}, {Sadowski}, {Sagrist{\`a}
  Sell{\'e}s}, {Sahlmann}, {Salgado}, {Salguero}, {Sarasso}, {Savietto},
  {Schnorhk}, {Schultheis}, {Sciacca}, {Segol}, {Segovia}, {Segransan},
  {Serpell}, {Shih}, {Smareglia}, {Smart}, {Smith}, {Solano}, {Solitro},
  {Sordo}, {Soria Nieto}, {Souchay}, {Spagna}, {Spoto}, {Stampa}, {Steele},
  {Steidelm{\"u}ller}, {Stephenson}, {Stoev}, {Suess}, {S{\"u}veges}, {Surdej},
  {Szabados}, {Szegedi-Elek}, {Tapiador}, {Taris}, {Tauran}, {Taylor},
  {Teixeira}, {Terrett}, {Tingley}, {Trager}, {Turon}, {Ulla}, {Utrilla},
  {Valentini}, {van Elteren}, {Van Hemelryck}, {van Leeuwen}, {Varadi},
  {Vecchiato}, {Veljanoski}, {Via}, {Vicente}, {Vogt}, {Voss}, {Votruba},
  {Voutsinas}, {Walmsley}, {Weiler}, {Weingrill}, {Werner}, {Wevers},
  {Whitehead}, {Wyrzykowski}, {Yoldas}, {{\v{Z}}erjal}, {Zucker}, {Zurbach},
  {Zwitter}, {Alecu}, {Allen}, {Allende Prieto}, {Amorim},
  {Anglada-Escud{\'e}}, {Arsenijevic}, {Azaz}, {Balm}, {Beck}, {Bernstein},
  {Bigot}, {Bijaoui}, {Blasco}, {Bonfigli}, {Bono}, {Boudreault}, {Bressan},
  {Brown}, {Brunet}, {Bunclark}, {Buonanno}, {Butkevich}, {Carret}, {Carrion},
  {Chemin}, {Ch{\'e}reau}, {Corcione}, {Darmigny}, {de Boer}, {de Teodoro}, {de
  Zeeuw}, {Delle Luche}, {Domingues}, {Dubath}, {Fodor}, {Fr{\'e}zouls},
  {Fries}, {Fustes}, {Fyfe}, {Gallardo}, {Gallegos}, {Gardiol}, {Gebran},
  {Gomboc}, {G{\'o}mez}, {Grux}, {Gueguen}, {Heyrovsky}, {Hoar}, {Iannicola},
  {Isasi Parache}, {Janotto}, {Joliet}, {Jonckheere}, {Keil}, {Kim},
  {Klagyivik}, {Klar}, {Knude}, {Kochukhov}, {Kolka}, {Kos}, {Kutka}, {Lainey},
  {LeBouquin}, {Liu}, {Loreggia}, {Makarov}, {Marseille}, {Martayan},
  {Martinez-Rubi}, {Massart}, {Meynadier}, {Mignot}, {Munari}, {Nguyen},
  {Nordlander}, {Ocvirk}, {O'Flaherty}, {Olias Sanz}, {Ortiz}, {Osorio},
  {Oszkiewicz}, {Ouzounis}, {Palmer}, {Park}, {Pasquato}, {Peltzer}, {Peralta},
  {P{\'e}turaud}, {Pieniluoma}, {Pigozzi}, {Poels}, {Prat}, {Prod'homme},
  {Raison}, {Rebordao}, {Risquez}, {Rocca-Volmerange}, {Rosen}, {Ruiz-Fuertes},
  {Russo}, {Sembay}, {Serraller Vizcaino}, {Short}, {Siebert}, {Silva},
  {Sinachopoulos}, {Slezak}, {Soffel}, {Sosnowska}, {Strai{\v{z}}ys}, {ter
  Linden}, {Terrell}, {Theil}, {Tiede}, {Troisi}, {Tsalmantza}, {Tur},
  {Vaccari}, {Vachier}, {Valles}, {Van Hamme}, {Veltz}, {Virtanen}, {Wallut},
  {Wichmann}, {Wilkinson}, {Ziaeepour}, \&
  {Zschocke}}]{Gaia2016A&A...595A...1G}
{Gaia Collaboration}, {Prusti}, T., {de Bruijne}, J.~H.~J., {et~al.} 2016,
  \aap, 595, A1, \dodoi{10.1051/0004-6361/201629272}

\bibitem[{{Gaia Collaboration} {et~al.}(2021){Gaia Collaboration}, {Brown},
  {Vallenari}, {Prusti}, {de Bruijne}, {Babusiaux}, {Biermann}, {Creevey},
  {Evans}, {Eyer}, {Hutton}, {Jansen}, {Jordi}, {Klioner}, {Lammers},
  {Lindegren}, {Luri}, {Mignard}, {Panem}, {Pourbaix}, {Randich}, {Sartoretti},
  {Soubiran}, {Walton}, {Arenou}, {Bailer-Jones}, {Bastian}, {Cropper},
  {Drimmel}, {Katz}, {Lattanzi}, {van Leeuwen}, {Bakker}, {Cacciari},
  {Casta{\~n}eda}, {De Angeli}, {Ducourant}, {Fabricius}, {Fouesneau},
  {Fr{\'e}mat}, {Guerra}, {Guerrier}, {Guiraud}, {Jean-Antoine Piccolo},
  {Masana}, {Messineo}, {Mowlavi}, {Nicolas}, {Nienartowicz}, {Pailler},
  {Panuzzo}, {Riclet}, {Roux}, {Seabroke}, {Sordo}, {Tanga}, {Th{\'e}venin},
  {Gracia-Abril}, {Portell}, {Teyssier}, {Altmann}, {Andrae}, {Bellas-Velidis},
  {Benson}, {Berthier}, {Blomme}, {Brugaletta}, {Burgess}, {Busso}, {Carry},
  {Cellino}, {Cheek}, {Clementini}, {Damerdji}, {Davidson}, {Delchambre},
  {Dell'Oro}, {Fern{\'a}ndez-Hern{\'a}ndez}, {Galluccio}, {Garc{\'\i}a-Lario},
  {Garcia-Reinaldos}, {Gonz{\'a}lez-N{\'u}{\~n}ez}, {Gosset}, {Haigron},
  {Halbwachs}, {Hambly}, {Harrison}, {Hatzidimitriou}, {Heiter},
  {Hern{\'a}ndez}, {Hestroffer}, {Hodgkin}, {Holl}, {Jan{\ss}en}, {Jevardat de
  Fombelle}, {Jordan}, {Krone-Martins}, {Lanzafame}, {L{\"o}ffler}, {Lorca},
  {Manteiga}, {Marchal}, {Marrese}, {Moitinho}, {Mora}, {Muinonen}, {Osborne},
  {Pancino}, {Pauwels}, {Petit}, {Recio-Blanco}, {Richards}, {Riello},
  {Rimoldini}, {Robin}, {Roegiers}, {Rybizki}, {Sarro}, {Siopis}, {Smith},
  {Sozzetti}, {Ulla}, {Utrilla}, {van Leeuwen}, {van Reeven}, {Abbas}, {Abreu
  Aramburu}, {Accart}, {Aerts}, {Aguado}, {Ajaj}, {Altavilla}, {{\'A}lvarez},
  {{\'A}lvarez Cid-Fuentes}, {Alves}, {Anderson}, {Anglada Varela}, {Antoja},
  {Audard}, {Baines}, {Baker}, {Balaguer-N{\'u}{\~n}ez}, {Balbinot}, {Balog},
  {Barache}, {Barbato}, {Barros}, {Barstow}, {Bartolom{\'e}}, {Bassilana},
  {Bauchet}, {Baudesson-Stella}, {Becciani}, {Bellazzini}, {Bernet}, {Bertone},
  {Bianchi}, {Blanco-Cuaresma}, {Boch}, {Bombrun}, {Bossini}, {Bouquillon},
  {Bragaglia}, {Bramante}, {Breedt}, {Bressan}, {Brouillet}, {Bucciarelli},
  {Burlacu}, {Busonero}, {Butkevich}, {Buzzi}, {Caffau}, {Cancelliere},
  {C{\'a}novas}, {Cantat-Gaudin}, {Carballo}, {Carlucci}, {Carnerero},
  {Carrasco}, {Casamiquela}, {Castellani}, {Castro-Ginard}, {Castro Sampol},
  {Chaoul}, {Charlot}, {Chemin}, {Chiavassa}, {Cioni}, {Comoretto}, {Cooper},
  {Cornez}, {Cowell}, {Crifo}, {Crosta}, {Crowley}, {Dafonte}, {Dapergolas},
  {David}, {David}, {de Laverny}, {De Luise}, {De March}, {De Ridder}, {de
  Souza}, {de Teodoro}, {de Torres}, {del Peloso}, {del Pozo}, {Delbo},
  {Delgado}, {Delgado}, {Delisle}, {Di Matteo}, {Diakite}, {Diener},
  {Distefano}, {Dolding}, {Eappachen}, {Edvardsson}, {Enke}, {Esquej}, {Fabre},
  {Fabrizio}, {Faigler}, {Fedorets}, {Fernique}, {Fienga}, {Figueras},
  {Fouron}, {Fragkoudi}, {Fraile}, {Franke}, {Gai}, {Garabato},
  {Garcia-Gutierrez}, {Garc{\'\i}a-Torres}, {Garofalo}, {Gavras}, {Gerlach},
  {Geyer}, {Giacobbe}, {Gilmore}, {Girona}, {Giuffrida}, {Gomel}, {Gomez},
  {Gonzalez-Santamaria}, {Gonz{\'a}lez-Vidal}, {Granvik},
  {Guti{\'e}rrez-S{\'a}nchez}, {Guy}, {Hauser}, {Haywood}, {Helmi}, {Hidalgo},
  {Hilger}, {H{\l}adczuk}, {Hobbs}, {Holland}, {Huckle}, {Jasniewicz},
  {Jonker}, {Juaristi Campillo}, {Julbe}, {Karbevska}, {Kervella}, {Khanna},
  {Kochoska}, {Kontizas}, {Kordopatis}, {Korn}, {Kostrzewa-Rutkowska},
  {Kruszy{\'n}ska}, {Lambert}, {Lanza}, {Lasne}, {Le Campion}, {Le Fustec},
  {Lebreton}, {Lebzelter}, {Leccia}, {Leclerc}, {Lecoeur-Taibi}, {Liao},
  {Licata}, {Lindstr{\o}m}, {Lister}, {Livanou}, {Lobel}, {Madrero Pardo},
  {Managau}, {Mann}, {Marchant}, {Marconi}, {Marcos Santos}, {Marinoni},
  {Marocco}, {Marshall}, {Martin Polo}, {Mart{\'\i}n-Fleitas}, {Masip},
  {Massari}, {Mastrobuono-Battisti}, {Mazeh}, {McMillan}, {Messina},
  {Michalik}, {Millar}, {Mints}, {Molina}, {Molinaro}, {Moln{\'a}r},
  {Montegriffo}, {Mor}, {Morbidelli}, {Morel}, {Morris}, {Mulone}, {Munoz},
  {Muraveva}, {Murphy}, {Musella}, {Noval}, {Ord{\'e}novic}, {Orr{\`u}},
  {Osinde}, {Pagani}, {Pagano}, {Palaversa}, {Palicio}, {Panahi}, {Pawlak},
  {Pe{\~n}alosa Esteller}, {Penttil{\"a}}, {Piersimoni}, {Pineau}, {Plachy},
  {Plum}, {Poggio}, {Poretti}, {Poujoulet}, {Pr{\v{s}}a}, {Pulone}, {Racero},
  {Ragaini}, {Rainer}, {Raiteri}, {Rambaux}, {Ramos}, {Ramos-Lerate}, {Re
  Fiorentin}, {Regibo}, {Reyl{\'e}}, {Ripepi}, {Riva}, {Rixon}, {Robichon},
  {Robin}, {Roelens}, {Rohrbasser}, {Romero-G{\'o}mez}, {Rowell}, {Royer},
  {Rybicki}, {Sadowski}, {Sagrist{\`a} Sell{\'e}s}, {Sahlmann}, {Salgado},
  {Salguero}, {Samaras}, {Sanchez Gimenez}, {Sanna}, {Santove{\~n}a},
  {Sarasso}, {Schultheis}, {Sciacca}, {Segol}, {Segovia}, {S{\'e}gransan},
  {Semeux}, {Shahaf}, {Siddiqui}, {Siebert}, {Siltala}, {Slezak}, {Smart},
  {Solano}, {Solitro}, {Souami}, {Souchay}, {Spagna}, {Spoto}, {Steele},
  {Steidelm{\"u}ller}, {Stephenson}, {S{\"u}veges}, {Szabados}, {Szegedi-Elek},
  {Taris}, {Tauran}, {Taylor}, {Teixeira}, {Thuillot}, {Tonello}, {Torra},
  {Torra}, {Turon}, {Unger}, {Vaillant}, {van Dillen}, {Vanel}, {Vecchiato},
  {Viala}, {Vicente}, {Voutsinas}, {Weiler}, {Wevers}, {Wyrzykowski}, {Yoldas},
  {Yvard}, {Zhao}, {Zorec}, {Zucker}, {Zurbach}, \&
  {Zwitter}}]{Gaia2021A&A...649A...1G}
{Gaia Collaboration}, {Brown}, A.~G.~A., {Vallenari}, A., {et~al.} 2021, \aap,
  649, A1, \dodoi{10.1051/0004-6361/202039657}

\bibitem[{{Gaia Collaboration} {et~al.}(2023){Gaia Collaboration}, {Vallenari},
  {Brown}, {Prusti}, {de Bruijne}, {Arenou}, {Babusiaux}, {Biermann},
  {Creevey}, {Ducourant}, {Evans}, {Eyer}, {Guerra}, {Hutton}, {Jordi},
  {Klioner}, {Lammers}, {Lindegren}, {Luri}, {Mignard}, {Panem}, {Pourbaix},
  {Randich}, {Sartoretti}, {Soubiran}, {Tanga}, {Walton}, {Bailer-Jones},
  {Bastian}, {Drimmel}, {Jansen}, {Katz}, {Lattanzi}, {van Leeuwen}, {Bakker},
  {Cacciari}, {Casta{\~n}eda}, {De Angeli}, {Fabricius}, {Fouesneau},
  {Fr{\'e}mat}, {Galluccio}, {Guerrier}, {Heiter}, {Masana}, {Messineo},
  {Mowlavi}, {Nicolas}, {Nienartowicz}, {Pailler}, {Panuzzo}, {Riclet}, {Roux},
  {Seabroke}, {Sordo}, {Th{\'e}venin}, {Gracia-Abril}, {Portell}, {Teyssier},
  {Altmann}, {Andrae}, {Audard}, {Bellas-Velidis}, {Benson}, {Berthier},
  {Blomme}, {Burgess}, {Busonero}, {Busso}, {C{\'a}novas}, {Carry}, {Cellino},
  {Cheek}, {Clementini}, {Damerdji}, {Davidson}, {de Teodoro}, {Nu{\~n}ez
  Campos}, {Delchambre}, {Dell'Oro}, {Esquej}, {Fern{\'a}ndez-Hern{\'a}ndez},
  {Fraile}, {Garabato}, {Garc{\'\i}a-Lario}, {Gosset}, {Haigron}, {Halbwachs},
  {Hambly}, {Harrison}, {Hern{\'a}ndez}, {Hestroffer}, {Hodgkin}, {Holl},
  {Jan{\ss}en}, {Jevardat de Fombelle}, {Jordan}, {Krone-Martins}, {Lanzafame},
  {L{\"o}ffler}, {Marchal}, {Marrese}, {Moitinho}, {Muinonen}, {Osborne},
  {Pancino}, {Pauwels}, {Recio-Blanco}, {Reyl{\'e}}, {Riello}, {Rimoldini},
  {Roegiers}, {Rybizki}, {Sarro}, {Siopis}, {Smith}, {Sozzetti}, {Utrilla},
  {van Leeuwen}, {Abbas}, {{\'A}brah{\'a}m}, {Abreu Aramburu}, {Aerts},
  {Aguado}, {Ajaj}, {Aldea-Montero}, {Altavilla}, {{\'A}lvarez}, {Alves},
  {Anders}, {Anderson}, {Anglada Varela}, {Antoja}, {Baines}, {Baker},
  {Balaguer-N{\'u}{\~n}ez}, {Balbinot}, {Balog}, {Barache}, {Barbato},
  {Barros}, {Barstow}, {Bartolom{\'e}}, {Bassilana}, {Bauchet}, {Becciani},
  {Bellazzini}, {Berihuete}, {Bernet}, {Bertone}, {Bianchi}, {Binnenfeld},
  {Blanco-Cuaresma}, {Blazere}, {Boch}, {Bombrun}, {Bossini}, {Bouquillon},
  {Bragaglia}, {Bramante}, {Breedt}, {Bressan}, {Brouillet}, {Brugaletta},
  {Bucciarelli}, {Burlacu}, {Butkevich}, {Buzzi}, {Caffau}, {Cancelliere},
  {Cantat-Gaudin}, {Carballo}, {Carlucci}, {Carnerero}, {Carrasco},
  {Casamiquela}, {Castellani}, {Castro-Ginard}, {Chaoul}, {Charlot}, {Chemin},
  {Chiaramida}, {Chiavassa}, {Chornay}, {Comoretto}, {Contursi}, {Cooper},
  {Cornez}, {Cowell}, {Crifo}, {Cropper}, {Crosta}, {Crowley}, {Dafonte},
  {Dapergolas}, {David}, {David}, {de Laverny}, {De Luise}, {De March}, {De
  Ridder}, {de Souza}, {de Torres}, {del Peloso}, {del Pozo}, {Delbo},
  {Delgado}, {Delisle}, {Demouchy}, {Dharmawardena}, {Di Matteo}, {Diakite},
  {Diener}, {Distefano}, {Dolding}, {Edvardsson}, {Enke}, {Fabre}, {Fabrizio},
  {Faigler}, {Fedorets}, {Fernique}, {Fienga}, {Figueras}, {Fournier},
  {Fouron}, {Fragkoudi}, {Gai}, {Garcia-Gutierrez}, {Garcia-Reinaldos},
  {Garc{\'\i}a-Torres}, {Garofalo}, {Gavel}, {Gavras}, {Gerlach}, {Geyer},
  {Giacobbe}, {Gilmore}, {Girona}, {Giuffrida}, {Gomel}, {Gomez},
  {Gonz{\'a}lez-N{\'u}{\~n}ez}, {Gonz{\'a}lez-Santamar{\'\i}a},
  {Gonz{\'a}lez-Vidal}, {Granvik}, {Guillout}, {Guiraud},
  {Guti{\'e}rrez-S{\'a}nchez}, {Guy}, {Hatzidimitriou}, {Hauser}, {Haywood},
  {Helmer}, {Helmi}, {Sarmiento}, {Hidalgo}, {Hilger}, {H{\l}adczuk}, {Hobbs},
  {Holland}, {Huckle}, {Jardine}, {Jasniewicz}, {Jean-Antoine Piccolo},
  {Jim{\'e}nez-Arranz}, {Jorissen}, {Juaristi Campillo}, {Julbe}, {Karbevska},
  {Kervella}, {Khanna}, {Kontizas}, {Kordopatis}, {Korn}, {K{\'o}sp{\'a}l},
  {Kostrzewa-Rutkowska}, {Kruszy{\'n}ska}, {Kun}, {Laizeau}, {Lambert},
  {Lanza}, {Lasne}, {Le Campion}, {Lebreton}, {Lebzelter}, {Leccia}, {Leclerc},
  {Lecoeur-Taibi}, {Liao}, {Licata}, {Lindstr{\o}m}, {Lister}, {Livanou},
  {Lobel}, {Lorca}, {Loup}, {Madrero Pardo}, {Magdaleno Romeo}, {Managau},
  {Mann}, {Manteiga}, {Marchant}, {Marconi}, {Marcos}, {Marcos Santos},
  {Mar{\'\i}n Pina}, {Marinoni}, {Marocco}, {Marshall}, {Martin Polo},
  {Mart{\'\i}n-Fleitas}, {Marton}, {Mary}, {Masip}, {Massari},
  {Mastrobuono-Battisti}, {Mazeh}, {McMillan}, {Messina}, {Michalik}, {Millar},
  {Mints}, {Molina}, {Molinaro}, {Moln{\'a}r}, {Monari}, {Mongui{\'o}},
  {Montegriffo}, {Montero}, {Mor}, {Mora}, {Morbidelli}, {Morel}, {Morris},
  {Muraveva}, {Murphy}, {Musella}, {Nagy}, {Noval}, {Oca{\~n}a}, {Ogden},
  {Ordenovic}, {Osinde}, {Pagani}, {Pagano}, {Palaversa}, {Palicio},
  {Pallas-Quintela}, {Panahi}, {Payne-Wardenaar}, {Pe{\~n}alosa Esteller},
  {Penttil{\"a}}, {Pichon}, {Piersimoni}, {Pineau}, {Plachy}, {Plum}, {Poggio},
  {Pr{\v{s}}a}, {Pulone}, {Racero}, {Ragaini}, {Rainer}, {Raiteri}, {Rambaux},
  {Ramos}, {Ramos-Lerate}, {Re Fiorentin}, {Regibo}, {Richards}, {Rios Diaz},
  {Ripepi}, {Riva}, {Rix}, {Rixon}, {Robichon}, {Robin}, {Robin}, {Roelens},
  {Rogues}, {Rohrbasser}, {Romero-G{\'o}mez}, {Rowell}, {Royer}, {Ruz Mieres},
  {Rybicki}, {Sadowski}, {S{\'a}ez N{\'u}{\~n}ez}, {Sagrist{\`a} Sell{\'e}s},
  {Sahlmann}, {Salguero}, {Samaras}, {Sanchez Gimenez}, {Sanna},
  {Santove{\~n}a}, {Sarasso}, {Schultheis}, {Sciacca}, {Segol}, {Segovia},
  {S{\'e}gransan}, {Semeux}, {Shahaf}, {Siddiqui}, {Siebert}, {Siltala},
  {Silvelo}, {Slezak}, {Slezak}, {Smart}, {Snaith}, {Solano}, {Solitro},
  {Souami}, {Souchay}, {Spagna}, {Spina}, {Spoto}, {Steele},
  {Steidelm{\"u}ller}, {Stephenson}, {S{\"u}veges}, {Surdej}, {Szabados},
  {Szegedi-Elek}, {Taris}, {Taylor}, {Teixeira}, {Tolomei}, {Tonello}, {Torra},
  {Torra}, {Torralba Elipe}, {Trabucchi}, {Tsounis}, {Turon}, {Ulla}, {Unger},
  {Vaillant}, {van Dillen}, {van Reeven}, {Vanel}, {Vecchiato}, {Viala},
  {Vicente}, {Voutsinas}, {Weiler}, {Wevers}, {Wyrzykowski}, {Yoldas}, {Yvard},
  {Zhao}, {Zorec}, {Zucker}, \&
  {Zwitter}}]{GaiaCollaboration2023A&A...674A...1G}
{Gaia Collaboration}, {Vallenari}, A., {Brown}, A.~G.~A., {et~al.} 2023, \aap,
  674, A1, \dodoi{10.1051/0004-6361/202243940}

\bibitem[{{Gavel} {et~al.}(2021){Gavel}, {Andrae}, {Fouesneau}, {Korn}, \&
  {Sordo}}]{Gavel2021A&A...656A..93G}
{Gavel}, A., {Andrae}, R., {Fouesneau}, M., {Korn}, A.~J., \& {Sordo}, R. 2021,
  \aap, 656, A93, \dodoi{10.1051/0004-6361/202141589}

\bibitem[{{Gehren} {et~al.}(2006){Gehren}, {Shi}, {Zhang}, {Zhao}, \&
  {Korn}}]{Gehren2006A&A...451.1065G}
{Gehren}, T., {Shi}, J.~R., {Zhang}, H.~W., {Zhao}, G., \& {Korn}, A.~J. 2006,
  \aap, 451, 1065, \dodoi{10.1051/0004-6361:20054434}

\bibitem[{{Gilmore} {et~al.}(2012){Gilmore}, {Randich}, {Asplund}, {Binney},
  {Bonifacio}, {Drew}, {Feltzing}, {Ferguson}, {Jeffries}, {Micela},
  {Negueruela}, {Prusti}, {Rix}, {Vallenari}, {Alfaro}, {Allende-Prieto},
  {Babusiaux}, {Bensby}, {Blomme}, {Bragaglia}, {Flaccomio}, {Fran{\c{c}}ois},
  {Irwin}, {Koposov}, {Korn}, {Lanzafame}, {Pancino}, {Paunzen},
  {Recio-Blanco}, {Sacco}, {Smiljanic}, {Van Eck}, {Walton}, {Aden}, {Aerts},
  {Affer}, {Alcala}, {Altavilla}, {Alves}, {Antoja}, {Arenou}, {Argiroffi},
  {Asensio Ramos}, {Bailer-Jones}, {Balaguer-Nunez}, {Bayo}, {Barbuy},
  {Barisevicius}, {Barrado y Navascues}, {Battistini}, {Bellas Velidis},
  {Bellazzini}, {Belokurov}, {Bergemann}, {Bertelli}, {Biazzo}, {Bienayme},
  {Bland-Hawthorn}, {Boeche}, {Bonito}, {Boudreault}, {Bouvier}, {Brandao},
  {Brown}, {de Bruijne}, {Burleigh}, {Caballero}, {Caffau}, {Calura},
  {Capuzzo-Dolcetta}, {Caramazza}, {Carraro}, {Casagrande}, {Casewell},
  {Chapman}, {Chiappini}, {Chorniy}, {Christlieb}, {Cignoni}, {Cocozza},
  {Colless}, {Collet}, {Collins}, {Correnti}, {Covino}, {Crnojevic}, {Cropper},
  {Cunha}, {Damiani}, {David}, {Delgado}, {Duffau}, {Edvardsson}, {Eldridge},
  {Enke}, {Eriksson}, {Evans}, {Eyer}, {Famaey}, {Fellhauer}, {Ferreras},
  {Figueras}, {Fiorentino}, {Flynn}, {Folha}, {Franciosini}, {Frasca},
  {Freeman}, {Fremat}, {Friel}, {Gaensicke}, {Gameiro}, {Garzon}, {Geier},
  {Geisler}, {Gerhard}, {Gibson}, {Gomboc}, {Gomez}, {Gonzalez-Fernandez},
  {Gonzalez Hernandez}, {Gosset}, {Grebel}, {Greimel}, {Groenewegen},
  {Grundahl}, {Guarcello}, {Gustafsson}, {Hadrava}, {Hatzidimitriou}, {Hambly},
  {Hammersley}, {Hansen}, {Haywood}, {Heber}, {Heiter}, {Held}, {Helmi},
  {Hensler}, {Herrero}, {Hill}, {Hodgkin}, {Huelamo}, {Huxor}, {Ibata},
  {Jackson}, {de Jong}, {Jonker}, {Jordan}, {Jordi}, {Jorissen}, {Katz},
  {Kawata}, {Keller}, {Kharchenko}, {Klement}, {Klutsch}, {Knude}, {Koch},
  {Kochukhov}, {Kontizas}, {Koubsky}, {Lallement}, {de Laverny}, {van Leeuwen},
  {Lemasle}, {Lewis}, {Lind}, {Lindstrom}, {Lobel}, {Lopez Santiago}, {Lucas},
  {Ludwig}, {Lueftinger}, {Magrini}, {Maiz Apellaniz}, {Maldonado}, {Marconi},
  {Marino}, {Martayan}, {Martinez-Valpuesta}, {Matijevic}, {McMahon},
  {Messina}, {Meyer}, {Miglio}, {Mikolaitis}, {Minchev}, {Minniti}, {Moitinho},
  {Momany}, {Monaco}, {Montalto}, {Monteiro}, {Monier}, {Montes}, {Mora},
  {Moraux}, {Morel}, {Mowlavi}, {Mucciarelli}, {Munari}, {Napiwotzki},
  {Nardetto}, {Naylor}, {Naze}, {Nelemans}, {Okamoto}, {Ortolani}, {Pace},
  {Palla}, {Palous}, {Parker}, {Penarrubia}, {Pillitteri}, {Piotto}, {Posbic},
  {Prisinzano}, {Puzeras}, {Quirrenbach}, {Ragaini}, {Read}, {Read}, {Reyle},
  {De Ridder}, {Robichon}, {Robin}, {Roeser}, {Romano}, {Royer}, {Ruchti},
  {Ruzicka}, {Ryan}, {Ryde}, {Santos}, {Sanz Forcada}, {Sarro Baro},
  {Sbordone}, {Schilbach}, {Schmeja}, {Schnurr}, {Schoenrich}, {Scholz},
  {Seabroke}, {Sharma}, {De Silva}, {Smith}, {Solano}, {Sordo}, {Soubiran},
  {Sousa}, {Spagna}, {Steffen}, {Steinmetz}, {Stelzer}, {Stempels},
  {Tabernero}, {Tautvaisiene}, {Thevenin}, {Torra}, {Tosi}, {Tolstoy}, {Turon},
  {Walker}, {Wambsganss}, {Worley}, {Venn}, {Vink}, {Wyse}, {Zaggia},
  {Zeilinger}, {Zoccali}, {Zorec}, {Zucker}, {Zwitter}, \& {Gaia-ESO Survey
  Team}}]{Gilmore2012Msngr.147...25G}
{Gilmore}, G., {Randich}, S., {Asplund}, M., {et~al.} 2012, The Messenger, 147,
  25

\bibitem[{{GRAVITY Collaboration} {et~al.}(2022){GRAVITY Collaboration},
  {Abuter}, {Aimar}, {Amorim}, {Ball}, {Baub{\"o}ck}, {Berger}, {Bonnet},
  {Bourdarot}, {Brandner}, {Cardoso}, {Cl{\'e}net}, {Dallilar}, {Davies}, {de
  Zeeuw}, {Dexter}, {Drescher}, {Eisenhauer}, {F{\"o}rster Schreiber},
  {Foschi}, {Garcia}, {Gao}, {Gendron}, {Genzel}, {Gillessen}, {Habibi},
  {Haubois}, {Hei{\ss}el}, {Henning}, {Hippler}, {Horrobin}, {Jochum}, {Jocou},
  {Kaufer}, {Kervella}, {Lacour}, {Lapeyr{\`e}re}, {Le Bouquin}, {L{\'e}na},
  {Lutz}, {Ott}, {Paumard}, {Perraut}, {Perrin}, {Pfuhl}, {Rabien},
  {Shangguan}, {Shimizu}, {Scheithauer}, {Stadler}, {Stephens}, {Straub},
  {Straubmeier}, {Sturm}, {Tacconi}, {Tristram}, {Vincent}, {von Fellenberg},
  {Widmann}, {Wieprecht}, {Wiezorrek}, {Woillez}, {Yazici}, \&
  {Young}}]{GRAVITY2022A&A...657L..12G}
{GRAVITY Collaboration}, {Abuter}, R., {Aimar}, N., {et~al.} 2022, \aap, 657,
  L12, \dodoi{10.1051/0004-6361/202142465}

\bibitem[{{Green}(2018)}]{Green2018JOSS....3..695M}
{Green}, G. 2018, The Journal of Open Source Software, 3, 695,
  \dodoi{10.21105/joss.00695}

\bibitem[{{Green} {et~al.}(2018){Green}, {Schlafly}, {Finkbeiner}, {Rix},
  {Martin}, {Burgett}, {Draper}, {Flewelling}, {Hodapp}, {Kaiser}, {Kudritzki},
  {Magnier}, {Metcalfe}, {Tonry}, {Wainscoat}, \&
  {Waters}}]{Green2018MNRAS.478..651G}
{Green}, G.~M., {Schlafly}, E.~F., {Finkbeiner}, D., {et~al.} 2018, \mnras,
  478, 651, \dodoi{10.1093/mnras/sty1008}

\bibitem[{{Guiglion} {et~al.}(2024){Guiglion}, {Nepal}, {Chiappini},
  {Khoperskov}, {Traven}, {Queiroz}, {Steinmetz}, {Valentini}, {Fournier},
  {Vallenari}, {Youakim}, {Bergemann}, {M{\'e}sz{\'a}ros}, {Lucatello},
  {Sordo}, {Fabbro}, {Minchev}, {Tautvai{\v{s}}ien{\.{e}}}, {Mikolaitis}, \&
  {Montalb{\'a}n}}]{Guiglion2024A&A...682A...9G}
{Guiglion}, G., {Nepal}, S., {Chiappini}, C., {et~al.} 2024, \aap, 682, A9,
  \dodoi{10.1051/0004-6361/202347122}

\bibitem[{Hastie {et~al.}(2001)Hastie, Tibshirani, \&
  Friedman}]{hastie01statisticallearning}
Hastie, T., Tibshirani, R., \& Friedman, J. 2001, The Elements of Statistical
  Learning, Springer Series in Statistics (New York, NY, USA: Springer New York
  Inc.)

\bibitem[{{Helmi} {et~al.}(2018){Helmi}, {Babusiaux}, {Koppelman}, {Massari},
  {Veljanoski}, \& {Brown}}]{Helmi2018Natur.563...85H}
{Helmi}, A., {Babusiaux}, C., {Koppelman}, H.~H., {et~al.} 2018, \nat, 563, 85,
  \dodoi{10.1038/s41586-018-0625-x}

\bibitem[{{Hinkle} {et~al.}(2000){Hinkle}, {Wallace}, {Valenti}, \&
  {Harmer}}]{Hinkle2000vnia.book.....H}
{Hinkle}, K., {Wallace}, L., {Valenti}, J., \& {Harmer}, D. 2000, {Visible and
  Near Infrared Atlas of the Arcturus Spectrum 3727-9300 A}

\bibitem[{{Hunter}(2007)}]{Hunter2007}
{Hunter}, J.~D. 2007, Computing in Science and Engineering, 9, 90,
  \dodoi{10.1109/MCSE.2007.55}

\bibitem[{{Jin} {et~al.}(2023){Jin}, {Trager}, {Dalton}, {Aguerri}, {Drew},
  {Falc{\'o}n-Barroso}, {G{\"a}nsicke}, {Hill}, {Iovino}, {Pieri}, {Poggianti},
  {Smith}, {Vallenari}, {Abrams}, {Aguado}, {Antoja}, {Arag{\'o}n-Salamanca},
  {Ascasibar}, {Babusiaux}, {Balcells}, {Barrena}, {Battaglia}, {Belokurov},
  {Bensby}, {Bonifacio}, {Bragaglia}, {Carrasco}, {Carrera}, {Cornwell},
  {Dom{\'\i}nguez-Palmero}, {Duncan}, {Famaey}, {Fari{\~n}a}, {Gonzalez},
  {Guest}, {Hatch}, {Hess}, {Hoskin}, {Irwin}, {Knapen}, {Koposov}, {Kuchner},
  {Laigle}, {Lewis}, {Longhetti}, {Lucatello}, {M{\'e}ndez-Abreu}, {Mercurio},
  {Molaeinezhad}, {Mongui{\'o}}, {Morrison}, {Murphy}, {Peralta de Arriba},
  {P{\'e}rez}, {P{\'e}rez-R{\`a}fols}, {Pic{\'o}}, {Raddi}, {Romero-G{\'o}mez},
  {Royer}, {Siebert}, {Seabroke}, {Som}, {Terrett}, {Thomas}, {Wesson},
  {Worley}, {Alfaro}, {Allende Prieto}, {Alonso-Santiago}, {Amos}, {Ashley},
  {Balaguer-N{\'u} nez}, {Balbinot}, {Bellazzini}, {Benn}, {Berlanas},
  {Bernard}, {Best}, {Bettoni}, {Bianco}, {Bishop}, {Blomqvist}, {Boeche},
  {Bolzonella}, {Bonoli}, {Bosma}, {Britavskiy}, {Busarello}, {Caffau},
  {Cantat-Gaudin}, {Castro-Ginard}, {Couto}, {Carbajo-Hijarrubia}, {Carter},
  {Casamiquela}, {Conrado}, {Corcho-Caballero}, {Costantin}, {Deason}, {de
  Burgos}, {De Grandi}, {Di Matteo}, {Dom{\'\i}nguez-G{\'o}mez}, {Dorda},
  {Drake}, {Dutta}, {Erkal}, {Feltzing}, {Ferr{\'e}-Mateu}, {Feuillet},
  {Figueras}, {Fossati}, {Franciosini}, {Frasca}, {Fumagalli}, {Gallazzi},
  {Garc{\'\i}a-Benito}, {Fusillo}, {Gebran}, {Gilbert}, {Gledhill},
  {Gonz{\'a}lez Delgado}, {Greimel}, {Guarcello}, {Guerra}, {Gullieuszik},
  {Haines}, {Hardcastle}, {Harris}, {Haywood}, {Helmi}, {Hernandez}, {Herrero},
  {Hughes}, {Irsic}, {Jablonka}, {Jarvis}, {Jordi}, {Kondapally}, {Kordopatis},
  {Krogager}, {La Barbera}, {Lam}, {Larsen}, {Lemasle}, {Lewis}, {Lhom{\'e}},
  {Lind}, {Lodi}, {Longobardi}, {Lonoce}, {Magrini}, {Ma{\'\i}z Apell{\'a}niz},
  {Marchal}, {Marco}, {Martin}, {Matsuno}, {Maurogordato}, {Merluzzi},
  {Miralda-Escud{\'e}}, {Molinari}, {Monari}, {Morelli}, {Mottram}, {Naylor},
  {Negueruela}, {Onorbe}, {Pancino}, {Peirani}, {Peletier}, {Pozzetti},
  {Rainer}, {Ramos}, {Read}, {Rossi}, {R{\"o}ttgering},
  {Rubi{\~n}o-Mart{\'\i}n}, {Sabater Montes}, {San Juan}, {Sanna}, {Schallig},
  {Schiavon}, {Schultheis}, {Serra}, {Shimwell}, {Sim{\'o}n-D{\'\i}az},
  {Smith}, {Sordo}, {Sorini}, {Soubiran}, {Starkenburg}, {Steele}, {Stott},
  {Stuik}, {Tolstoy}, {Tortora}, {Tsantaki}, {Van der Swaelmen}, {van Weeren},
  {Vergani}, {Verheijen}, {Verro}, {Vink}, {Vioque}, {Walcher}, {Walton},
  {Wegg}, {Weijmans}, {Williams}, {Wilson}, {Wright}, {Xylakis-Dornbusch},
  {Youakim}, {Zibetti}, \& {Zurita}}]{Jin2023MNRAS.tmp..715J}
{Jin}, S., {Trager}, S.~C., {Dalton}, G.~B., {et~al.} 2023, \mnras,
  \dodoi{10.1093/mnras/stad557}

\bibitem[{Johnson(2024)}]{Johnson2024}
Johnson, R.~A. 2024, Journal of Open Source Software, 9, 5976,
  \dodoi{10.21105/joss.05976}

\bibitem[{{Jones} {et~al.}(2001){Jones}, {Oliphant}, \& {Peterson}}]{Jones2001}
{Jones}, E., {Oliphant}, T., \& {Peterson}, P., e.~a. 2001, {SciPy}: Open
  source scientific tools for {Python}.
\newblock \url{http://www.scipy.org/}

\bibitem[{{Kramida} {et~al.}(2024){Kramida}, {Ralchenko}, {Reader}, \& {NIST
  ASD Team}}]{NIST2024}
{Kramida}, A., {Ralchenko}, Y., {Reader}, J., \& {NIST ASD Team}. 2024, NIST
  Atomic Spectra Database (version 5.12), [Online].,  National Institute of
  Standards and Technology, Gaithersburg, MD.,
  \dodoi{https://doi.org/10.18434/T4W30F}

\bibitem[{{Laroche} \& {Speagle}(2023)}]{Laroche2023arXiv230706378L}
{Laroche}, A., \& {Speagle}, J.~S. 2023, arXiv e-prints, arXiv:2307.06378,
  \dodoi{10.48550/arXiv.2307.06378}

\bibitem[{{Lee} {et~al.}(2011){Lee}, {Beers}, {An}, {Ivezi{\'c}}, {Just},
  {Rockosi}, {Morrison}, {Johnson}, {Sch{\"o}nrich}, {Bird}, {Yanny},
  {Harding}, \& {Rocha-Pinto}}]{Lee2011ApJ...738..187L}
{Lee}, Y.~S., {Beers}, T.~C., {An}, D., {et~al.} 2011, \apj, 738, 187,
  \dodoi{10.1088/0004-637X/738/2/187}

\bibitem[{{Leung} \& {Bovy}(2023)}]{Leung2023MNRAS.tmp.2896L}
{Leung}, H.~W., \& {Bovy}, J. 2023, \mnras, \dodoi{10.1093/mnras/stad3015}

\bibitem[{{Li} {et~al.}(2022){Li}, {Aoki}, {Matsuno}, {Xing}, {Suda},
  {Tominaga}, {Chen}, {Honda}, {Ishigaki}, {Shi}, {Zhao}, \&
  {Zhao}}]{Li2022ApJ...931..147L}
{Li}, H., {Aoki}, W., {Matsuno}, T., {et~al.} 2022, \apj, 931, 147,
  \dodoi{10.3847/1538-4357/ac6514}

\bibitem[{{Li} {et~al.}(2023){Li}, {Wong}, {Hogg}, {Rix}, \&
  {Chandra}}]{Li2023arXiv230914294L}
{Li}, J., {Wong}, K. W.~K., {Hogg}, D.~W., {Rix}, H.-W., \& {Chandra}, V. 2023,
  arXiv e-prints, arXiv:2309.14294, \dodoi{10.48550/arXiv.2309.14294}

\bibitem[{{Lindegren} {et~al.}(2021){Lindegren}, {Bastian}, {Biermann},
  {Bombrun}, {de Torres}, {Gerlach}, {Geyer}, {Hern{\'a}ndez}, {Hilger},
  {Hobbs}, {Klioner}, {Lammers}, {McMillan}, {Ramos-Lerate},
  {Steidelm{\"u}ller}, {Stephenson}, \& {van
  Leeuwen}}]{Lindegren2021A&A...649A...4L}
{Lindegren}, L., {Bastian}, U., {Biermann}, M., {et~al.} 2021, \aap, 649, A4,
  \dodoi{10.1051/0004-6361/202039653}

\bibitem[{{Mackereth} {et~al.}(2019){Mackereth}, {Schiavon}, {Pfeffer},
  {Hayes}, {Bovy}, {Anguiano}, {Allende Prieto}, {Hasselquist}, {Holtzman},
  {Johnson}, {Majewski}, {O'Connell}, {Shetrone}, {Tissera}, \&
  {Fern{\'a}ndez-Trincado}}]{Mackereth2019MNRAS.482.3426M}
{Mackereth}, J.~T., {Schiavon}, R.~P., {Pfeffer}, J., {et~al.} 2019, \mnras,
  482, 3426, \dodoi{10.1093/mnras/sty2955}

\bibitem[{{Majewski} {et~al.}(2016){Majewski}, {APOGEE Team}, \& {APOGEE-2
  Team}}]{Majewski2016AN....337..863M}
{Majewski}, S.~R., {APOGEE Team}, \& {APOGEE-2 Team}. 2016, Astronomische
  Nachrichten, 337, 863, \dodoi{10.1002/asna.201612387}

\bibitem[{{Martin} {et~al.}(2023){Martin}, {Starkenburg}, {Yuan}, {Fouesneau},
  {Arentsen}, {De Angeli}, {Gran}, {Montelius}, {Andrae}, {Bellazzini},
  {Montegriffo}, {Esselink}, {Zhang}, {Venn}, {Viswanathan}, {Aguado},
  {Battaglia}, {Bayer}, {Bonifacio}, {Caffau}, {C{\^o}t{\'e}}, {Carlberg},
  {Fabbro}, {Fern{\'a}ndez Alvar}, {Gonz{\'a}lez Hern{\'a}ndez}, {Gonz{\'a}lez
  Rivera de La Vernhe}, {Hill}, {Ibata}, {Jablonka}, {Kordopatis}, {Lardo},
  {McConnachie}, {Navarrete}, {Navarro}, {Recio-Blanco}, {S{\'a}nchez Janssen},
  {Sestito}, {Thomas}, {Vitali}, \& {Youakim}}]{Martin2023arXiv230801344M}
{Martin}, N.~F., {Starkenburg}, E., {Yuan}, Z., {et~al.} 2023, arXiv e-prints,
  arXiv:2308.01344, \dodoi{10.48550/arXiv.2308.01344}

\bibitem[{{McMillan}(2017)}]{McMillan2017}
{McMillan}, P.~J. 2017, \mnras, 465, 76, \dodoi{10.1093/mnras/stw2759}

\bibitem[{{Montegriffo} {et~al.}(2023){Montegriffo}, {De Angeli}, {Andrae},
  {Riello}, {Pancino}, {Sanna}, {Bellazzini}, {Evans}, {Carrasco}, {Sordo},
  {Busso}, {Cacciari}, {Jordi}, {van Leeuwen}, {Vallenari}, {Altavilla},
  {Barstow}, {Brown}, {Burgess}, {Castellani}, {Cowell}, {Davidson}, {De
  Luise}, {Delchambre}, {Diener}, {Fabricius}, {Fr{\'e}mat}, {Fouesneau},
  {Gilmore}, {Giuffrida}, {Hambly}, {Harrison}, {Hidalgo}, {Hodgkin},
  {Holland}, {Marinoni}, {Osborne}, {Pagani}, {Palaversa}, {Piersimoni},
  {Pulone}, {Ragaini}, {Rainer}, {Richards}, {Rowell}, {Ruz-Mieres}, {Sarro},
  {Walton}, \& {Yoldas}}]{Montegriffo2023A&A...674A...3M}
{Montegriffo}, P., {De Angeli}, F., {Andrae}, R., {et~al.} 2023, \aap, 674, A3,
  \dodoi{10.1051/0004-6361/202243880}

\bibitem[{{Rix} {et~al.}(2022){Rix}, {Chandra}, {Andrae}, {Price-Whelan},
  {Weinberg}, {Conroy}, {Fouesneau}, {Hogg}, {De Angeli}, {Naidu}, {Xiang}, \&
  {Ruz-Mieres}}]{Rix2022ApJ...941...45R}
{Rix}, H.-W., {Chandra}, V., {Andrae}, R., {et~al.} 2022, \apj, 941, 45,
  \dodoi{10.3847/1538-4357/ac9e01}

\bibitem[{{Sanders} \& {Matsunaga}(2023)}]{Sanders2023MNRAS.521.2745S}
{Sanders}, J.~L., \& {Matsunaga}, N. 2023, \mnras, 521, 2745,
  \dodoi{10.1093/mnras/stad574}

\bibitem[{{Schlafly} \& {Finkbeiner}(2011)}]{Schlafly2011ApJ...737..103S}
{Schlafly}, E.~F., \& {Finkbeiner}, D.~P. 2011, \apj, 737, 103,
  \dodoi{10.1088/0004-637X/737/2/103}

\bibitem[{{Schlegel} {et~al.}(1998){Schlegel}, {Finkbeiner}, \&
  {Davis}}]{Schlegel1998ApJ...500..525S}
{Schlegel}, D.~J., {Finkbeiner}, D.~P., \& {Davis}, M. 1998, \apj, 500, 525,
  \dodoi{10.1086/305772}

\bibitem[{{Steinmetz} {et~al.}(2006){Steinmetz}, {Zwitter}, {Siebert},
  {Watson}, {Freeman}, {Munari}, {Campbell}, {Williams}, {Seabroke}, {Wyse},
  {Parker}, {Bienaym{\'e}}, {Roeser}, {Gibson}, {Gilmore}, {Grebel}, {Helmi},
  {Navarro}, {Burton}, {Cass}, {Dawe}, {Fiegert}, {Hartley}, {Russell},
  {Saunders}, {Enke}, {Bailin}, {Binney}, {Bland-Hawthorn}, {Boeche}, {Dehnen},
  {Eisenstein}, {Evans}, {Fiorucci}, {Fulbright}, {Gerhard}, {Jauregi}, {Kelz},
  {Mijovi{\'c}}, {Minchev}, {Parmentier}, {Pe{\~n}arrubia}, {Quillen}, {Read},
  {Ruchti}, {Scholz}, {Siviero}, {Smith}, {Sordo}, {Veltz}, {Vidrih}, {von
  Berlepsch}, {Boyle}, \& {Schilbach}}]{Steinmetz2006AJ....132.1645S}
{Steinmetz}, M., {Zwitter}, T., {Siebert}, A., {et~al.} 2006, \aj, 132, 1645,
  \dodoi{10.1086/506564}

\bibitem[{{Takada} {et~al.}(2014){Takada}, {Ellis}, {Chiba}, {Greene},
  {Aihara}, {Arimoto}, {Bundy}, {Cohen}, {Dor{\'e}}, {Graves}, {Gunn},
  {Heckman}, {Hirata}, {Ho}, {Kneib}, {Le F{\`e}vre}, {Lin}, {More},
  {Murayama}, {Nagao}, {Ouchi}, {Seiffert}, {Silverman}, {Sodr{\'e}},
  {Spergel}, {Strauss}, {Sugai}, {Suto}, {Takami}, \&
  {Wyse}}]{Takada2014PASJ...66R...1T}
{Takada}, M., {Ellis}, R.~S., {Chiba}, M., {et~al.} 2014, \pasj, 66, R1,
  \dodoi{10.1093/pasj/pst019}

\bibitem[{{Tolstoy} {et~al.}(2009){Tolstoy}, {Hill}, \&
  {Tosi}}]{Tolstoy2009ARA&A..47..371T}
{Tolstoy}, E., {Hill}, V., \& {Tosi}, M. 2009, \araa, 47, 371,
  \dodoi{10.1146/annurev-astro-082708-101650}

\bibitem[{{van der Walt} {et~al.}(2011){van der Walt}, {Colbert}, \&
  {Varoquaux}}]{vanderWalt2011}
{van der Walt}, S., {Colbert}, S.~C., \& {Varoquaux}, G. 2011, Computing in
  Science Engineering, 13, 22, \dodoi{10.1109/MCSE.2011.37}

\bibitem[{{Vasiliev}(2019)}]{Vasiliev2019_AGAMA}
{Vasiliev}, E. 2019, \mnras, 482, 1525, \dodoi{10.1093/mnras/sty2672}

\bibitem[{{Wang} {et~al.}(2022){Wang}, {Huang}, {Yuan}, {Zhang}, {Xiang}, \&
  {Liu}}]{Wang2022ApJS..259...51W}
{Wang}, C., {Huang}, Y., {Yuan}, H., {et~al.} 2022, \apjs, 259, 51,
  \dodoi{10.3847/1538-4365/ac4df7}

\bibitem[{{Wang} \& {Chen}(2019)}]{Wang2019ApJ...877..116W}
{Wang}, S., \& {Chen}, X. 2019, \apj, 877, 116,
  \dodoi{10.3847/1538-4357/ab1c61}

\bibitem[{{Witten} {et~al.}(2022){Witten}, {Aguado}, {Sanders}, {Belokurov},
  {Evans}, {Koposov}, {Allende Prieto}, {De Angeli}, \&
  {Irwin}}]{Witten2022MNRAS.516.3254W}
{Witten}, C. E.~C., {Aguado}, D.~S., {Sanders}, J.~L., {et~al.} 2022, \mnras,
  516, 3254, \dodoi{10.1093/mnras/stac2273}

\bibitem[{{Xylakis-Dornbusch} {et~al.}(2024){Xylakis-Dornbusch}, {Christlieb},
  {Hansen}, {Nordlander}, {Webber}, \&
  {Marshall}}]{Xylakis-Dornbusch2024arXiv240308454X}
{Xylakis-Dornbusch}, T., {Christlieb}, N., {Hansen}, T.~T., {et~al.} 2024,
  arXiv e-prints, arXiv:2403.08454, \dodoi{10.48550/arXiv.2403.08454}

\bibitem[{{Yang} {et~al.}(2022){Yang}, {Yuan}, {Xiang}, {Duan}, {Huang}, {Liu},
  {Beers}, {Galarza}, {Daflon}, {Fern{\'a}ndez-Ontiveros}, {Cenarro},
  {Crist{\'o}bal-Hornillos}, {Hern{\'a}ndez-Monteagudo}, {L{\'o}pez-Sanjuan},
  {Mar{\'\i}n-Franch}, {Moles}, {Varela}, {V{\'a}zquez Rami{\'o}}, {Alcaniz},
  {Dupke}, {Ederoclite}, {Sodr{\'e}}, \& {Angulo}}]{Yang2022A&A...659A.181Y}
{Yang}, L., {Yuan}, H., {Xiang}, M., {et~al.} 2022, \aap, 659, A181,
  \dodoi{10.1051/0004-6361/202142724}

\bibitem[{{Yanny} {et~al.}(2009){Yanny}, {Rockosi}, {Newberg}, {Knapp},
  {Adelman-McCarthy}, {Alcorn}, {Allam}, {Allende Prieto}, {An}, {Anderson},
  {Anderson}, {Bailer-Jones}, {Bastian}, {Beers}, {Bell}, {Belokurov},
  {Bizyaev}, {Blythe}, {Bochanski}, {Boroski}, {Brinchmann}, {Brinkmann},
  {Brewington}, {Carey}, {Cudworth}, {Evans}, {Evans}, {Gates}, {G{\"a}nsicke},
  {Gillespie}, {Gilmore}, {Nebot Gomez-Moran}, {Grebel}, {Greenwell}, {Gunn},
  {Jordan}, {Jordan}, {Harding}, {Harris}, {Hendry}, {Holder}, {Ivans},
  {Ivezi{\v{c}}}, {Jester}, {Johnson}, {Kent}, {Kleinman}, {Kniazev},
  {Krzesinski}, {Kron}, {Kuropatkin}, {Lebedeva}, {Lee}, {French Leger},
  {L{\'e}pine}, {Levine}, {Lin}, {Long}, {Loomis}, {Lupton}, {Malanushenko},
  {Malanushenko}, {Margon}, {Martinez-Delgado}, {McGehee}, {Monet}, {Morrison},
  {Munn}, {Neilsen}, {Nitta}, {Norris}, {Oravetz}, {Owen}, {Padmanabhan},
  {Pan}, {Peterson}, {Pier}, {Platson}, {Re Fiorentin}, {Richards}, {Rix},
  {Schlegel}, {Schneider}, {Schreiber}, {Schwope}, {Sibley}, {Simmons},
  {Snedden}, {Allyn Smith}, {Stark}, {Stauffer}, {Steinmetz}, {Stoughton},
  {SubbaRao}, {Szalay}, {Szkody}, {Thakar}, {Sivarani}, {Tucker}, {Uomoto},
  {Vanden Berk}, {Vidrih}, {Wadadekar}, {Watters}, {Wilhelm}, {Wyse}, {Yarger},
  \& {Zucker}}]{Yanny2009AJ....137.4377Y}
{Yanny}, B., {Rockosi}, C., {Newberg}, H.~J., {et~al.} 2009, \aj, 137, 4377,
  \dodoi{10.1088/0004-6256/137/5/4377}

\bibitem[{{Yao} {et~al.}(2024){Yao}, {Ji}, {Koposov}, \&
  {Limberg}}]{Yao2024MNRAS.52710937Y}
{Yao}, Y., {Ji}, A.~P., {Koposov}, S.~E., \& {Limberg}, G. 2024, \mnras, 527,
  10937, \dodoi{10.1093/mnras/stad3775}

\bibitem[{{Zhang} {et~al.}(2023){Zhang}, {Green}, \&
  {Rix}}]{Zhang2023MNRAS.524.1855Z}
{Zhang}, X., {Green}, G.~M., \& {Rix}, H.-W. 2023, \mnras, 524, 1855,
  \dodoi{10.1093/mnras/stad1941}

\bibitem[{{Zhao} {et~al.}(2012){Zhao}, {Zhao}, {Chu}, {Jing}, \&
  {Deng}}]{Zhao2012RAA....12..723Z}
{Zhao}, G., {Zhao}, Y.-H., {Chu}, Y.-Q., {Jing}, Y.-P., \& {Deng}, L.-C. 2012,
  Research in Astronomy and Astrophysics, 12, 723,
  \dodoi{10.1088/1674-4527/12/7/002}

\end{thebibliography}

\appendix

\begin{figure*}
\centering 
\includegraphics[width=0.48\textwidth ]{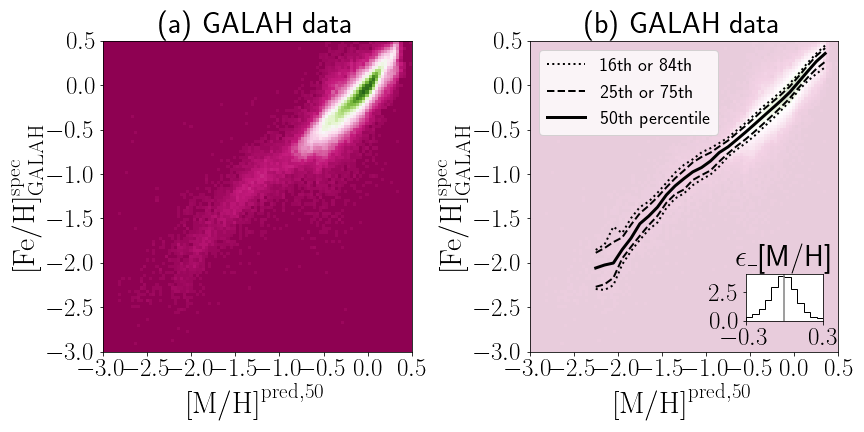} 
\includegraphics[width=0.48\textwidth ]{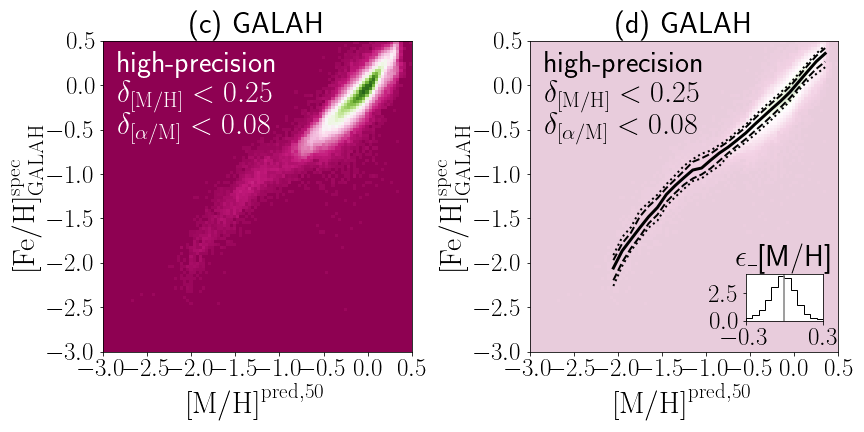} \\
\includegraphics[width=0.48\textwidth ]{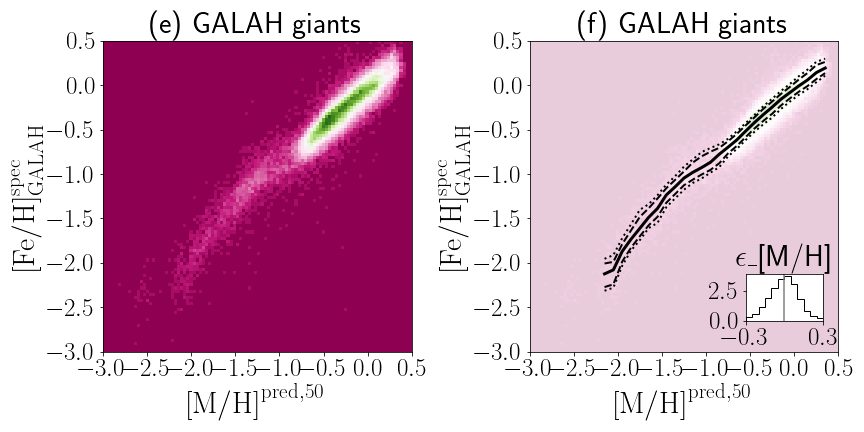} 
\includegraphics[width=0.48\textwidth ]{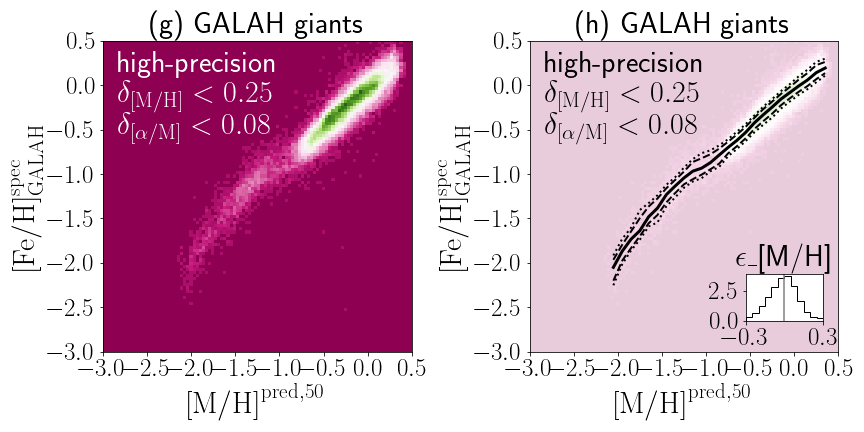} \\
\includegraphics[width=0.48\textwidth ]{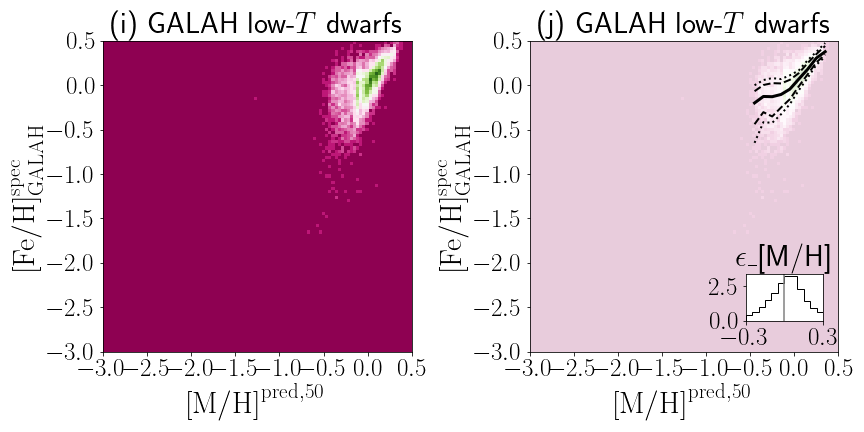} 
\includegraphics[width=0.48\textwidth ]{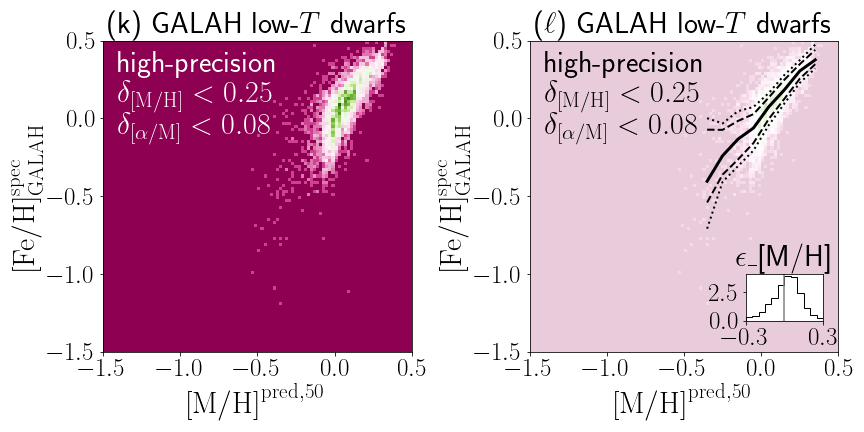} \\
\includegraphics[width=0.48\textwidth ]{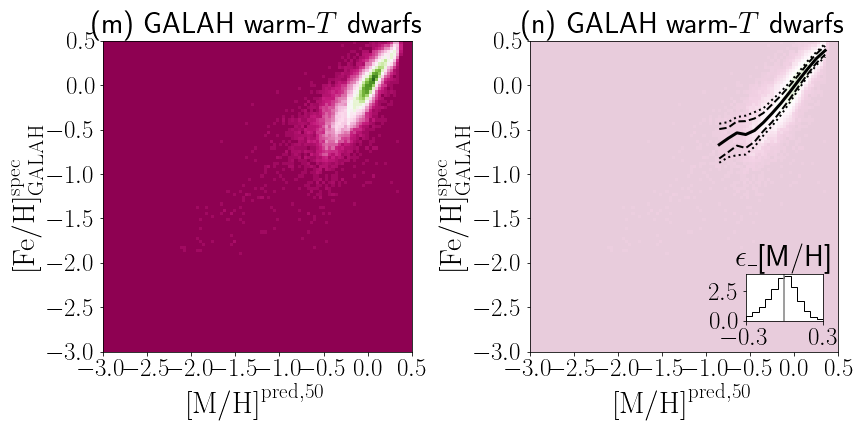} 
\includegraphics[width=0.48\textwidth ]{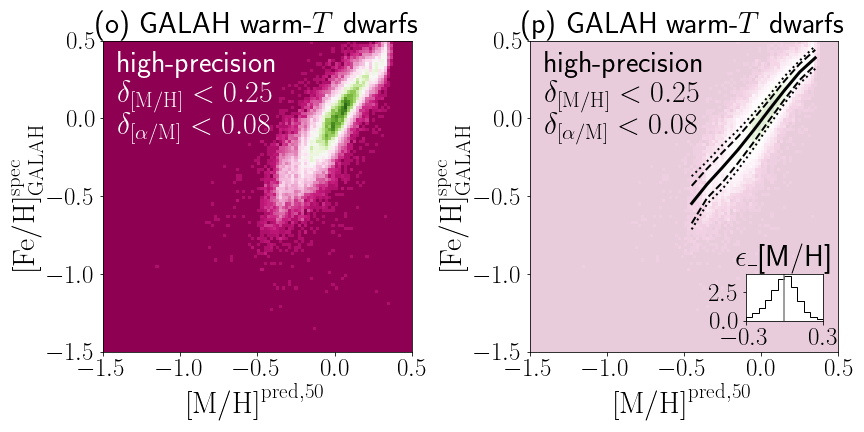} 
\caption{
The same as Fig.~\ref{fig:test_data_mh}, but showing 
the detailed performance of the QRF-MH model (our model for estimating \mh) 
applied to the GALAH data. 
}
\label{fig:GALAH_data_mh}
\end{figure*}

\begin{figure*}
\centering 
\includegraphics[width=0.48\textwidth ]{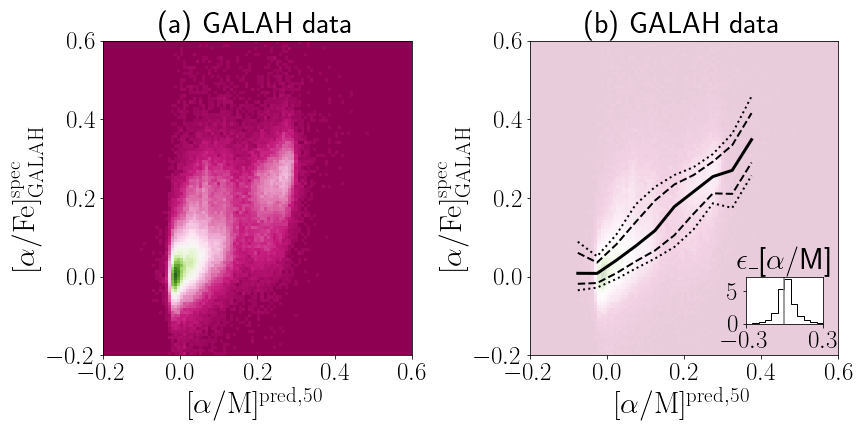} 
\includegraphics[width=0.48\textwidth ]{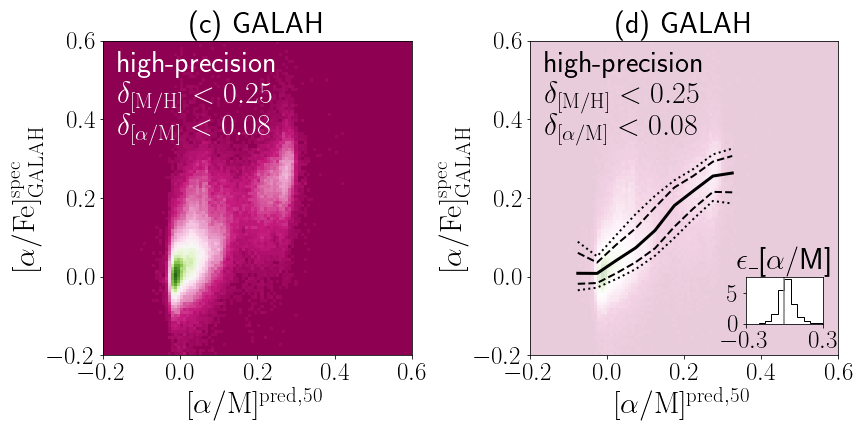} \\
\includegraphics[width=0.48\textwidth ]{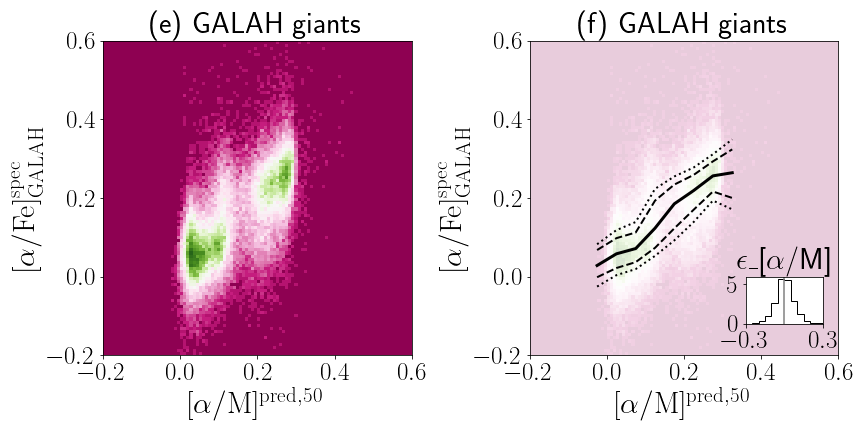} 
\includegraphics[width=0.48\textwidth ]{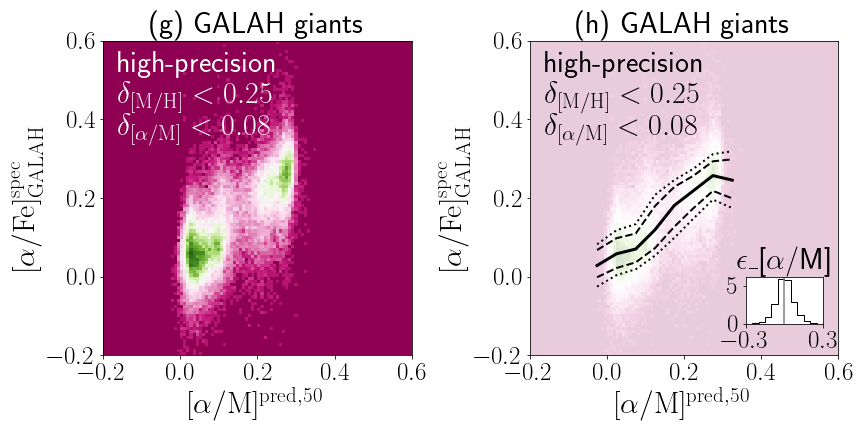} \\
\includegraphics[width=0.48\textwidth ]{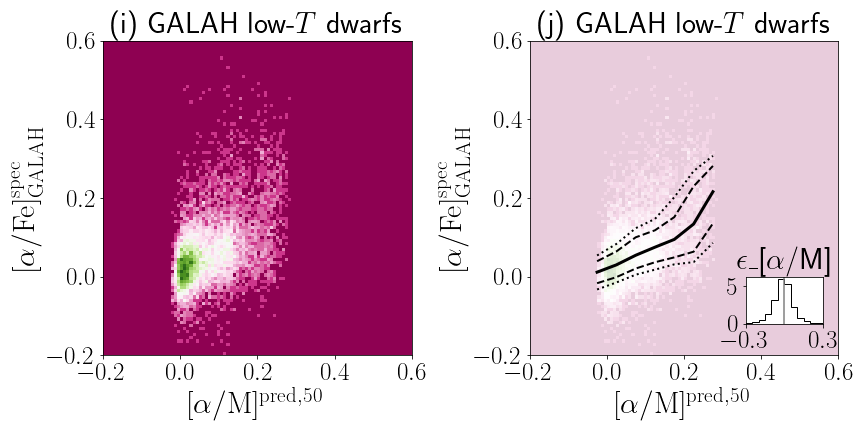} 
\includegraphics[width=0.48\textwidth ]{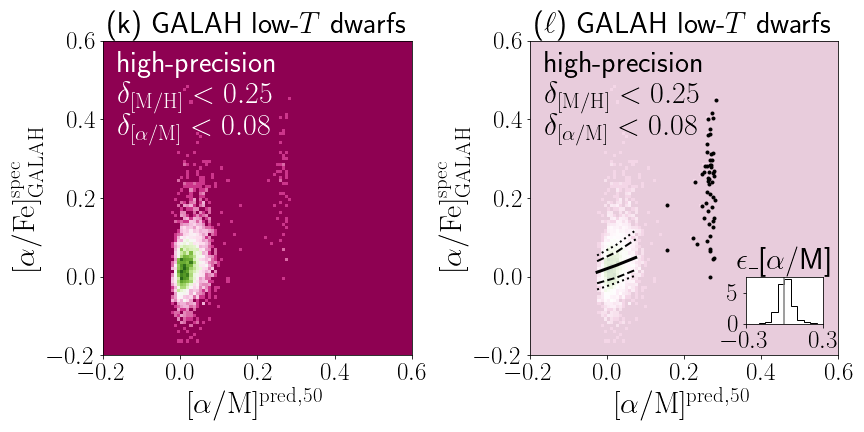} \\
\includegraphics[width=0.48\textwidth ]{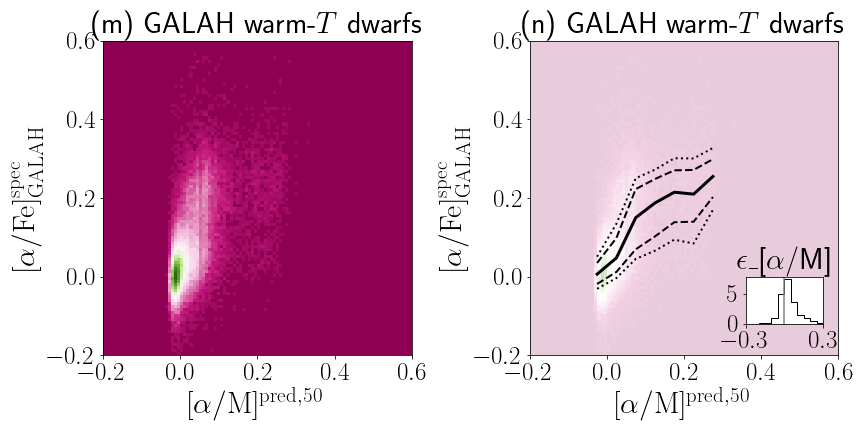} 
\includegraphics[width=0.48\textwidth ]{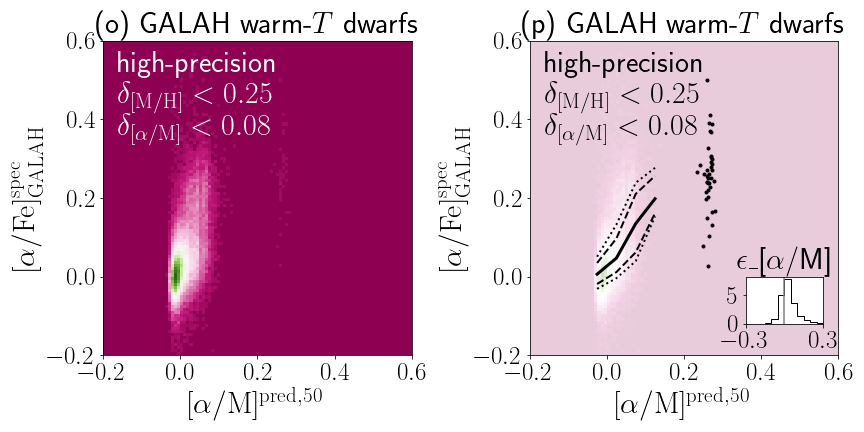} 
\caption{
The same as Fig.~\ref{fig:test_data_mh}, but showing 
the detailed performance of the QRF-AM model (our model for estimating \am) 
applied to the GALAH data. 
In panels ($\ell$) and (p), 
we plot the stars with $\am^\mathrm{pred,50}> 0.15$ with black dots 
to highlight that the majority of them  
are genuine high-$\alpha$ stars ($\am^\mathrm{spec}_\mathrm{GALAH} > 0.15$). 
}
\label{fig:GALAH_data_am}
\end{figure*}

\section{Validation of our model using the external test data from GALAH DR3}
\label{sec:test_GALAH}

Here we conduct an additional test of our QRF-MH and QRF-AM models 
with the external test data taken from the GALAH DR3 
described in Section \ref{sec:GALAH_DR3_data}. 
The analyses are the same as those in 
Sections \ref{sec:accuracy_mh_giant_dwarf} and \ref{sec:accuracy_am_giant_dwarf}, 
but using the GALAH data set. 
The results are shown in Figs.~
\ref{fig:GALAH_data_mh} and \ref{fig:GALAH_data_am}. 
We note that 
we treat [Fe/H] and \afe\ in GALAH DR3 
as the proxies for \mh\ and \am, respectively.

\section{Correlation between Na and Mg} \label{sec:Na_Mg}

\begin{figure*}
\centering 
\includegraphics[width=0.9\textwidth ]{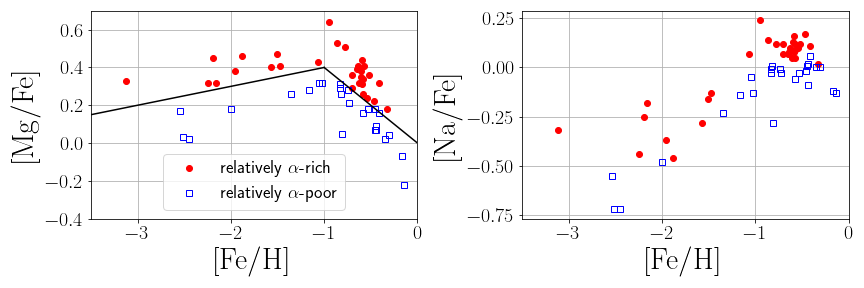} 
\caption{
The observed chemical distribution of nearby metal-poor stars 
taken from \cite{Gehren2006A&A...451.1065G}. 
We see a correlation between Na and Mg, 
such that [Mg/Fe]-enhanced stars are likely [Na/Fe]-enhanced for a fixed value of [Fe/H]. 
}
\label{fig:Na_and_Mg}
\end{figure*}

Here we briefly describe the observed 
correlation between Na and Mg abundances. 
Fig.~\ref{fig:Na_and_Mg} shows the distribution of ([Fe/H], [Mg/Fe]) 
and ([Fe/H], [Na/Fe]) of nearby stars taken from 
the non-local thermodynamic equilibrium analysis by 
\cite{Gehren2006A&A...451.1065G}. 
To guide the eyes, 
we divide the sample stars into two groups, 
by using a simple boundary shown in Fig.~\ref{fig:Na_and_Mg}(a). 
By definition, for a given [Fe/H], 
the relatively $\alpha$-rich group is always 
have higher [Mg/Fe] than 
the relatively $\alpha$-poor group. 
We see from Fig.~\ref{fig:Na_and_Mg}(b) that, 
for a given [Fe/H], 
the relatively $\alpha$-rich group tend to 
have higher [Na/Fe] than 
the relatively $\alpha$-poor group.

\section{Coordinate system} \label{appendix:coordinate}

We adopt a right-handed Galactocentric Cartesian coordinate system $(x,y,z)$, in which the $(x,y)$-plane is the Galactic disk plane. 
The position of the Sun is assumed to be $\vector{x}_\odot = (x_\odot,y_\odot,z_\odot) = (-R_\odot,0,z_\odot)$, with $R_\odot = 8.277 \kpc$ \citep{GRAVITY2022A&A...657L..12G} and 
$z_\odot = 0.0208 \kpc$ \citep{Bennett2019MNRAS.482.1417B}. 
The velocity of the Sun with respect to the Galactic rest frame is assumed to be 
$\vector{v}_\odot = (v_{x,\odot},v_{y,\odot},v_{z,\odot}) = (9.3, 251.5, 8.59) \kms$. 

\revise{
In this paper, we use Gaia DR3 data to derive kinematical properties of stars. 
We correct the parallax $\varpi$ by using a constant zero-point offset, 
$\varpi_\mathrm{corrected} = \varpi - (-0.017 \; \mathrm{mas})$ 
following \citep{Lindegren2021A&A...649A...4L}. 
The heliocentric distance is then simply calculated as  $d=1/\varpi_\mathrm{corrected}$. 
For sky coordinates, proper motion, and radial velocity, 
we use the cataloged values and neglect the observational uncertainties. 
We combine the 6D stellar data with the assumed Solar position and velocity 
to compute the 3D position $\vector{x}$ and 3D velocity $\vector{v}$ of the stars. 
}
When we evaluate the orbital eccentricity of stars, 
we assume the Galactic potential model in \cite{McMillan2017}.

\section{A comment on the flag in our catalog}
\label{appendix:flag}

\revise{
To demonstrate the usefulness of \texttt{{bool\_flag\_cmd\_good}} in our catalog, 
here we consider two example stars: 
a giant star A with 
$(G_\mathrm{BP}-G_\mathrm{RP}), M_\mathrm{G})=(1,0)$ 
and a dwarf star B with 
$(G_\mathrm{BP}-G_\mathrm{RP}, M_\mathrm{G})=(1,5)$. 
These color and magnitude values are representative of the stars in our catalog. 
If these stars appear in our catalog, 
we can safely assume that their parallax uncertainty is 
at most 0.12 mas (or smaller), 
ased on two factors: 
(i) most stars with Gaia XP spectra have $G<18$, 
and (ii) stars with $G<18$ typically have parallax uncertainty smaller than 0.12 mas \citep{Lindegren2021A&A...649A...4L}. 
Given their color, 
these stars are classified as \texttt{{bool\_flag\_cmd\_good=True}} 
if the point-estimate of the $G$-band absolute magnitude 
is brighter than 8. 
In the following, we show that $M_\mathrm{G,0}<8$ 
is satisfied even under extreme conditions. 
}

\revise{
Let us begin with giant star A. 
If this star has $G=18$, 
its heliocentric distance is 39.8 \kpc, and  
the true parallax is $0.025$ mas. 
By using the relationship 
$M_\mathrm{G,0} = G - 5 \log_{10} (1000/(\varpi / \mathrm{mas})) + 5$, 
we observe that 
$M_\mathrm{G,0}<8$ is satisfied as long as 
the point-estimate of $\varpi$ is less than 1 mas. 
Given the true parallax of 0.025 mas and 
the parallax uncertainty of 0.12 mas, 
the condition $\varpi < 1$ mas is violated 
only in the case of an extreme outlier 
(about an $8 \sigma$ overestimate of the parallax). 
For brighter stars, where the parallax uncertainty is smaller, an even more extreme outlier would be needed to violate $\varpi < 1$ mas. 
Therefore, a typical giant star, such as giant A, 
is almost always flagged as
\texttt{{bool\_flag\_cmd\_good}=True} in our catalog. 
}

\revise{
Next let us consider dwarf star B. 
If this star has $G=18$, 
its heliocentric distance is 3.98 \kpc, and  
the true parallax is $0.25$ mas. 
Following the same reasoning, 
the condition $M_\mathrm{G,0}<8$ is satisfied for dwarf B as long as 
the point-estimate of $\varpi$ is less than 1 mas. 
Given the true parallax of 0.25 mas and 
the parallax uncertainty of 0.12 mas, 
the condition $\varpi < 1$ mas is violated 
is violated only in the case of an extreme outlier 
(around a 6 $\sigma$ overestimate of the parallax).
Therefore, a typical dwarf star, such as dwarf B, 
is almost always flagged as 
\texttt{{bool\_flag\_cmd\_good}=True} in our catalog. 
}

\section{Binary effect}
\label{appendix:binary}

\revise{
As hinted from the binary sequence (a secondary stripe above the main sequence)
in the CMD in Fig.~\ref{fig:local_disc_dwarf_lowT}(a), 
some fraction of stars in our catalog are binary systems. 
If the secondary star in a stellar binary is bright enough to significantly contribute to the XP spectrum, 
we expect that our model prediction 
(for the chemical abundances of the primary star) 
may be negatively impacted. 
To assess the effect of binarity on our predictions, 
we perform additional tests in which we divide the test data in terms of 
\texttt{ruwe} and
\texttt{ipd\_frac\_multi\_peak}. 
}

\revise{
First, we divide the test data into two subsamples 
with (i) \texttt{ruwe}$<2$ (95\% of the test data) and (ii) \texttt{ruwe}$>2$ (5\% of the test data). 
Here, the subsample (ii) corresponds to objects with astrometric evidence of binarity. 
We find that:
\begin{itemize}
\item
The RMSE values for \mh\ are 0.0876 dex for subsample (i) 
and 0.1124 dex for subsample (ii).  
\item 
The RMSE values for \am\ are 0.0433 dex for subsample (i) 
and 0.0474 dex for subsample (ii). 
\end{itemize}
}

\revise{
Next, we divide the test data into two subsamples 
with (iii) \texttt{ipd\_frac\_multi\_peak}$<2$ (96\% of the test data) and (iv) \texttt{ipd\_frac\_multi\_peak}$\geq2$ (4\% of the test data). 
The subsample (iv) corresponds to binary systems in which the two stars are barely resolved. 
We find that 
\begin{itemize}
\item
The RMSE values for \mh\ are 0.0847 dex for subsample (iii) 
and 0.1569 dex for subsample (iv). 
\item 
The RMSE values for \am\ are 0.0432 dex for subsample (iii) 
and 0.0498 dex for subsample (iv). 
\end{itemize}
}

\revise{
By comparing these numbers with the RMSE values 
in Table~\ref{table:rmse} (0.0890 dex for \mh\ and 0.0436 dex for \am), 
we infer that our model prediction deteriorates 
for stars in binary systems. 
}

\section{Dimensionality of the input data}
\label{appendix:dimension}

\revise{
The information of Gaia XP spectra 
is provided by 110 dimensional coefficients $\vector{C}$ in Gaia DR3. 
As described in Fig.~\ref{fig:XPschematic}, 
we convert these coefficients to reconstruct the XP spectra 
and sample the flux at 1200 points in the pseudo-wavelength domain, 
$(\vector{F}_\mathrm{BP}, \vector{F}_\mathrm{RP})$. 
In the main part of this paper, 
we use the combined 1310 dimensional data 
$(\vector{F}_\mathrm{BP}, \vector{F}_\mathrm{RP}, \vector{C})$ 
as the input for our ML models. 
Here, we summarize how the performance of these models varies with different input data dimensionalities.
}

\revise{
A simple and likely the most natural choice for the input data 
is to use $\vector{C}$, 
because all information about the XP spectra is derived from these coefficients. 
We construct the QRF models to infer \mh\ and \am\ 
by using $\vector{C}$ as the input. 
In this case, the RMSE values are 
$0.0924$ dex for \mh\ and 
$0.0438$ dex for \am. 
}

\revise{
To better interpret how our ML models work, it is practical to use the flux data as a function of wavelength as input. 
In Section \ref{sec:interpret}, 
we use $(\vector{F}_\mathrm{BP}, \vector{F}_\mathrm{RP})$ as the input. 
In this case, the RMSE values are 
$0.119$ dex for \mh\ and 
$0.0556$ dex for \am. 
We observe that the performance with this input is slightly worse compared to using $\vector{C}$. 
This result is intriguing, 
considering that the sampled spectrum has much larger dimensionality. 
However, it makes some sense because 
all the information in $(\vector{F}_\mathrm{BP}, \vector{F}_\mathrm{RP})$ 
is encapsulated in $\vector{C}$.
}

\revise{
In the main part of this paper, 
we use $(\vector{F}_\mathrm{BP}, \vector{F}_\mathrm{RP}, \vector{C})$ as the input. 
This configuration yields the best performance, with RMSE values of 
$0.0890$ dex for \mh\ and 
$0.0436$ dex for \am. 
}

\revise{
The improvement in RMSE values by including 
$(\vector{F}_\mathrm{BP}, \vector{F}_\mathrm{RP})$ 
in addition to $\vector{C}$ 
suggests that the flux data 
$(\vector{F}_\mathrm{BP}, \vector{F}_\mathrm{RP})$ 
are valuable for estimating \mh\ and \am. 
This might be counterintuitive, 
because $(\vector{F}_\mathrm{BP}, \vector{F}_\mathrm{RP})$ 
are derived from $\vector{C}$ 
and therefore these data are redundant. 
}

\revise{
We hypothesize that this counterintuitive result stems from the way QRF models sort information to make inferences. 
Specifically, QRF is a tree-based method that makes splits along feature dimensions. 
For QRF to effectively capture information related to \mh\ or \am\ in the input data, that information must be `easily accessible.' 
Therefore, the structure of the input data is crucial, and presenting the same information in multiple formats can improve performance.
}

\end{document}